%% file: main.tex
\documentclass[format=acmsmall,screen=true,review=false]{acmart}
\acmJournal{TOSEM}
\AtBeginDocument{%
  }
\input{tablefigures}
\setcopyright{acmlicensed}
\copyrightyear{2025}
\acmYear{2025}
\acmDOI{XXXXXXX.XXXXXXX}
\acmISBN{978-1-4503-XXXX-X/2025/11}

\usepackage{siunitx}
\usepackage{hyperref}
\usepackage{pgf-pie}
\usepackage{graphicx}
\usepackage{blkarray}
\usepackage{standalone}
\usepackage{amsmath}
\usepackage{csquotes}
\usepackage{xcolor} 
\usepackage{subfig}
\usepackage{pgfplots}
\usepackage{pgfplotstable}
\usepackage{adjustbox}
\usepackage{booktabs}
\usepackage{csquotes}
\usepackage{multirow}
\usepackage[figuresright]{rotating}
\usepackage[most]{tcolorbox}
\usepackage{colortbl}
\usepackage{xcolor}
\usepackage{balance}
\usepackage{url}
\usepackage{ifxetex,ifluatex}
\usepackage{etoolbox}
\usepackage{tikz}
\usepackage{enumitem}
\usepackage{framed}

\DeclareMathAlphabet\mathbfcal{OMS}{cmsy}{b}{n}
\pgfplotsset{compat=newest}

\usepackage{pifont}

\definecolor{steel}{rgb}{0.1, 0.2, 0.8} 
\definecolor{lightblue}{rgb}{0.4, 0.6, 1.0}
\definecolor{mediumblue}{rgb}{0.2, 0.4, 0.8}
\definecolor{lightgray}{rgb}{0.8, 0.8, 0.8}
\definecolor{lightred}{rgb}{1.0, 0.6, 0.6}
\definecolor{mediumred}{rgb}{0.9, 0.3, 0.3}
\definecolor{darkred}{rgb}{0.7, 0.1, 0.1}
\definecolor{gre}{RGB}{101, 191, 127}
\definecolor{gree}{RGB}{7, 135, 44}

\newtcolorbox{quotebox}{colback=black!10,boxrule=0.4pt,colframe=black,fonttitle=\bfseries,top=2pt,bottom=2pt}

\newcommand{\keybox}[1]{
\begin{tcolorbox}[colback=white,leftrule=1mm,toprule=0mm,bottomrule=0mm,left=1pt,right=2pt,top=2pt,bottom=2pt]
\em #1
\end{tcolorbox}
}

\definecolor{mycolor}{rgb}{0.122, 0.435, 0.698}
\newtcbox{\mytag}{nobeforeafter,colframe=mycolor,colback=mycolor!30!white,boxrule=0.7pt,arc=0pt,
 boxsep=-3pt,left=6pt,right=6pt,top=4pt,bottom=5pt,tcbox raise base}

\newcounter{findingcount}
\newcounter{insightcount}
\newcommand{\revisioncolor}{black} 
\newcommand{\revision}[1]{{\color{\revisioncolor}#1}}

\newtcolorbox{keytheory}{colback=green!10,enhanced,title=Theory of Model Usefulness for Tuning,
	attach boxed title to top left={xshift=0mm},boxrule=0pt,after skip=1cm,before skip=1cm,right skip=0cm,breakable,fonttitle=\bfseries,toprule=0pt,bottomrule=0pt,rightrule=0pt,leftrule=4pt,arc=0mm,skin=enhancedlast jigsaw,sharp corners,colframe=gree,colbacktitle=gre,boxed title style={
		frame code={ 
			\fill[gre](frame.south west)--(frame.north west)--(frame.north east)--([xshift=3mm]frame.east)--(frame.south east)--cycle;
			\draw[line width=1mm,gre]([xshift=2mm]frame.north east)--([xshift=5mm]frame.east)--([xshift=2mm]frame.south east);
			
			\draw[line width=1mm,gre]([xshift=5mm]frame.north east)--([xshift=8mm]frame.east)--([xshift=5mm]frame.south east);
		}
	}
}

\makeatletter
\tcbset{
    myhbox/.style 2 args={%
        enhanced, 
        breakable,
        colback=white,
        colframe=black!60,
        attach boxed title to top left={yshift*=-\tcboxedtitleheight}, 
        title={#2},
        boxed title size=title,
        boxed title style={%
            sharp corners, 
            rounded corners=northwest, 
            colback=tcbcolframe, 
            boxrule=0pt,
        },
        underlay boxed title={%
            \path[fill=tcbcolframe] (title.south west)--(title.south east) 
                to[out=0, in=180] ([xshift=5mm]title.east)--
                (title.center-|frame.east)
                [rounded corners=\kvtcb@arc] |- 
                (frame.north) -| cycle; 
        },
        #1
    }
}   
\makeatother

\newtcolorbox{myhbox}[2][]{%
    myhbox={#1}{#2}
}

\newcommand{\approach}{\texttt{Model4Tune}} %

\mathchardef\mhyphen="2D
\newcommand{\vect}[1]{\boldsymbol{#1}}
\DeclareMathAlphabet\mathbfcal{OMS}{cmsy}{b}{n}

\newcolumntype{P}[1]{>{\centering\arraybackslash}m{#1}}
\newcolumntype{Y}{>{\centering\arraybackslash}X}

   \usepackage[utf8]{inputenc}
   \usepackage[T1]{fontenc}
   \newcommand*\quotefont{\fontfamily{LinuxLibertineT-LF}} 

\newcommand*\quotesize{30} 
\newcommand*{\openquote}
   {\tikz[remember picture,overlay,xshift=-3ex,yshift=-0.5ex]
   \node (OQ) {\quotefont\fontsize{\quotesize}{\quotesize}\selectfont``};\kern0pt}

\newcommand*{\closequote}[1]
  {\tikz[remember picture,overlay,xshift=3ex,yshift={#1}]
   \node (CQ) {\quotefont\fontsize{\quotesize}{\quotesize}\selectfont''};}

\colorlet{shadecolor}{white}

\renewcommand{\mytag}[1]{}

\newcommand*\shadedauthorformat{\emph} 

\newcommand*\authoralign[1]{%
  \if#1l
    \def\authorfill{}\def\quotefill{\hfill}
  \else
    \if#1r
      \def\authorfill{\hfill}\def\quotefill{}
    \else
      \if#1c
        \gdef\authorfill{\hfill}\def\quotefill{\hfill}
      \else\typeout{Invalid option}
      \fi
    \fi
  \fi}
{\authoralign{#1}
\ifblank{#2}
   {\def\shadequoteauthor{}\def\yshift{-2ex}\def\quotefill{\hfill}}
   {\def\shadequoteauthor{\par\authorfill\shadedauthorformat{#2}}\def\yshift{1ex}}
\begin{snugshade}\begin{quote}\openquote}
{\shadequoteauthor\quotefill\closequote{\yshift}\end{quote}\end{snugshade}}

\begin{document}


\title{Evaluating Useful Surrogate Models for Configuration Tuning Beyond Accuracy: A Fitness Landscape Analysis Perspective}

\author{Pengzhou Chen}
\authornote{Pengzhou Chen and Hongyuan Liang are also supervised in the IDEAS Lab.}
\affiliation{%
  \institution{University of Electronic Science and Technology of China}
  \city{Chengdu}
  \country{China}
}

\author{Hongyuan Liang}
\authornotemark[1]
\affiliation{%
  \institution{University of Electronic Science and Technology of China}
  \city{Chengdu}
  \country{China}
}

\author{Tao Chen}
\authornote{Corresponding author: Tao Chen, t.chen@bham.ac.uk.}
\affiliation{%
  \institution{IDEAS Lab, University of Birmingham}
  \city{Birmingham}
  \country{UK}
}
\email{t.chen@bham.ac.uk} 
\renewcommand{\shortauthors}{Chen et al.}

\begin{abstract}
     To efficiently tune configuration for better software system performance (e.g., latency) at the deployment and maintenance stage, many tuners have leveraged a surrogate model to expedite the process instead of solely relying on the profoundly expensive system measurement. As such, it is naturally believed that we need more accurate models. However, the fact of ``\textit{accuracy can lie}''---a somewhat surprising finding from prior work---has left us many unanswered questions regarding what role the surrogate model plays in configuration tuning. This paper provides the very first systematic exploration and discussion, together with a resolution proposal, to disclose the many faces of useful surrogate models for configuration tuning beyond accuracy, through the novel perspective of fitness landscape analysis. We present a theory as an alternative to accuracy for assessing the model usefulness in tuning, based on which we conduct an extensive empirical study involving up to $27,000$ cases. Drawing on the above, we propose \approach, an automated predictive tool that estimates which model-tuner pairs are the best for an unforeseen system without expensive tuner profiling. Our results suggest that \approach, as one of the first of its kind, performs significantly better than random guessing in $79\%-82\%$ of the cases, hence greatly mitigating the required efforts in engineering configuration for software systems. Our results not only shed light on the possible future research directions but also offer a practical resolution that can assist practitioners in evaluating the most useful model for configuration tuning.
\end{abstract}



\maketitle

\input{introduction}

\input{background}

\input{theory}

\input{methodology}

\input{results}

\input{assessment}

\input{approach}

\input{threats}

\input{related}

\input{conclusion}

\begin{acks}
This work was supported by an NSFC Grant (62372084).
\end{acks}

\bibliographystyle{ACM-Reference-Format}
\bibliography{reference}

\end{document}
\endinput

%% file: tablefigures.tex
\usepackage{algorithm2e}
\usepackage{xcolor}

\SetAlFnt{\footnotesize}

\SetAlCapSty{xAlCapSty}

\SetCommentSty{xCommentSty}

\SetNlSty{mynlfont}{}{} 

\LinesNumbered

\SetSideCommentRight

\DontPrintSemicolon

\RestyleAlgo{algoruled}

\usepackage[autolanguage]{numprint}

\newcommand*\np[2][z]{
\ifx z#1%
$\numprint{#2}$%
\else%
$\numprint[#1]{#2}$%
\fi\xspace%
}

\usepackage{fp}
\newcommand{\ShowAbsoluteNumber}[1]{%
\ifnum #1<10%
{\hspace*{0pt}#1}%
\else%
\ifnum #1<100%
{\hspace*{0pt}#1}%
\else%
\ifnum #1<1000%
{\hspace*{0pt}#1}%
\else%
{\numprint{#1}}%
\fi%
\fi%
\fi%
}

\newcommand{\ShowPercentage}[2]{%
\FPeval\percentage{round(#1/#2*100,0)}%
\FPeval\percentageOneDecimal{round(#1/#2*100,1)}%
\ifnum \percentage=0%
{\np[\%]{\FPprint{percentageOneDecimal}}}%
\else%
\ifnum \percentage<10%
{\np[\%]{\FPprint{percentageOneDecimal}}}%
\else%
{\np[\%]{\FPprint{percentageOneDecimal}}}%
\fi%
\fi%
\xspace
}

\newcommand{\ShowPercentageTwo}[2]{%
\FPeval\percentagetwo{round(#1/#2*100,0)}%
\FPeval\percentageTwoDecimal{round(#1/#2*100,2)}%
\ifnum \percentagetwo=0%
{\np[\%]{\FPprint{percentageTwoDecimal}}}%
\else%
\ifnum \percentagetwo<10%
{\np[\%]{\FPprint{percentageTwoDecimal}}}%
\else%
{\np[\%]{\FPprint{percentageTwoDecimal}}}%
\fi%
\fi%
\xspace
}

\newlength\BARSIZE  
\newcommand{\inlinechart}[2]{%
\FPeval{\BLACKBARSIZE}{#1/#2}\textcolor{black!80}{\rule{\BLACKBARSIZE\BARSIZE}{1.6ex}}%
\FPeval{\BLACKBARSIZE}{1 - (#1/#2)}\textcolor{black!10}{\rule{\BLACKBARSIZE\BARSIZE}{1.6ex}}%
}

\newcommand{\pinlinechart}[3]{%
\FPeval{\BLACKBARSIZE}{#1/#2}\textcolor{gree!#3}{\rule{\BLACKBARSIZE\BARSIZE}{1.6ex}}%
\FPeval{\BLACKBARSIZE}{1 - (#1/#2)}\textcolor{black!10}{\rule{\BLACKBARSIZE\BARSIZE}{1.6ex}}%
}

\newcommand{\ninlinechart}[3]{%
\FPeval{\BLACKBARSIZE}{1 - (#1/#2)}\textcolor{black!10}{\rule{\BLACKBARSIZE\BARSIZE}{1.6ex}}%
\FPeval{\BLACKBARSIZE}{#1/#2}\textcolor{red!#3}{\rule{\BLACKBARSIZE\BARSIZE}{1.6ex}}%
}

\newcommand*\percent[5][v]{%
\ifx q#1%
    \np{#2}/\np{#3}(\ShowPercentage{#2}{#3})\else%
\ifx w#1%
    \np{#2}(\ShowPercentage{#2}{#3})\else%
\ifx m#1%
    \np{#2}%
    \inlinechart{#2}{#3}\else%
\ifx t#1%
    \FPprint{#2}%
    \hspace*{0.5ex}%
    \inlinechart{#2}{#3}\else%
\ifx p#1%
    \FPprint{#4}%
    \hspace*{0.5ex}%
    \pinlinechart{#2}{#3}{#5}\else%
\ifx n#1%
    \FPprint{#4}%
    \hspace*{0.5ex}%
    \ninlinechart{#2}{#3}{#5}\else%
\ifx c#1%
    \inlinechart{#2}{#3}{#5}%
\else%
    \np{#2}%
    \ifx r#1%
        /\np{#3}%
    \fi%
    \hspace*{0.5ex}(\ShowPercentage{#2}{#3}) %
    \inlinechart{#2}{#3}%
    \xspace
\fi\fi\fi\fi\fi%
}

%% file: introduction.tex
\section{Introduction}
\label{sec:introduction}

Configurations are pervasive in software systems. Setting those configurations requires careful engineering efforts since they can profoundly influence the performance, such as latency, throughput or accuracy. For example, some poorly tuned configurations can be $480\times$ worse than the optimal one~\cite{DBLP:conf/mascots/JamshidiC16}. As such, configuration tuning---a process that automatically searches the configuration for the best performance via an intelligent heuristic tuner---has attracted increasing attention in recent years~\cite{flash,DBLP:journals/corr/abs-2112-07303,DBLP:journals/tosem/ChenL23a}. Indeed, tuning configuration has been an important practice and phase in software engineering, as highlighted in a recent TSE survey~\cite{DBLP:journals/tse/SayaghKAP20}.



\input{Tables/test2}

A key challenge in configuration tuning is its expensiveness when measuring the performance achieved by configurations~\cite{DBLP:conf/icse/MaCL25}. For example, it takes more than $1,536$ hours to fully sample the configurations from only $11$ options for a system~\cite{DBLP:conf/wosp/ValovPGFC17}. To tackle this, existing tuners have relied on the surrogate model: a data-driven model that learns the correlation between configuration and performance from past data. Much progress along this line of research has been made in the software engineering community, some examples of which have been shown in Table~\ref{tab:automatic_tuning_methods}.




Indeed, tuners that rely on the model-centric Bayesian optimization (and its variants) are overwhelming~\cite{accuracy_can_lie,k2vtune,DBLP:journals/corr/abs-2112-07303,DBLP:conf/ijcai/ChenCDS4RAG} and even those tuners that are model-free by design can be seamlessly paired with a surrogate model~\cite{conex, bestconfig, lopez2016irace, paramils, Sway}, including those that support multi-fidelity~\cite{DBLP:conf/ase/Ye0L26}. Here, a surrogate model correlates configurations with their likely performance (often achieved by some machine learning methods), serving as a prediction proxy for the costly system measurement. The core benefits of those model-based tuners is intuitive: with a model, the tuner can obtain a quick (estimated) evaluation of the configurations' performance to guide the tuning, hence saving a huge amount of resources that would otherwise be required to measure the system. For example, it has been shown that predicting a configuration's performance using the neural network-based model \texttt{DaL} \cite{DaL,DBLP:journals/tse/GongCB25} only take $\approx18$ seconds, which is trivial compared with the maximum of $166$ minutes required for measuring only one configuration directly on the database system \textsc{MariaDB}~\cite{DBLP:conf/sigsoft/0001L24}. Therefore, while some works concentrate on how to integrate the model with a tuner, many have also exclusively focused on proposing a more accurate model~\cite{DBLP:journals/corr/abs-2403-03322}, such as \texttt{HINNPerf} \cite{HINNPerf}, \texttt{SeMPL} \cite{DBLP:journals/pacmse/Gong024}, and \texttt{DHDA}~\cite{DBLP:conf/icse/XiangChen26}, which can be paired with diverse heuristic tuners~\cite{accuracy_can_lie}.



Despite the importance, we have not yet obtained a comprehensive understanding on what role the models play in configuration tuning in the community. A long-standing belief over decades is that ``\textit{the more accurate the model (e.g., on MAPE \cite{DeepPerf, DaL} and $\mu$RD \cite{flash, DBLP:conf/sigsoft/NairMSA17}), the more useful for tuning, and vice versa}''. However in early 2025, for the first time, our prior work published in TSE demonstrated that such a belief can be misleading \cite{accuracy_can_lie}---the model accuracy has no positive monotonic correlation to the quality of tuning, e.g., a more accurate model does not help or can even harm the tuning results. Yet still, it has neither provided a systematic/comprehensive study on the root causes behind nor the resolution of how the community might move forward, when it is known that the model accuracy can be misleading. An important, but challenging and unanswered question is: how can we infer and assess what models are useful for tuning a system without comprehensively profiling them under the possible tuners?


\begin{figure}[t!]
\centering
\subfloat[Fitness landscape]{
\includegraphics[width=.3\textwidth]{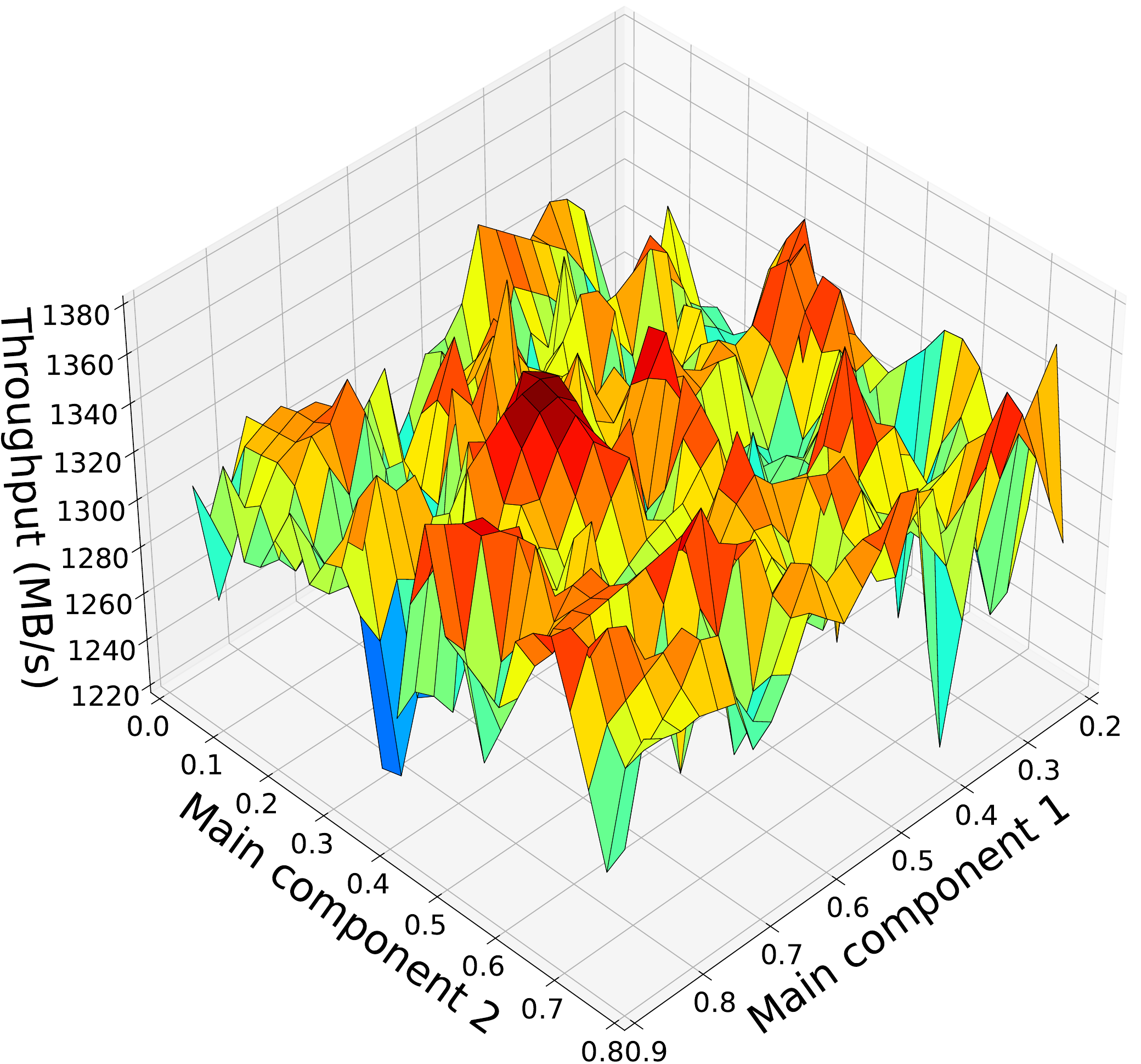}
}
\hspace{.1\textwidth} 
\subfloat[Purely statistical distribution]{
\includegraphics[width=.3\textwidth]{Figures/2D_perf.tex}
}
\caption{Comparing fitness landscape analysis with purely statistical distribution analysis for \textsc{Hadoop}.} 
\label{fig:fla-vs-sda}
\end{figure}

To explore a potential resolution for the above gap, in this paper, we unveil the many faces of useful surrogate models for configuration tuning beyond accuracy from the perspective of fitness landscape analysis \cite{1932The}---a well-established paradigm in the optimization community. In a nutshell, studying the fitness landscape provides insights into the problem's characteristics and how those can influence the behaviors of the optimization algorithms~\cite{DBLP:series/sci/PitzerA12,DBLP:journals/isci/MalanE13}. As shown in Figure~\ref{fig:fla-vs-sda}, compared with the purely statistical distribution analysis used for configurable systems~\cite{DBLP:conf/msr/GongC22,muhlbauer2023analyzing}, which mainly focuses on the performance/option variation, fitness landscape analysis additionally provides spatial information for:

\begin{itemize}
    \item the strength of the global guidance provided to steer the tuning process, i.e., the problem difficulty;
    \item and the topologically local neighborhood structure of the paths that can be explored by a tuner, including ruggedness, funnel/basin size, and gradient degree, etc. This often indicate the severity of local optima, i.e., the smoother the better.
\end{itemize}

It has been known that the fitness landscape analysis often helps stimulate more advanced and domain-specific algorithm designs~\cite{DBLP:series/sci/PitzerA12,DBLP:journals/isci/MalanE13,10.1145/3728954,DBLP:conf/icse/Ye0L25}. What makes our work unique is that we systematically analyze the landscape emulated by the model, with respect to that of the real system when appropriate, thus allowing us to obtain insights into the model usefulness for tuning that goes beyond the standard accuracy. Specifically, we make several contributions:

\begin{itemize}
    \item \textbf{A New Theory.} We propose \textbf{\textit{landscape dominance}}, a simple yet effective theory by discriminating global and local landscape features under the notion of Pareto dominance~\cite{DBLP:journals/ec/VeldhuizenL00}, to assess the model's usefulness for tuning in a coarse-grained manner. Here, the global and local landscape features are two complementary properties of how the model emulate the tuning problem: the former measures the model's ability to retain the overall difficulty of the problem; the latter, in contrast, measures the severity of local optima traps that the model emulates (a common issue of tuning stagnation). In simpler words, we expect a more useful model $\mathcal{A}$ for tuning Pareto dominant model $\mathcal{B}$ on both the global and local landscape features over the landscape they emulated: compared with the landscape of $\mathcal{B}$, $\mathcal{A}$'s landscape should be closer to the real one on the global landscape features (hence the overall problem difficulty is more consistent) while being smoother on the local landscape features (hence the local optima issues are less problematic). Beyond accuracy, for the first time, this provides an alternative way to evaluate and compare the usefulness of models for configuration tuning without expensive tuner profiling.
    \item \textbf{An Empirical Study.} Drawing on the landscape dominance, we conduct an empirical study with $18$ configurable systems, $10$ surrogate models, $16$ tuners, $8$ landscape features, and $2$ accuracy metrics, resulting in a range between $1,980$ and $27,000$ cases of investigation. We found that:
    \begin{itemize}
        \item The proposed landscape dominance can correctly identify the more useful model for tuning in $78\%$ cases overall.
        \item The accuracy metrics and landscape features are hardly correlated in a positive manner ($<30\%$ cases), meaning that they generally provide different aspects of information on the surrogate model for tuning.
        \item For $93\%$ of the cases, the current models are still far away from correctly emulating the landscape of the real systems, but they might be helpful to relax the issue of local optima for a tuner, even considering a relatively low fidelity.
        \item No single model is generally the most preferred for tuning, considering all landscape features according to the landscape dominance. 
        \item Memory and queue relevant configuration options can be $\approx 2\times$ more influential to the landscape emulated by the models than options related to the others (e.g., CPU and utility), but the actual case can vary depending on the system and feature.
    \end{itemize}
    \item \textbf{A Predictive Tool.} The landscape dominance and insights from the empirical study have led us to design \approach---a fine-grained, predictive tool that leverages learning-to-rank---for identifying the most useful model-tuner pairs. We show that, when predicting on an unforeseen system, \approach~achieves remarkably better results than random guessing on up to $82\%$ of the cases with a maximum of $244\%$ improvement. This can greatly help to mitigate the required efforts in engineering configuration for software systems.
\end{itemize}

To promote open science, all data, source code, and supplementary materials can be found at our repository: \textcolor{blue}{\texttt{\href{https://github.com/ideas-labo/model4tune}{https://github.com/ideas-labo/model4tune}}}.

The remainder of this paper is structured as follows: Section~\ref{sec:background} overviews the necessary preliminaries. Section~\ref{Sec:Theory} presents the theory---a new alternative to assess the usefulness of models for tuning beyond accuracy---followed by the details of our empirical study in Section~\ref{sec:methodology}. Section~\ref{sec:results} presents the findings with an in-depth analysis. Section~\ref{Sec:Insights} discusses the insights learned from the observations and findings. Section~\ref{sec:approach} presents \approach, a predictive tool that assists practitioners in the tuner design by inferring the model usefulness for tuning via a fine-grained manner. Sections~\ref{sec:threats} and \ref{sec:related} present the threats to validity and related work, respectively. Section~\ref{sec:conclusion} concludes the paper.

%% file: Tables/test2.tex
\begin{table*}[t!]
  \centering
  \caption{Representative papers on model-based configuration tuning published in major software engineering venues during the last 5 years.}
  \resizebox{\linewidth}{!}{
    \begin{tabular}{l l l l l}
      \toprule
      \textbf{Tuner} & \textbf{Key Model-based Optimization Algorithm} & \textbf{Domain} & \textbf{Year} & \textbf{Venue} \\
      \midrule
      \texttt{PromiseTune} & Bayesian optimization with Random Forest and causal inference. & Any systems & 2026 & ICSE \cite{PromiseTune} \\
      \texttt{CSAT} & Bayesian optimization with Guassion Process and ANFIS. & Any systems & 2025 & JSS \cite{CSAT} \\
      \texttt{PDCAT} & Bayesian optimization with preference-driven selection. & Compiler & 2025 & FSE \cite{PDCAT} \\
      \texttt{CompTuner} &Improved particle swarm optimization with Random Forest. & Compiler & 2024 & TOSEM \cite{10.1145/3640330}\\
      \texttt{ETune} & Bayesian optimization with Random Forest and parameter selection. & Big data system & 2024 & JSS \cite{ETune} \\
      \texttt{Tsoa} & Evolutionary optimization with Random Forest at two stages. & Compiler & 2024 & ASE \cite{Tsoa} \\
      \texttt{AutoConf} & Bayesian optimization with Tree Parzen Estimator and metamorphic testing. & Machine learning model & 2023 & ASE \cite{DBLP:conf/kbse/SharGHSTK23} \\
      \texttt{CFSCA} & Bayesian optimization with Random Forest and critical flag selection. & Compiler & 2023 & ASE \cite{CFSCA} \\
      \texttt{DAT} & Bayesian optimization with Tree Parzen Estimator and a new distance metric. & AST analyzer & 2023 & TSE \cite{DAT} \\
      \texttt{VEER} & Bayesian optimization with CART regression tree and explanation. & Any systems & 2023 & ESE \cite{DBLP:journals/corr/abs-2106-02716} \\
      \texttt{SafeTune} & Classification and prediction with Random Forest & Any systems & 2022 & ICSE \cite{9793958} \\
      \texttt{ROME} & Bayesian optimization with Decision Trees. & Effort estimation tool & 2022 & TSE \cite{ROME} \\
      \texttt{BOCA} & Bayesian optimization with Random Forest and dimensionality reduction. & Compiler & 2021 & ICSE \cite{BOCA} \\
      \texttt{TPE} & Bayesian optimization with Tree Parzen Estimator. & Microservice & 2021 & ASE \cite{TPE} \\
      \texttt{SATune} & Bayesian optimization with Random Forest and simulated annealing. & Program verification tool & 2021 & ASE \cite{SATune} \\
      \texttt{FLASH} & Bayesian optimization with Decision Tree. & Any systems & 2020 & TSE \cite{flash} \\
      \bottomrule
    \end{tabular}
  }
  \label{tab:automatic_tuning_methods}
\end{table*}

%% file: background.tex
\section{Preliminaries}
\label{sec:background}

In this section, we discuss the necessary preliminaries that underpin this work.

\subsection{Configuration Tuning with Surrogate Model}


The general goal of configuration tuning is to optimize a performance metric, e.g., latency or throughput, subject to a budget:
\begin{equation}
		\arg\min f(\vect{c}) \text{ or } \arg\max f(\vect{c})
\end{equation}
where $\vect{c} = \{o_{1}, o_{2}, \dots, o_{n}\}$ is a configuration such that $o_n$ is the $n$th configuration option, which can be a binary, integer, or enumerated value. $f$ denotes measuring the system for evaluating the performance obtained by setting a certain configuration.

A key challenge thereof is that measuring $f$ is profoundly expensive, limiting the ability to explore the configuration space. As such, model-based tuners, which leverage a surrogate model that mimics the real system, have become increasingly important, since the evaluations of the model can be done in seconds versus hours/days for measuring the system. Generally, there are two types of model-based tuners depending on how the model is used~\cite{accuracy_can_lie}:

\begin{itemize}
  \item \textbf{Sequential model-based tuners} instead iteratively update the model during the tuning as newly measured configurations become available, most commonly following a Bayesian optimization procedure. Among the tuners, the main distinction is the way in which the model is updated and how to leverage the model to select the next configuration to be measured on the real system \cite{flash, PromiseTune, BOCA, Ottertune, SMAC, restune}.

    \item \textbf{Batch model-based tuners} pre-train a fixed model using a set of past measurements; such a model is then served as a delegate of the real system. In this way, the surrogate model can be seamlessly paired with any heuristics tuners \cite{conex, bestconfig, lopez2016irace, paramils, Sway}.

\end{itemize}

\subsection{Fitness Landscape Analysis}


Fitness landscape~\cite{1932The,DBLP:series/sci/PitzerA12,DBLP:journals/isci/MalanE13} serves as the foundation for characterizing the spatial structure of search spaces in optimization problems, particularly well-studied in the communities of machine learning \cite{DBLP:conf/automl/MohanBWDL23, DBLP:conf/kdd/HuangL25}, evolutionary algorithms \cite{thousand2017, doi:10.1126/science.adh3860}, and computational optimization \cite{DBLP:conf/gecco/ThomsonGHB24, DBLP:conf/gecco/AdairOM19}. Over the past decades \cite{1932The}, analyzing landscapes has helped researchers to understand the complexity of various optimization problems and their relationships to the behaviors of the algorithms. Since like other Search-based Software Engineering problems~\cite{DBLP:conf/laser/HarmanMSY10,DBLP:journals/tse/LiCY22,DBLP:journals/tosem/ChenL23}, configuration tuning is essentially an optimization problem, examining its fitness landscape is naturally promising.


An important notion in fitness landscape is the local optimum: suppose that we seek to minimize a performance metric, a configuration $\vect{c}$ is a local optimum if its performance value is no worse than all other configurations in its neighborhood radius within the landscape:
\begin{equation}
    f(\vect{c}) \leq f(\vect{c}') \quad \forall \vect{c'} \in \mathcal{N}(\vect{c})
\end{equation}
wherein $f(\vect{c})$ is the performance value (fitness) of configuration $\vect{c}$; $\mathcal{N}(\vect{c})$ is the neighborhood of configuration $\vect{c}$, which is defined depending on the problems/systems~\cite{1932The,DBLP:series/sci/PitzerA12,DBLP:journals/isci/MalanE13}; and $\vect{c}'$ denotes any configuration in the neighborhood $\mathcal{N}(\vect{c})$ of $\vect{c}$. In this work, we adopt Hamming distance to measure the neighborhood radio, which is the most common practice for configurable systems \cite{10.1145/3728954, DBLP:conf/seams/Chen22}. Understanding the local optima is important, as prior studies have demonstrated that the configuration landscape for many configurable systems can contain tough local optima that would easily trap the tuners~\cite{DBLP:journals/corr/abs-2112-07303,PromiseTune,flash}---a root cause of tuning stagnation. As such, many existing tuner designs have primarily focused on how to jump out from those local optima whenever possible~\cite{DBLP:journals/corr/abs-2112-07303,PromiseTune}.



The benefits of fitness landscape analysis for configuration tuning is vital: for example, precisely and spatially understanding the complexity of local optima can enable us to design tailored tuners: those that favor tuning exploration would be more suitable for systems with more challenging local optima \cite{conex, DBLP:journals/corr/abs-2112-07303}; in contrast, preferring exploitation can be more appropriate with a strongly-guided and smoother configuration landscape~\cite{PromiseTune}.



\begin{figure}[t!]
\centering
\subfloat[\texttt{HINNPerf} landscape]{
\includegraphics[width=.3\textwidth]{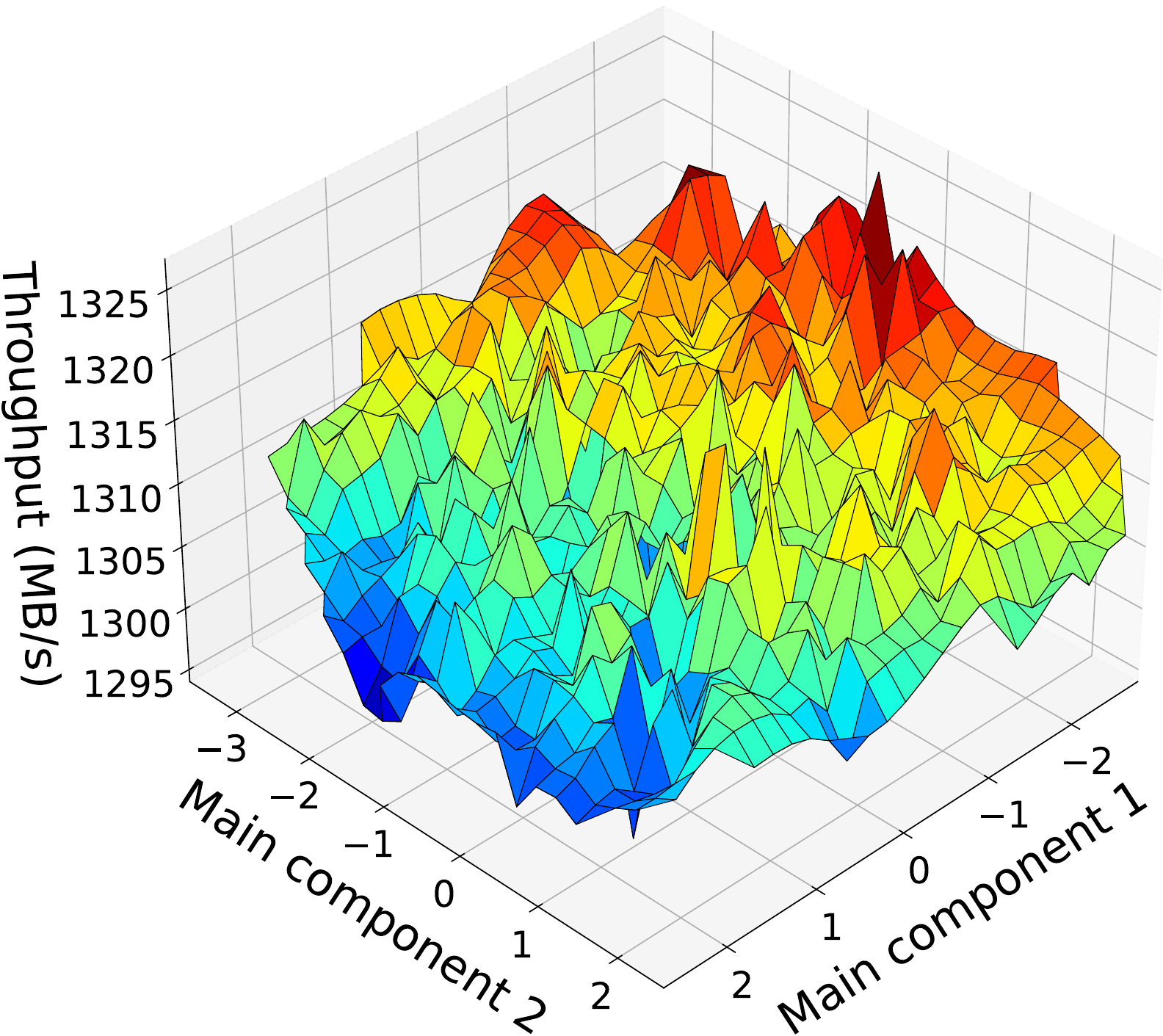}
\label{fig:model-a}
}
\hspace{.1\textwidth} 
\subfloat[\texttt{LR} landscape]{
\includegraphics[width=.3\textwidth]{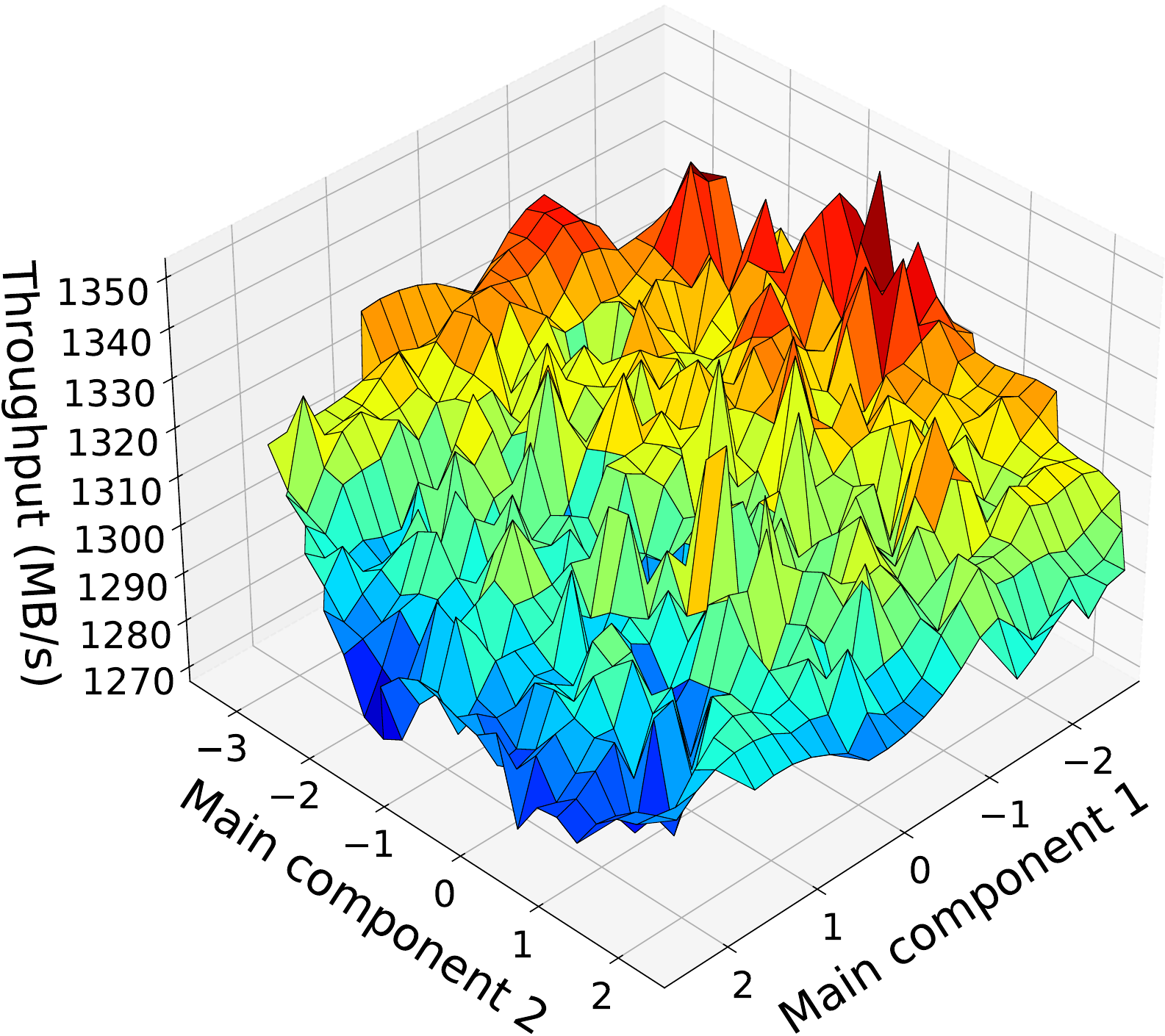}
\label{fig:model-b}
}
\caption{A projected configuration-throughput landscape derived from two models, \texttt{HINNPerf} and \texttt{LR}, for the system \textsc{Hadoop} (dimensionality-reduced by PCA \cite{DBLP:journals/corr/Shlens14}).} 

\label{fig:models}
\end{figure}

\subsection{Motivation: Accuracy can Lie}

For configuration tuning, our previous study \cite{accuracy_can_lie} has indicated that the best/worst accuracy of the surrogate model is generally inconsistent with the best/worst tuning results under sequential model-based tuners---only $\leq20\%$ of the cases are matched. In contrast, the batch model-based tuners tend to have slightly better consistency, yet this still merely covers no more than 45\% of the cases. The above empirical findings have already evidenced that the accuracy of surrogate models tends to be highly misleading and cannot serve as a reliable reflection of the usefulness of a model for tuning.


Indeed, since a model-based tuner would explore configurations in the model landscape instead of that of the real system, it is important to investigate the spatial information of the model-emulated landscape. However, this is not a straightforward task: even visually similar landscapes may have rather different underlying properties, hence affecting the tuner drastically. For example, in Figure~\ref{fig:models}, we show the landscapes emulated by \texttt{HINNPerf} \cite{HINNPerf} and Linear Regression (\texttt{LR}) \cite{montgomery2021LR} for the \textsc{Hadoop}. Visually, both emulated landscapes are almost identical, yet mysteriously, running the same tuner on them leads to rather different tuning results. As such, the visualization of the landscape still cannot accurately portray the usefulness of the model.




The above is what motivates us to study and investigate new ways to assess the model's usefulness beyond accuracy while mitigating the drawback of directly visualizing the landscape.


%% file: theory.tex
\section{Theory: What are Useful Models Beyond Accuracy?} 
\label{Sec:Theory}


The key of this work is to leverage the \textbf{\textit{landscape features}} for explaining, assessing, and predicting the usefulness of surrogate models for configuration tuning\footnote{By the usefulness of a model for configuration tuning, we refer to the extent to which it helps to achieve better tuning results.} (we leave detailed elaborations of these features in Section~\ref{sec:land-metrics}).
Landscape features are core spatial properties of the configuration landscape~\cite{1932The,DBLP:series/sci/PitzerA12,DBLP:journals/isci/MalanE13}, which helps to quantitatively reason the behaviors of tuners when searching therein. Other than simple visualization, those features quantify the landscape from various spatial aspects, e.g., measuring the ruggedness~\cite{stadler1996landscapes} and the difficulty to be optimized~\cite{DBLP:conf/evoW/MullerS11}, shedding light on understanding the usefulness of models for configuration tuning. Recall the example from Figure~\ref{fig:models}, our posterior investigation suggests that the overall guidance provided (measured by Fitness Distance Correlation \cite{DBLP:conf/evoW/MullerS11}) and the general ruggedness of the local optima (measured by Correlation Length \cite{stadler1996landscapes}) of the two emulated landscapes differ by at least $1.63\times$, which means they are of diverse difficulty levels to a tuner---the root cause of the different results.

Broadly, landscape features can be divided into two categories depending on their granularities of information provided: \textbf{\textit{global}} and \textbf{\textit{local}} features \cite{8832171, DBLP:conf/gecco/LiefoogheVLZM21, DBLP:conf/gecco/MersmannBTPWR11}. The former refers to the features that reflect the overall structure of the landscape, e.g., difficulty or guidance; the latter measures the properties related to local optima, e.g., their density and ruggedness.

\begin{figure}[t!]
\centering

\subfloat[\texttt{DT}]{\includegraphics[width=0.33\columnwidth]{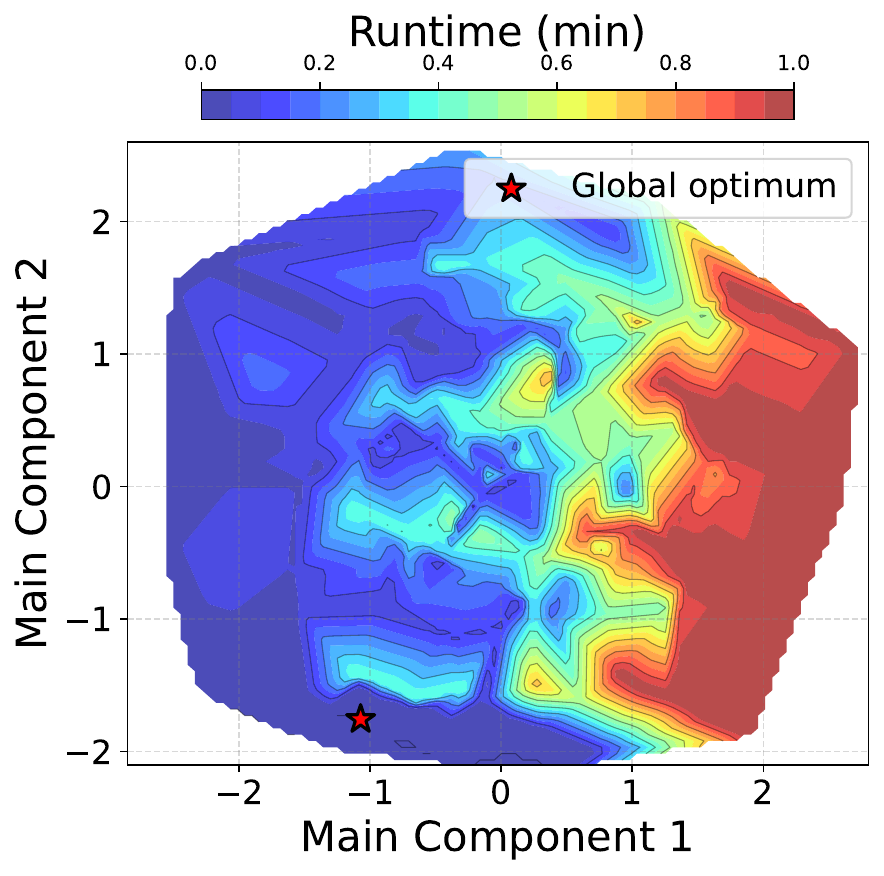}}
~\hfill  
\subfloat[\texttt{GP}]{\includegraphics[width=0.33\columnwidth]{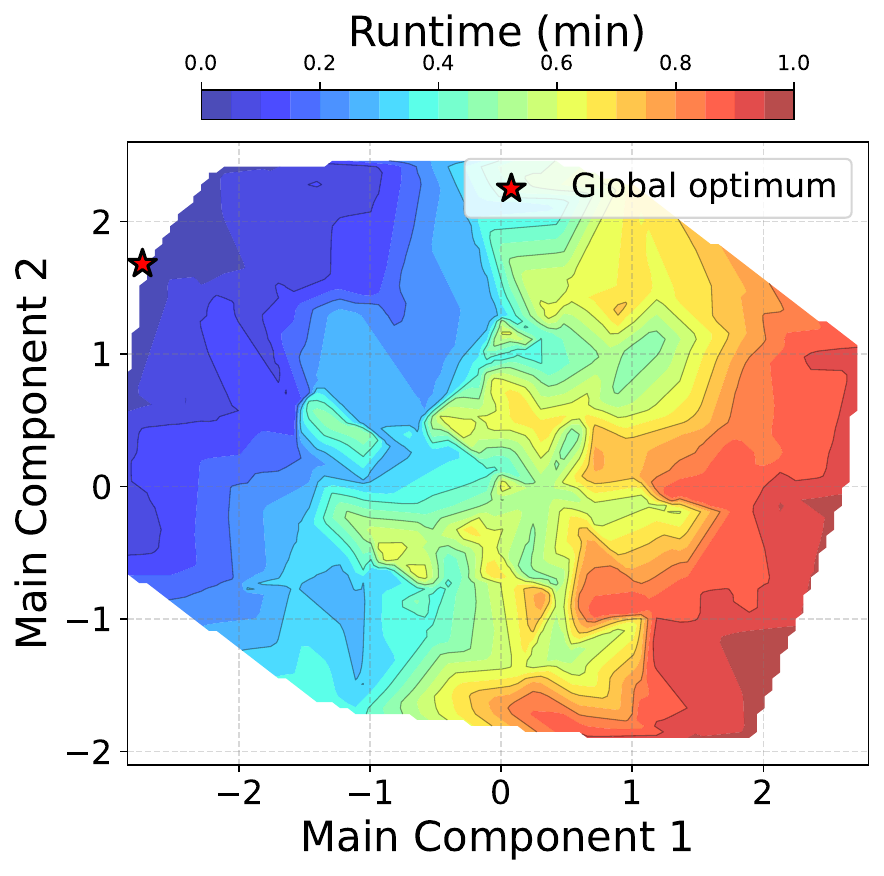}}
~\hfill
\subfloat[Real landscape]{\includegraphics[width=0.33\columnwidth]{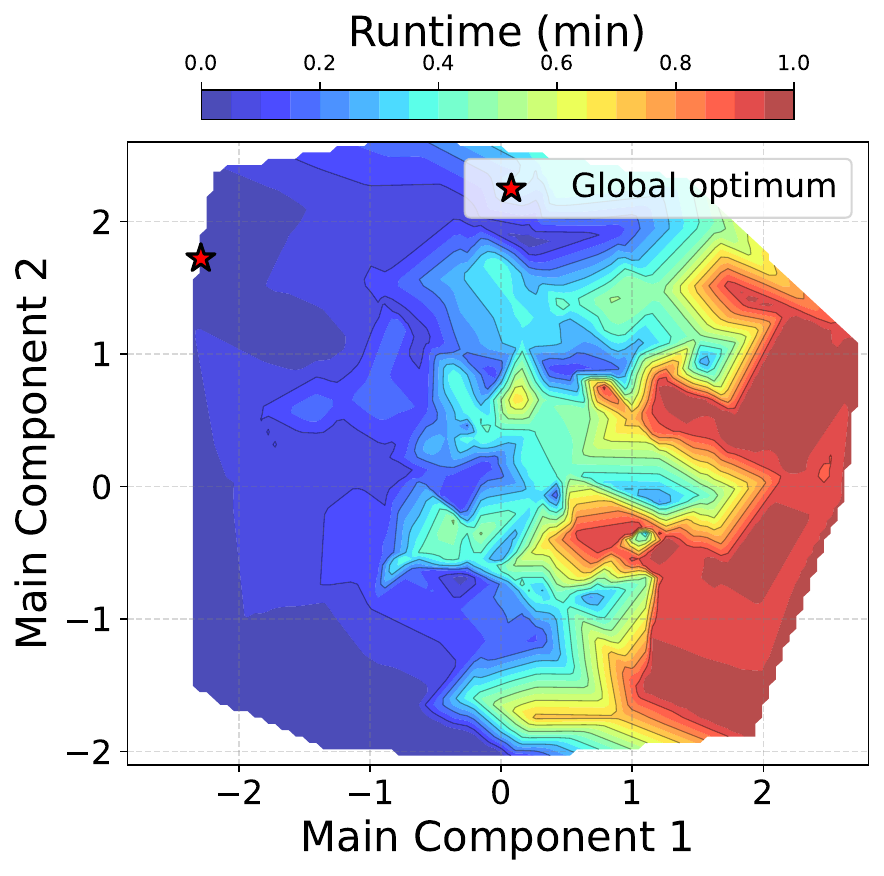}}

\caption{Configuration landscapes emulated by the model \texttt{GP} and \texttt{DT}, together with the real landscape for system \textsc{XGBoost}. All are processed by PCA.} 
\label{fig: example_2}
\end{figure}

With accuracy, the general belief is that the closer the predicted performance to the actual performance, the more useful the model for configuration tuning. While this has been demonstrated as misleading~\cite{accuracy_can_lie} and given the promising landscape features, a key question remains is that \textit{how do we know whether a model is more useful than the other for tuning even by leveraging the landscape features?} Indeed, it is natural to believe that a good/useful model should emulate a landscape that is as close to the real landscape as possible, which resembles to the principle behind accuracy. However, we have found that this is not always the case in practice. Consider the example in Figure~\ref{fig: example_2}, we see that the landscape emulated by the Decision Tree (\texttt{DT}) in (a) is clearly much more similar to the real one in (c) than the one by Gaussian Process (\texttt{GP}) in (b). Yet, the fact is that a tuner using the \texttt{GP} achieves much better results than using \texttt{DT} in general. This is because, in this case, although \texttt{GP} emulates a landscape that appears to be rather distinct form the real landscape, it still has the similar location of the global optimum/overall structure and, more importantly, it is smoother (than the real one and the one emulated by \texttt{DT}), hence less likely causing a tuner to be trapped at local optima; the importance of landscape smoothness has also recently been demonstrated by Yedida and Menzies~\cite{10919477}. As such, we propose a theoretical hypothesis:



\begin{tcolorbox}[enhanced,boxrule=0.2mm, title=\textbf{\textit{Theory of Model Usefulness for Tuning}},
colframe=black,colback=white,colbacktitle=white,
coltitle=black,attach boxed title to top center=
{yshift=-0.25mm-\tcboxedtitleheight/2,yshifttext=2mm-\tcboxedtitleheight/2},
boxed title style={boxrule=-0.2mm,
frame code={ \path[tcb fill frame] ([xshift=-3mm]frame.west)
-- (frame.north west) -- (frame.north east) -- ([xshift=3mm]frame.east)
-- (frame.south east) -- (frame.south west) -- cycle; },
interior code={ \path[tcb fill interior] ([xshift=-2.1mm]interior.west)
-- (interior.north west) -- (interior.north east)
-- ([xshift=2.1mm]interior.east) -- (interior.south east) -- (interior.south west)
-- cycle;} }]
A \textbf{more useful} surrogate model for configuration tuning is the one that emulates a landscape with

\begin{itemize}
    \item \textbf{closer} global feature(s) to that(those) of the real landscape;
    \item while exhibiting \textbf{less severe} local optima property from the local feature(s).
\end{itemize}

\end{tcolorbox}
The intention behind the theory is intuitive: if the global features of the model-emulated landscape deviated too much from the real one, then we are tuning for a different, irrelevant problem. Meanwhile, if the emulated landscape is of a similar global structure to the real one, then it would be more beneficial if its local optima issues are easier to tackle.

We call the above definition \textbf{\textit{landscape dominance}}.

In what follows, we will derive our empirical study and propose new ways to assess/predict the usefulness of surrogate models according to the above theoretical understandings.


%% file: methodology.tex

\section{Empirical Study Methodology}
\label{sec:methodology}

We now delineate the methodology of our empirical study.

\input{Tables/systems}

\subsection{Research Questions}




We investigate the many faces of surrogate models for tuning via several research questions (RQs). The most fundamental goal is to verify our \textit{theory of model usefulness for tuning}:


\keybox{
\textbf{RQ1:} How effective is the proposed theory of landscape dominance using global and local landscape features in assessing model usefulness for configuration tuning?
}

Then, to understand why the accuracy of surrogate models can be misleading for configuration tuning, we ask:

\keybox{
\textbf{RQ2:} What is the correlation between accuracy and landscape features of surrogate models?
}

Given our theory of the useful model for configuration tuning and that the landscape features can help to explain the behaviors of the tuners when exploring the landscape emulated by a model, understanding the above is essential to study why the accuracy is misleading when being used to reflect the quality of configuration tuning. Next, a natural question to answer is:


\keybox{
\textbf{RQ3:} To what extent does the landscape emulated by a surrogate model differ from the real landscape of configurable systems?
}

This enables us to explore the current fidelity of the surrogate models in terms of the landscape features.



While \textbf{RQ1}-\textbf{RQ3} provide a general overview of the landscape of the surrogate model for configuration tuning, they merely treat the system as a black box. As such, in \textbf{RQ4}, we ask:


\keybox{
\textbf{RQ4:} How does each individual option influence the emulated landscape of the surrogate model?
}

With domain knowledge and understanding of the configuration options, we seek to provide insights on how the specific options, and their categories, of the systems can change the landscape being built by a model, if each of them is omitted in the modeling process. \mytag{R2C4}\revision{Prior work has already conducted similar analysis on accuracy~\cite{muhlbauer2023analyzing}, we wish to confirm if that is the same for the landscape features and aid explanation to some observations from the previous RQs.}

\mytag{R2C1}\revision{Notably, these RQs differ from our prior study~\cite{accuracy_can_lie} in the following aspects:}

\begin{itemize}
    \item \revision{Our prior work~\cite{accuracy_can_lie} reveals the misleading issue of solely relying on accuracy to evaluate the model's uselessness for configuration tuning, together with qualitative analysis of the potential causes. In contrast, \textbf{RQ1} in this work provides a concrete and quantifiable resolution to address the issue identified.}
    \item \revision{The landscape features have not been studied previously~\cite{accuracy_can_lie} while the \textbf{RQ2} and \textbf{RQ3} in this work explore the correlation between accuracy and landscape features, as well as how those features emulated the real landscape.}
    \item \revision{\textbf{RQ4} explicitly links the landscape features to the domain-specific information of the configurable systems, i.e., the options, which have not been studied before.}
\end{itemize}




\subsection{Configurable Systems}
\label{sec:sys}

\subsubsection{System Selection}

As shown in Table~\ref{tb:systems}, we select the software systems and their workloads used by the most notable works from the key venues in software modeling system engineering. After extensively surveying the systems, we conduct a screening of them following two criteria:

\begin{itemize}
    \item We excluded systems with fewer than five options, as in those cases the quality difference among surrogate models appears to be insignificant.
    \item For the same systems where different sets of relevant configuration options have been used, we select the one with the highest number of options. For example, \textsc{STORM} is a configurable system that has been studied in many prior studies, and we use the instance with 12 options to tune, which is the most complicated case.
\end{itemize}


We have also omitted systems with a too small configuration space (i.e., $<10^7$) while designed for the same domain as a larger system. For example, \textsc{HSQLDB} and \textsc{PostgreSQL} are both database systems, but the former only has a space of $2.62\times10^5$ while the latter has $1.42\times10^9$ (but both would be selected if they have a search space $>10^7$). Additionally, certain systems (e.g., \textsc{SaC}, \textsc{Tomcat}) are removed because their calculation of certain landscape features (e.g., \texttt{CL}) become too sparse to identify neighbors, rendering calculations impossible.
\input{Tables/surrogate}


Table~\ref{tb:systems} shows the final set of $18$ configurable systems considered in this study. It is clear that those systems come from various domains with diverse performance attributes that are of concern, ranging from $9$ options (\textsc{PostgreSQL}) to $39$ options (\textsc{Polly}). The configuration space also covers a wide range: from $2.04\times10^3$ to $2.67\times 10^{41}$. It is worth noting that, even for those with a relatively small search space, the measurement of a single configuration can still be time-consuming. For instance, it can take up to hours to measure one configuration on \textsc{MariaDB}~\cite{DBLP:conf/sigsoft/0001L24}, and a single run of measurement for \textsc{DeepArch} might even be half a day \cite{DBLP:conf/sigsoft/JamshidiVKS18}.

Note that the configuration performance data is collected by using standard benchmarks~\cite{DBLP:conf/icse/WeberKSAS23,DeepPerf,flash, muhlbauer2023analyzing, DBLP:journals/corr/abs-2106-02716, DBLP:conf/sigsoft/JamshidiVKS18, DBLP:journals/tse/KrishnaNJM21,DBLP:journals/jss/CaoBWZLZ23}. For instance, when tuning \textsc{SQLite}, \textsc{Sysbench}---a standard performance benchmark---is used to generate the workloads and for measuring the performance. To avoid measurement noise, the performance of each configuration is the average or median of $3$ to $23$ measurement repeats, as stated in the prior work. 




\subsubsection{Option Selection}
\label{option}

To ensure the validity and comparability of our research outcomes, we adopt a rigorous approach that adheres strictly to the established norms within the research community. Instead of devising our own selection criteria, we have meticulously matched the configuration options, corresponding values, workloads/benchmarks, and target performance metrics with those employed in previous studies of the same systems \cite{DBLP:conf/icse/WeberKSAS23,DeepPerf,flash, muhlbauer2023analyzing, DBLP:journals/corr/abs-2106-02716, DBLP:conf/sigsoft/JamshidiVKS18, DBLP:journals/tse/KrishnaNJM21,DBLP:journals/jss/CaoBWZLZ23}.

\subsection{Surrogate Models}


In this study, we examine surrogate models commonly used by research on software configuration \cite{chen2019all,DBLP:conf/msr/GongC22,DBLP:conf/kbse/JamshidiSVKPA17,DBLP:conf/splc/0003APJ21}, including both the general machine learning models (e.g., \texttt{GP}) and those that are proposed for configuration performance learning (e.g., \texttt{DaL}~\cite{DaL}). In particular, the models are of diverse categories, e.g., Gaussian distribution, tree-based, and deep learning based. The results have been shown in Table~\ref{tb:models}.



It is worth noting that we have to omit certain models due to their slow runtime on training and/or prediction. For example, \texttt{Perf-AL}~\cite{PerfAL}, a deep learning model specifically designed for learning configuration performance, relies on an adversarial neural network with hyperparameter tuning that requires an extremely long training time.





\subsection{Accuracy Metrics of Surrogate Models}
\label{sec:acc-metrics}





To comprehensively assess accuracy, we include two of the most widely used metrics that are derived from a completely different nature: residual and ranked. 

\textbf{MAPE}~\cite{bestconfig,DeepPerf,DaL,DBLP:conf/asplos/YuBQ18}, a.k.a. MRE or MMRE, is a popular residual metric for measuring accuracy, which is calculated as:


\begin{equation}
\text{MAPE} = \frac{1}{n} \sum_{i=1}^{n} \left|\frac{y_i - \hat{y}_i}{y_i}\right| \cdot 100\% 
\label{eq:MAPE}
\end{equation}
where $n$ is the number of testing samples; $\hat{y}_i$ and $y_i$ correspond to the model-predicted and true performance value of a configuration, respectively. Since MAPE measures the error, the smaller the MAPE, the better the accuracy. 



It is not hard to understand that MAPE is independent of the performance scale, therefore enabling cross-comparisons of models built for different systems and performance attributes. In contrast to other common metrics such as root mean square error (RMSE) and mean absolute error (MAE), MPAE has its unique advantage: it is more scale-robust compared with MAE, while it is more resilient to outliers and is more interpretable than RMSE.

Unlike MAPE, \textbf{$\mathbf{\mu}$RD} is a rank-based metric~\cite{flash} that provides the coarse-grained quantification of the accuracy. The motivation is that tuning configuration might care less about the extent of performance deviation between the configurations, but only whether they are better or worse.

In a nutshell, $\mu$RD computes the number of pairs in the test data where the performance model ranks them incorrectly and measures the average rank difference. It can be calculated using equation~(\ref{eq:MURD}) :

\begin{equation}
\mu RD = \frac{1}{n} \sum_{i=1}^{n} |rank(y_i) - rank(\hat{y}_i)|
\label{eq:MURD}
\end{equation}
in which $rank$ denotes the function that produces the rank value of a performance value among the others in the testing data; $n$ signifies the number of testing samples. $y_i$ corresponds to the true performance value of the configuration, and $\hat{y}_i$ stands for the predicted performance value for the same configuration.

\subsection{Landscape Features}
\label{sec:land-metrics}



Indeed, from the literature, there exist hundreds of landscape features~\cite{8832171, DBLP:conf/gecco/LiefoogheVLZM21, DBLP:conf/gecco/MersmannBTPWR11}. However, not all of them are suitable for our case following the criteria below: 

\begin{itemize}

    \item \textbf{Using representative from distinct categories.} For example, both Maximum Information Entropy (\texttt{MIE}) and ratio of partial information measure the amount of information contained by the fitness distribution, but the former takes a more holistic view in the entropy computation, and hence it is chosen as a representative. Similarly, many features assess the difficulty of the configuration landscape, e.g., Fitness Distance Correlation (\texttt{FDC}), fitness mean, and distance mean, from which we use \texttt{FDC} since it is more widely applied for fitness landscape analysis \cite{10.1145/3728954}.
    


    \item \textbf{Omitting multi-objective features.} Since we are focusing on tuning a single performance objective, landscape features that are designed for multi-objective optimization are omitted~\cite{8832171}. 
    

    
    

\end{itemize}

The above apply to both the global and local landscape features.  A brief summary of all landscape features considered can be found in Table~\ref{tb: landscape}.

\input{Tables/metrics}

\subsubsection{Global Features}


\textbf{Fitness Distance Correlation (\texttt{FDC})} \cite{DBLP:conf/evoW/MullerS11} measures the correlation between a configuration's performance value and its distance to the global optimum, with a value range of $[-1, 1]$. 
\begin{equation}
    FDC = \frac{\sum_{i=1}^{n}(f_i - \bar{f})(d_i - \bar{d})}{\sqrt{\sum_{i=1}^{n}(f_i - \bar{f})^2 \sum_{i=1}^{n}(d_i - \bar{d})^2}}
\end{equation}
where $f_i$ is the performance value, $d_i$ is the distance to the best solution, $\bar{f}$ and $\bar{d}$ are their respective means. Assuming the performance is to be minimized, when the \texttt{FDC} value is close to $1$, the performance value becomes better since the distance reduces, indicating an easier tuning problem as the guidance provided by closeness of configurations is stronger. In contrast, if the \texttt{FDC} is close to $-1$, the performance value becomes worse despite reduced distance, suggesting that the guidance is more misleading and hence more challenging for tuning. A \texttt{FDC} of $0$ implies no information can be found. As a global metric, \texttt{FDC} effectively reflects the general search difficulty of the problem.

\textbf{Fitness Best Distance (\texttt{FBD})} \cite{10.1145/3728954} quantifies the Hamming distance between the best configuration in the current configuration set---which is often sampled via random walk---and the known global optimum, with a value range of $[0, n]$, where $n$ is the dimension of options. 
\begin{equation}
    FBD = \min_{x^* \in S^*} \min_{x \in X_{best}} d(x, x^*)
\end{equation}
where $S^*$ is the set of global optima, $X_{best}$ is the set of best configuration found, and $d(\cdot, \cdot)$ denotes the Hamming distance. \texttt{FBD} directly indicates the proximity of the current best configuration to the global optimum: the smaller value means that the global optimum is not far from a best configuration found, hence easier for a tuner. 


\textbf{Skewness (\texttt{Ske})} \cite{pearson1895contributions} and \textbf{Kurtosis (\texttt{Kur})} \cite{decarlo1997meaning} are statistical measures that describe the shape of a fitness/performance distribution in the tuning problem from distinct perspectives. Specifically, \texttt{Ske} quantifies the asymmetry: for minimizing performance, a positive skewness indicates a longer right tail and a negative skewness indicates a longer left tail, providing insights into the distribution's non-uniformity for the performance. As such, a smaller \texttt{Ske} value is better, since it is less common to have worse performance values. \texttt{Kur} measures the peakedness or flatness of a landscape's performance, where a value greater than $3$ indicates a more peaked distribution and a value less than $3$ suggests a flatter one, revealing the concentration of solutions and the complexity of the configuration space. Thus, a smaller \texttt{Kur} implies a relatively easier tuning problem. Formally, they are calculated as:
\begin{equation}
    Ske = \frac{\mathbb{E}[(Y - \mu)^3]}{\sigma^3}
\end{equation}
\begin{equation}    
    Kur = \frac{\mathbb{E}[(Y - \mu)^4]}{\sigma^4}
\end{equation}
where $Y$ is the fitness/performance distribution; $\mu$ and $\sigma$ are the mean and standard deviation of $Y$, respectively.  




\subsubsection{Local Features}


\textbf{Proportion of Local Optima (\texttt{PLO})} \cite{8832171} measures the ratio of local optima to the total number of configurations in the configuration space, with a value range of $[0, 1]$ (assume minimization). 
\begin{equation}
    PLO = \frac{1}{|X|} \sum_{x \in X} \mathbb{I}[f(x) \leq f(y) \text{ for all } y \in N(x)]
\end{equation}
where $N(x)$ is the neighborhood of $x$, and $\mathbb{I}[\cdot]$ is the indicator function, calculated as the number of local optima divided by the total number of configurations. \texttt{PLO} reflects the multi-modal nature of the configuration space, where it is possible to have multiple global optima. A high \texttt{PLO} value indicates a large number of local optima, potentially increasing the risk of a tuner being trapped in local optima. 


\textbf{Correlation Length (\texttt{CL})} \cite{stadler1996landscapes} is a key metric that quantifies the ruggedness of the configuration landscape.
\begin{align}
    CL &= -\frac{1}{\log|r_1|} \\
    r_1 &= \frac{\sum_{i=1}^{n-1}(y_i - \bar{y})(y_{i+1} - \bar{y})}{\sum_{i=1}^{n}(y_i - \bar{y})^2}
\end{align}
wherein $r_1$ is the autocorrelation. Practically, \texttt{CL} is computed by sampling the configuration space through a random walk to generate a sequence, calculating the autocorrelation coefficients of the sequence, and applying a logarithmic transformation to obtain the final correlation length. A bigger absolute \texttt{CL} indicates a smoother landscape, while a smaller absolute \texttt{CL} suggests a more rugged landscape, which implies severe local optima traps. \texttt{CL} is valuable for guiding the selection of local search step sizes, evaluating the effectiveness of neighborhood structures, and predicting the efficiency of local search strategies.

\textbf{Maximum Information Entropy (\texttt{MIE})} \cite{6719480} reflects the information complexity of the configuration space. 
\begin{align}
    MIE &= \max_{\epsilon} H(\epsilon) \\
    H(\epsilon) &= -\sum_{i,j} p_{ij}(\epsilon) \log_6 p_{ij}(\epsilon)
\end{align}
and $p_{ij}(\epsilon)$ represents the transition probabilities in the $\epsilon$-based information content analysis. Drawing on the information entropy theory, it accounts for local variations in the configuration space, serving as an indication of the information quality of local optima. A higher \texttt{MIE} indicates the landscape contains local optima of more complex structure. This metric is instrumental in assessing problem complexity, evaluating the difficulty of algorithm convergence, and guiding the selection of appropriate search strategies.

\textbf{Nearest Better Clustering (\texttt{NBC})} \cite{10.1145/2739480.2754642} analyzes the local clustering characteristics of the configuration space. 
\begin{equation}
    NBC = \frac{1}{|X|} \sum_{x \in X} \frac{d_{nb}(x)}{d_{nn}(x)}
\end{equation}
in which $d_{nb}(x)$ is the distance to the nearest better configuration and $d_{nn}(x)$ is the distance to the nearest neighbor. \texttt{NBC} is calculated by identifying the nearest neighbor with a better fitness/performance value for each configuration, computing the mean of the distance ratios. \texttt{NBC} reflects the degree of clustering in the configuration space, describes the distribution of local optimal configurations, and reveals the local structure of the landscape. If \texttt{NBC} is higher, it indicates that it is easy to find a better configuration in a localized region, as the size of the local optima cluster is smaller.


\subsection{Metric for Tuning Quality}

In this work, we are concerned with the most popular aspect of the tuning quality, namely the performance, which is system-dependent. Naturally, we use the measured performance of the best measured configuration. Notably, according to Table~\ref{tb:systems}, some of those performance metrics are to be minimized (e.g., the latency for \textsc{MariaDB}) while others are to be maximized (e.g., the throughput for \textsc{Hadoop}). \revision{\mytag{R1C1} Yet, for the simplicity of exposition and interpretation, in this work, we convert all maximizing performance to be minimized via additive inversion\footnote{\revision{Note that additive inversion changes neither the ranking nor density/scale of the optimization problem, but merely flip the distribution and the side of optimality (but the shape of distribution remains identical)~\cite{kemple1991stratigraphic}.}}: if we have an optimization problem to solve as $\text{argmax }f(x)$, we invert it as $\text{argmin }-f(x)$.}


\subsection{Statistical Validation}
\label{sec:sta}


To mitigate stochastic bias, we use several statistical methods.

\subsubsection{Wilcoxon Signed Rank/Rank Sum Test}

To verify pairwise comparisons, we use Wilcoxon test---a non-parametric test that has been recommended for software engineering research~\cite{DBLP:conf/icse/ArcuriB11}. We apply both of its paired and non-paired versions, i.e., Wilcoxon signed-rank and Wilcoxon rank-sum test, depending on the RQs and needs. We set a confidence level of $\alpha=95\%$, meaning that if the comparison results in $p<0.05$ then the difference is statistically significant. 


\subsubsection{Scott-Knott Effect Size Difference (ESD) Test}

When comparing multiple models, we leverage the non-parametric version of the Scott-Knott Effect Size Difference (ESD) test\footnote{We use the non-parametric version of the Scott-Knott ESD test.}~\cite{tantithamthavorn2017mvt}. In a nutshell, Scott-Knott ESD test sorts the list of treatments (e.g., models) by their median values of the metric. Next, it splits the list into two sub-lists with the largest expected difference~\cite{xia2018hyperparameter}. For example, suppose that we compare $A$, $B$, and $C$, a possible split could be $\{A, B\}$, $\{C\}$, with the rank ($r$) of $1$ and $2$, respectively. This means that, in the statistical sense, $A$ and $B$ perform similarly, but they are significantly better than $C$. 



In contrast to other non-parametric statistical tests on multiple comparisons (e.g., Kruskal-Wallis test~\cite{mckight2010kruskal}), Scott-Knott ESD test offers the following unique advantages:

\begin{itemize}
    \item It does not require posterior correction, as the comparisons are essentially conducted in a pairwise manner.
    \item It not only shows whether some treatments are statistically different or not, but also indicates which one is better than another, i.e., by means of ranking, while avoiding overlapping groups.
    \item It overcomes the confounding factor of overlapping groups that are produced by other post hoc tests.
\end{itemize}

\subsubsection{Spearman's Rank Correlation}

To quantify the relationship between two metrics, we leverage Spearman's rank correlation ($\rho$)~\cite{myers2004spearman}, which is a widely used indicator in software engineering~\cite{chen2019all, DBLP:journals/tse/WattanakriengkraiWKTTIM23}. Specifically, Spearman's rank correlation measures the nonlinear monotonic relation between two random variables, and we have $-1\leq \rho \leq1$. $|\rho|$ represents the strength of monotonic correlation, and $\rho=0$ means that the two metrics do not correlate with each other in any way; $-1\leq \rho<0$ and $0< \rho \leq 1$ denote that the monotonic correlation is negative and positive, respectively. 

As widely used in software engineering~\cite{DBLP:journals/infsof/SamoladasAS10,DBLP:journals/tse/WattanakriengkraiWKTTIM23}, the Spearman's rank correlation can be interpreted as: \textbf{negligible:} $0 \leq |\rho| \leq 0.09$; \textbf{weak:} $0.09 < |\rho| \leq 0.39$; \textbf{moderate:} $0.39 < |\rho| \leq 0.69$; \textbf{strong:} $0.69 < |\rho| \leq 1$. The statistical significance of the $\rho$ value is verified by z-score under a confidence level of $95\%$, i.e., the correlation is significant only when $p<0.05$. Commonly, only moderate level or higher ($|\rho| > 0.39$) with $p<0.05$ are significant enough to create practical implication~\cite{accuracy_can_lie,DBLP:journals/tse/WattanakriengkraiWKTTIM23}.


\subsection{Settings for Model Training and Testing}

To train the surrogate models, for each system, we randomly sample a small set of measured configuration-performance pairs from the available one, following exactly what has been done in prior work \cite{DaL, SPL}: for binary systems, the sample size would be $5n$ where $n$ is the number of configuration options; for mixed/non-binary systems, we use \texttt{SPLConqueror} \cite{SPL} to estimate the size. All the remaining samples in the systems' datasets serve as the testing data, which would be used to compute both the landscape features and accuracy metrics\footnote{The training data is randomly chosen in each repeated run.}. The results are shown in Table~\ref{tb:systems}.


\begin{figure}[t!]
\centering
\includegraphics[width=0.6\columnwidth]{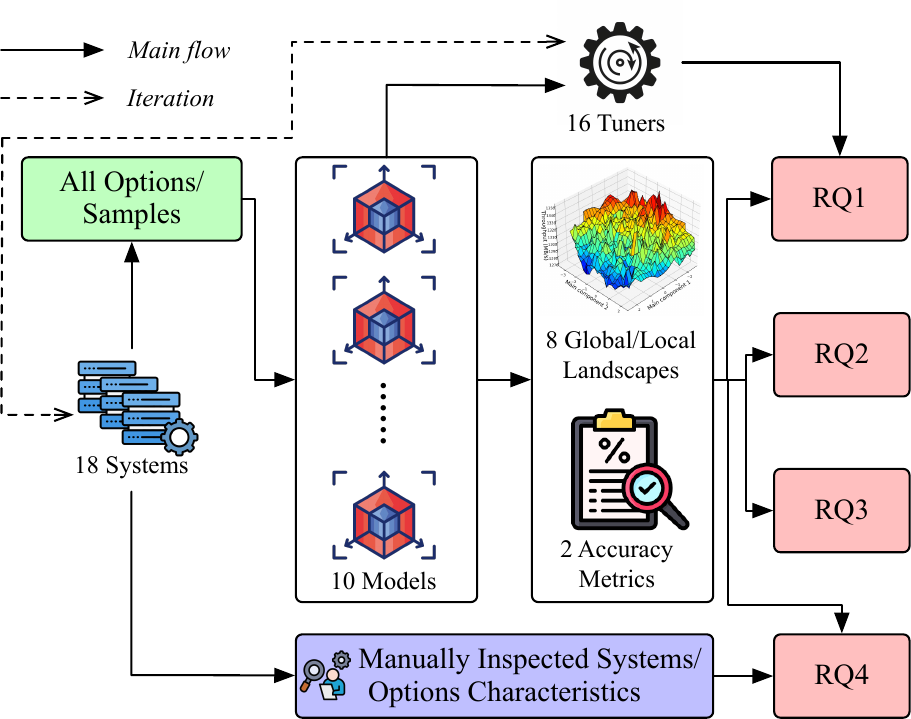}
\caption{Methodology of the empirical study.}
\label{fig:overview}
\end{figure}

\subsection{Procedure of Empirical Study}

The methodology of the empirical study has been depicted in Figure~\ref{fig:overview}. In total, our study covers $18$ configurable systems, $10$ surrogate models, $16$ tuners, $8$ landscape features, and $2$ accuracy metrics, ranging between $1,980$ and $27,000$ cases for the corresponding RQs. To mitigate the stochastic noise, we repeat each experiment $30$ times with varying seeds while verifying their statistical significance. Note that answering \textbf{RQ1} and \textbf{RQ4} additionally requires conducting iterated tuning under diverse model-tuner pairs (see Section~\ref{sec:rq1}) and manually inspecting the characteristics of configuration options (see Section~\ref{sec: rq4}), respectively. 

\mytag{R1C6}\revision{It is worth noting that for model training (\textbf{RQ1-4}), the $30$ different seeds lead to different samples for training and testing samples in the split (but the number of samples remains the same), and different stochastic behaviors in the model's training. This is also reflected in the accuracy and landscape features since they are based on the testing samples. When running the tuners (\textbf{RQ1}), we also use $30$ different seeds, which would create different samples for the surrogate initialization (if applicable) and the different stochastic behaviors in the tuning.}

All experiments were run in parallel on two high-performance servers each with Ubuntu 20.04.6 LTS, two Intel(R) Xeon(R) Platinum 50 cores CPU@2.30GHz (with two Nvidia A100 80GB GPU support for training the deep learning-based models), and 500GB memory over the period of $14$ months, $24 \times 7$---roughly $10,080$ GPU/CPU hours. Furthermore, manually categorizing the characteristics of systems' options has also taken $\approx6.4$ person months.

%% file: Tables/systems.tex
\begin{table*}[t!]
\centering
\footnotesize
\caption{Details of the subject systems. ($|\mathbfcal{B}|$/$|\mathbfcal{N}|$) denotes the number of binary/numeric options. $\mathbfcal{S}_{space}$ is the full configuration/search space. $\mathbfcal{S}_{train}$ and $\mathbfcal{S}_{test}$ denote the training and testing size for a surrogate model, respectively.}
\label{tb:systems}

\begin{adjustbox}{width=\linewidth,center}

\begin{tabular}{llllllllll}
\toprule
\textbf{System} & \textbf{Version} & \textbf{Workload} & \textbf{Domain} & \textbf{Performance} & \textbf{$|\mathbfcal{B}|$/$|\mathbfcal{N}|$}  & \textbf{$\mathbfcal{S}_{space}$} &\textbf{$\mathbfcal{S}_{train}$} & \textbf{$\mathbfcal{S}_{test}$} & \textbf{Ref.} \\ \midrule
\textsc{Apache} & 21.0 & ApacheBench 2.3 & Web Server & Maximum load ($\#$ users) & 14/2  & $2.35 \times 10^8$ & 507 & 12933 & \cite{DBLP:conf/icse/WeberKSAS23} \\

\rowcolor{gree!15}\textsc{7z} & 9.20 & Compressing a 3 GB directory & File Compressor & Runtime (ms) & 11/3  & $1.68 \times 10^8$ & 3600 & 65040 & \cite{DBLP:conf/icse/WeberKSAS23} \\


\textsc{DConvert} & 3.0 & Transform resources at different scales & Image Scaling & Runtime (s) & 17/1  & $1.05 \times 10^7$ & 578 & 6186& \cite{muhlbauer2023analyzing} \\

\rowcolor{gree!15}\textsc{DeepArch} & 2.2.4 & UCR Archive time series dataset & Deep Learning Tool & Runtime (min) & 12/0  & $4.10 \times 10^3$ & 2867 & 1229& \cite{DBLP:conf/sigsoft/JamshidiVKS18} \\

\textsc{ExaStencils} & 1.2 & Three default benchmarks & Code Generator & Runtime (ms) & 7/5  & $1.61 \times 10^9$ & 2401& 83675& \cite{DBLP:conf/icse/WeberKSAS23} \\

\rowcolor{gree!15}\textsc{Hadoop} & 3.0 & HiBench & Data Analytics & Throughput (MB/s) & 2/7  & $1.27 \times 10^{29}$  & 250 & 3744& \cite{DBLP:journals/jss/CaoBWZLZ23}\\

\textsc{MariaDB} & 10.5 & Sysbench & Database & Runtime (ms) & 8/3  & $5.31 \times 10^{11}$ & 196& 776& \cite{DBLP:journals/corr/abs-2106-02716} \\

\rowcolor{gree!15}\textsc{MongoDB} & 4.4 & Yahoo! cloud serving benchmark & Database & Runtime (ms) & 14/2  & $3.77 \times 10^8$ & 1008 &5832 & \cite{DBLP:journals/corr/abs-2106-02716}  \\

\textsc{PostgreSQL} & 22.0 & PolePosition 0.6.0 & Database & Runtime (ms) & 6/3  & $1.42 \times 10^9$ & 108 & 18900& \cite{DBLP:conf/icse/WeberKSAS23} \\

\rowcolor{gree!15}\textsc{Redis} & 6.0 & Sysbench & Database & Requests per second & 1/8  & $5.78 \times 10^{16}$ & 1633& 2767& \cite{DBLP:journals/jss/CaoBWZLZ23} \\

\textsc{Spark} & 3.0 & HiBench & Data Analytics & Throughput (MB/s) & 5/8  & $2.55 \times 10^{12}$ &252 & 2448& \cite{DBLP:journals/jss/CaoBWZLZ23}\\ 

\rowcolor{gree!15}\textsc{Storm} & 0.9.5 & Randomly generated benchmark & Data Analytics & Latency (s) & 0/12  & $2.83 \times 10^{23}$ & 543 & 1004 &\cite{DBLP:journals/tse/KrishnaNJM21} \\

\textsc{HSMGP} & 1.0 & Perform one V-cycle & Stencil-Grid Solver & Latency (ms) & 11/3  & $2.97 \times 10^8$ &2419 & 1037 & \cite{DeepPerf} \\

\rowcolor{gree!15}\textsc{XGBoost} & 12.0 & Two standard datasets & Machine Learning Tool & Runtime (min) & 11/0 & $2.04 \times 10^3$&121 & 1927 & \cite{DBLP:conf/sigsoft/JamshidiVKS18} \\

\textsc{HIPAcc} & 0.8 & A set of partial differential equations & Image Processing & Latency (ms) & 31/2  & $3.30 \times 10^{12}$ & 9439& 4046& \cite{DeepPerf} \\

\rowcolor{gree!15}\textsc{SQLite} & 3.0 & Benchmark provided by vendor & Database & Runtime (s) & 39/0  & $5.50 \times 10^{11}$ & 3187& 1366& \cite{DeepPerf} \\

\textsc{JavaGC} & 7.0 & DaCapo benchmark suite & Java Runtime & Runtime (ms) & 12/23  & $2.67 \times 10^{41}$ & 3718& 163257& \cite{DeepPerf} \\

\rowcolor{gree!15}\textsc{Polly} & 3.9 & The gemm program from polybench & Code Optimizer & Runtime (s) & 39/0  & $5.50 \times 10^{11}$ & 1521& 58479& \cite{HINNPerf} \\

\bottomrule
\end{tabular}

\end{adjustbox}

\end{table*}

%% file: Tables/surrogate.tex
\begin{table*}[t!]
\centering
\footnotesize
\caption{Specification of the surrogate models considered. ML and DL denotes classic Machine Learning and Deep Learning, respectively.}
\begin{adjustbox}{width=\linewidth,center}
\begin{tabular}{lllllll}
\toprule
\textbf{Model} & \textbf{Abbr.} & \textbf{Type} & \textbf{Domain} & \textbf{Characteristics} & \textbf{Ref.} & \textbf{Used by} \\ \midrule

Support Vector Regression & \texttt{SVR}&ML&General&Good for high dimensionality but can be sensitive to its parameters&\cite{SVM}&\cite{DaL}\\
\rowcolor{gree!15}Linear Regression & \texttt{LR}&ML&General&Fast execution but only suitable for independent linear relationships&\cite{montgomery2021LR}&\cite{DaL}\\
Gaussian Process & \texttt{GP}&ML&General&Providing uncertainty estimates but only suitable for low dimension&\cite{GPR}&\cite{llamatune}\\
\rowcolor{gree!15}Decision Tree & \texttt{DT}&ML&General&Easy to interpret but can be overfitting and instable&\cite{DT}&\cite{flash}\\
Random Forests & \texttt{RF}&ML&General&High accuracy and stable but can be complex and difficult to explain&\cite{RF}&\cite{BOCA}\\
\rowcolor{gree!15}\texttt{DECART} & \texttt{DCT}&ML&Configurable systems&Handling non-linear relationships with tuned parameters but can overfit&\cite{DECART}&\cite{DeepPerf}\\
\texttt{SPLConqueror} & \texttt{SPL}&ML&Configurable systems&Handling option interactions but mainly work linearly&\cite{SPL}&\cite{HINNPerf}\\
\rowcolor{gree!15}\texttt{DaL}& \texttt{DaL} &ML/DL&Configurable systems&Efficiently addressing sample
sparsity but can be expensive to train&\cite{DaL}&\cite{DaL}\\
\texttt{DeepPerf}& \texttt{DeP}&DL&Configurable systems&Accurate for sparse options but sensitive to hyperparameters&\cite{DeepPerf}&\cite{HINNPerf}\\
\rowcolor{gree!15}\texttt{HINNPerf} &\texttt{HIP}&DL&Configurable systems&High accuracy through hierarchical modeling but hard to tune&\cite{HINNPerf}&\cite{DaL}\\

\bottomrule

%
\end{tabular}
\end{adjustbox}
\label{tb:models}
\end{table*}

%% file: Tables/metrics.tex
\begin{table*}[t!]
\centering
\footnotesize
\setlength{\tabcolsep}{1mm}
\caption{The landscape features considered. $\uparrow$ and $\downarrow$ denote the interpretations for bigger and smaller value, respectively (assuming that the performance metric is to be minimized).}
\begin{adjustbox}{width=\linewidth,center}
\begin{tabular}{lllp{8.4cm}ll}
\toprule
\textbf{Metric} & \textbf{Abbr.} &\textbf{Domain} & \textbf{Characteristics} &  $\uparrow$ \textbf{Bigger Value} &  $\downarrow$ \textbf{Smaller Value}   \\ \midrule

Fitness Distance Correlation \cite{DBLP:conf/evoW/MullerS11} & \texttt{FDC} & Global  & Correlation between configurations' distances  and the distance to the optimal configuration. & more straightforward guidance & more misleading guidance\\
\rowcolor{gree!15}From Best Distance \cite{10.1145/3728954} &\texttt{FBD} & Global  & Distance between the configuration and the global optimum. & more dense optimum & more sparse optimum \\
Skewness \cite{pearson1895contributions} &\texttt{Ske} & Global & Describes the skewness of the landscape.& worse performance is common & better performance is common\\
\rowcolor{gree!15}Kurtosis \cite{decarlo1997meaning} &\texttt{Kur} & Global & Describes the sharpness or flatness of the landscape's peak. & more peaked performance & flatter performance\\ \midrule
Proportion of Local Optima \cite{8832171} &\texttt{PLO} & Local & Proportion of local optima in the configuration space. & more local optima & less local optima\\
\rowcolor{gree!15}Correlation Length \cite{stadler1996landscapes} &\texttt{CL} & Local & The rate at which landscape correlation decays with distance. & more smoother landscape & more rugged landscape\\
Maximum Information Entropy \cite{6719480} &\texttt{MIE} & Local  & Logarithmic operation based on information content.& more complex local optima & simpler local optima \\
\rowcolor{gree!15}Nearest Better Cluster \cite{10.1145/2739480.2754642}  &\texttt{NBC}& Local  & Relationship between nearest neighbor and nearest better neighbor. & larger local optima cluster & smaller local optima cluster\\

\bottomrule

%
\end{tabular}
\end{adjustbox}
\label{tb: landscape}
\end{table*}

%% file: results.tex
\section{Results and Analysis}
\label{sec:results}

In this section, we summarize and discuss the results obtained.

\subsection{RQ1: Verifying New Theory of Model Usefulness}
\label{sec:rq1}

\subsubsection{Method}

To verify our theoretical hypothesis in \textbf{RQ1}, we propose a simple method to instantiate landscape dominance based on \textit{Pareto dominance}~\cite{DBLP:journals/ec/VeldhuizenL00}---a common concept in multi-objective optimization---for investigating whether the global and local landscape features are helpful. \mytag{R2C3}\revision{Indeed, Pareto dominance is common for multi-objective decision making where there are no explicit preferences on any (conflicting) objective to make trade-off~\cite{DBLP:journals/tse/LiCY22}. In our case, the global and local landscape features can be seen as independent indicators of different aspects of the landscape, which might not be consistent with each other, thus this naturally fits the purpose of Pareto dominance.}




Traditionally, the Pareto dominance relationship is defined as a binary relationship between two solutions $\mathcal{A}$ and $\mathcal{B}$ with respect to a set of objectives $\mathbf{F}$ (assuming minimization), from which $\mathcal{A}$ dominates $\mathcal{B}$ is denoted as:
\begin{equation}
\begin{split}
\mathcal{A} \succ \mathcal{B} \iff &\ (f_{i,\mathcal{A}} \leq f_{i,\mathcal{B}},\forall f_i \in \mathbf{F}) \land (f_{j,\mathcal{A}} < f_{j,\mathcal{B}},\exists f_j \in \mathbf{F}) 
\end{split}
\end{equation}
whereby $f_{i,\mathcal{A}}$ and $f_{i,\mathcal{B}}$ are the values of the $i$th objective for solution $\mathcal{A}$ and $\mathcal{B}$, respectively. 



However, since it is unclear about the relative importance within/between the global and local features, while it is difficult for us to precisely define their preferences, it is natural to see every global and local feature as an independent objective. Yet, this would lead to too many Pareto non-dominated and incomparable models, since we have a total of eight features~\cite{DBLP:journals/csur/LiLTY15}. \revision{\mytag{R1C3} To mitigate this and verify our theory, we examine three alternative variants:}

\begin{itemize}
    \item \revision{\textbf{No aggregation:} Pick a combination of landscape features at a time (one from global features and the other from the local features) without aggregation over the combinations of global and local features.}
    \item \revision{\textbf{Majority vote:} Conduct majority vote over all possible combinations of global and local features.}
    \item \revision{\textbf{Weighted sum:} perform weighted sum aggregation over all possible combinations of global and local features.}
\end{itemize}

\revision{We examine all the possible comparisons between any two models under each of the above options.} Specifically, according to our definition of the \textbf{\textit{landscape dominance}} for ``useful'' model from Section~\ref{Sec:Theory}, we define and instantiate two important components (the models are trained and tested using the sample size in Table~\ref{tb:systems} for calculating the landscape features):

\begin{itemize}
    \item \textbf{Absolute deviation of a global feature $\Delta g$:} We measure the distance between the feature of the model emulated landscape ($g_{model}$) and that of the real landscape ($g_{system}$): $\Delta g= |g_{system} - g_{model}|$. Note that $g_{model}$ and $g_{system}$ are computed using the same points in the testing dataset.
    \item \textbf{A local feature $l$:} We directly measure the landscape features ($l$) of the landscape built by the models. For the ease of exposition, we additively convert all features such that a smaller value represents easier local optima case.
\end{itemize}
We expect both $\Delta g$ and $l$ to be smaller. Yet, the above variants also lead to diverse actual realization of the landscape dominance.

For the ``no aggregation'' variant, given two surrogate models $\mathcal{A}$ and $\mathcal{B}$, the proposed {landscape dominance} can be formalized by directly leveraging the Pareto dominance on the two objectives of a chosen global and local feature combination as:
\begin{equation}
\mathcal{A} \succ \mathcal{B} \iff  (\Delta g_{i,\mathcal{A}} \leq \Delta g_{i,\mathcal{B}} \land l_{i,\mathcal{A}} < l_{i,\mathcal{B}}) \lor
(\Delta g_{i,\mathcal{A}} < \Delta g_{i,\mathcal{B}} \land l_{i,\mathcal{A}} \leq l_{i,\mathcal{B}})
\end{equation}
whereby $\Delta g_{i,\mathcal{A}}$ and $\Delta g_{i,\mathcal{B}}$ are the value of the $i$th chosen global feature deviation for model $\mathcal{A}$ and $\mathcal{B}$, respectively; the same applies to $l_{i,\mathcal{A}}$ and $l_{i,\mathcal{B}}$ which are for the $i$th selected local feature (one combination of $\Delta g_i$ and $l_i$ at a time). Here, $\mathcal{A}$ is the \textbf{\textit{landscape dominating}} model when $\mathcal{B}$ is the \textbf{\textit{landscape dominated}} one, if $\mathcal{A}\succ\mathcal{B}$, hence $\mathcal{A}$ is better and more useful. \revision{\mytag{R1C3} We examine the combinations of the global and local features as mutually exclusive cases. For example, $\mathcal{A}$ might dominate $\mathcal{B}$ under \texttt{FDC} and \texttt{CL}, but it could become $\mathcal{B}$ dominates $\mathcal{A}$ under \texttt{Kur} and \texttt{PLO}; yet both are valid and individual samples of comparison.}

\revision{\mytag{R1C3} For the ``majority vote'' variant, the landscape dominance can be realized by aggregating the outcomes under all combinations of global and local features by counting on how many times a model is Pareto dominated by the others in total. That is, we have the following for any pair of models $\mathcal{A}$ and $\mathcal{B}$:
\begin{equation}
\begin{split}
\mathcal{A} \rhd \mathcal{B} \iff &\ \sum_{i,\mathcal{X}} |\mathcal{X} \succ \mathcal{A}: (\Delta g_{i,\mathcal{X}} \leq \Delta g_{i,\mathcal{A}} \land l_{i,\mathcal{X}} < l_{i,\mathcal{A}}) \lor (\Delta g_{i,\mathcal{X}} < \Delta g_{i,\mathcal{A}} \land l_{i,\mathcal{X}} \leq l_{i,\mathcal{A}})| <\\
&\ \sum_{i,\mathcal{X}} | \mathcal{X} \succ \mathcal{B}: (\Delta g_{i,\mathcal{X}} \leq \Delta g_{i,\mathcal{B}} \land l_{i,\mathcal{X}} < l_{i,\mathcal{B}}) \lor (\Delta g_{i,\mathcal{X}} < \Delta g_{i,\mathcal{B}} \land l_{i,\mathcal{X}} \leq l_{i,\mathcal{B}})|
\end{split}
\end{equation}
where $i$ denote the arbitrary $i$th combination of global and local features; $\mathcal{X}$ is any other possible model. Here, $\mathcal{A}$ is the \textbf{\textit{landscape dominating}} model when $\mathcal{B}$ is the \textbf{\textit{landscape dominated}} one, if $\mathcal{A}\rhd\mathcal{B}$ (over all combinations of global and local features), hence $\mathcal{A}$ is better and more useful. For example, if $\mathcal{B}$ has $\mathcal{C}\succ\mathcal{B}$ and $\mathcal{A}\succ\mathcal{B}$ under \texttt{FDC} and \texttt{CL}; and $\mathcal{A}\succ\mathcal{B}$ under \texttt{FDC} and \texttt{PLO}, and it is Pareto non-dominated under all other combinations, we say it has a count of 3. Model $\mathcal{A}$ only has $\mathcal{D}\succ\mathcal{A}$ under \texttt{Kur} and \texttt{PLO}; it is Pareto non-dominated under all other combinations. This would lead to a count of 1 for $\mathcal{A}$. Hence, we have $\mathcal{A}\rhd\mathcal{B}$ and $\mathcal{A}$ should be preferred. We consider the comparison between any pair of models as the sample.}

\revision{\mytag{R1C3} For the ``weighted sum'' variant, the variant of landscape domains become the aggregation of all global and local features via normalized weighted sum. Formally, for any pair of models $\mathcal{A}$ and $\mathcal{B}$:
\begin{equation}
\begin{split}
\mathcal{A} \gg \mathcal{B} \iff &\ \hat{\Delta g}_{FDC,\mathcal{A}} + \hat{\Delta g}_{FBD,\mathcal{A}} + \hat{\Delta g}_{Ske,\mathcal{A}} + \hat{\Delta g}_{Kur,\mathcal{A}} + \hat{l}_{PLO,\mathcal{A}} + \hat{l}_{CL,\mathcal{A}} + \hat{l}_{MIE,\mathcal{A}} + \hat{l}_{NBC,\mathcal{A}} <\\
&\ \hat{\Delta g}_{FDC,\mathcal{B}} + \hat{\Delta g}_{FBD,\mathcal{B}} + \hat{\Delta g}_{Ske,\mathcal{B}} + \hat{\Delta g}_{Kur,\mathcal{B}} + \hat{l}_{PLO,\mathcal{B}} + \hat{l}_{CL,\mathcal{B}} + \hat{l}_{MIE,\mathcal{B}} + \hat{l}_{NBC,\mathcal{B}}
\end{split}
\end{equation}
where $\hat{\Delta g}$ and $\hat{l}$ denote the normalized landscape feature values. $\mathcal{A}$ is the \textbf{\textit{landscape dominating}} model when $\mathcal{B}$ is the \textbf{\textit{landscape dominated}} one, if $\mathcal{A}\gg\mathcal{B}$ (over all combinations of global and local features), hence $\mathcal{A}$ is better and more useful. Again, we consider the comparison between any pair of models as the sample.}

\input{Tables/tuners}
\input{Tables/budget}

Next, we pair the models with 16 state-of-the-art tuners, which have been used in prior work~\cite{accuracy_can_lie}, and run them on all the systems. We follow the same method to identify the budgets that ensure all tuners can reasonably converge. For the hot-start sample size of sequential model-based tuners, we use the same setting of $20$ measurements as used by Chen et al.~\cite{accuracy_can_lie}. The training size from Table~\ref{tb:systems} is used for the batch model-based tuners. The results can be found in Tables~\ref{tb:tuners} and~\ref{tb:budgets}. \revision{\mytag{R1C2} Note that for sequential model-based tuners, the budget denotes the hot-start sample size and the system measurements, which allows them to achieve generally reasonable convergence. For batch model-based tuners, the budget only means the number of model evaluations, as the model is pre-trained using the training samples (again, those are system measurements too) from Table~\ref{tb:systems}, which have been shown to be sufficient for training a good model in advance. This is because our goal is not to directly and individually compare the sequential and batch model-based tuners, but to examine how they, as the two key alternative model-based tuning paradigms, can be sensitive to the surrogate model. It is important to examine them under their respective best/most representative conditions. As a result, the budget difference for them is purposely designed to enable their respective best/most representative states.}


\input{Tables/rq1-example}

Drawing on the above, for each system, we perform the following:


\begin{enumerate}[label=(\roman*)]
    \item Pick a tuner (from the total of $16$ tuners).
    \item \revision{\mytag{R1C3} Examine all combinations of global and local landscape features (total of $4\times4=16$ combinations).}
    \item \revision{\mytag{R1C3} Compute the necessary $\Delta g$ and $l$ (or $\hat{\Delta g}$ and $\hat{l}$) for each of the 10 models considered.}
    \item \revision{\mytag{R1C3} Put any pair of models ($\mathcal{A}$ and $\mathcal{B}$), such that it satisfies $\mathcal{A}\succ\mathcal{B}$/$\mathcal{A}\rhd\mathcal{B}$/$\mathcal{A}\gg\mathcal{B}$, into the landscape dominating (DG) group and landscape dominated (DD) group, respectively.}
    \item Repeat from (i) until all tuners have been investigated.
\end{enumerate}

\revision{\mytag{R1C3} Therefore, for a system, in total we need to extract the pairs with dominance relations for $16 \times 16=256$ tuner and global/local feature combination, under each of which there might be $0$ or more dominating/dominated pairs (for ``no aggregation''); for ``majority vote'' and ``weighted sum'', there can be a maximum of 45 pairs since we compare all 10 models by aggregating all combinations of global and local features. In general, to confirm our theory, we expect that the \% of wins of the models from the DG group in the pairs is higher than the \% of their losses.}

An example of a given system has been shown in Table~\ref{tb:rq1-exp} with the possible valid samples. Clearly, the models come in pairs from a tuner under a particular/all combination of global and local landscape features, hence the size of both the dominating and dominated groups would always be equal. Note that it is possible for the two groups to have overlapping models even under the same tuner and global/local feature combination, e.g., for the ``no aggregation'' variant, model $\mathcal{B}$ is in both the dominating and dominated group for \texttt{GA} when using \texttt{FDC} as the $\Delta g$ and \texttt{CL} as the $l$, since $\mathcal{B}$ is both dominating $\mathcal{C}$ while being dominated by $\mathcal{A}$. As such, we have two pairs: $\mathcal{A} \succ \mathcal{B}$ and $\mathcal{B} \succ \mathcal{C}$. This makes sense, as according to our theory, $\mathcal{B}$ is indeed both more and less useful than one other model. The same applies to the ``majority vote'' and ``weighted sum'' variants.


We report on the mean tuning performance (e.g., latency) deviation of all the $n$ pairs between the dominating and dominated group of a system (we additively invert maximizing performance into minimizing for simple exposition):
\begin{equation}
\Delta p ={1\over n}\sum^n_{i=1} (p_{i,DG} - p_{i,DD})
\end{equation}
where $p_{i,DG}$ is the system's tuned performance when using the $i$th model-tuner pair that belongs to the dominating group; $p_{i,DD}$ is that under the corresponding model-tuner from the dominated group. Therefore, a negative $\Delta p$ generally confirms our theory; otherwise, it is violated. We apply the Wilcoxon signed-rank test to verify the statistical significance of such deviations between the dominating and dominated groups, since we have paired data. The percentages of pairs for which a model in the dominating group wins/loses (i.e., it has better/worse performance) are also reported.


\subsubsection{Results}

As can be seen in Table~\ref{tab:domination_performance_merged}, there are consistent patterns over all three variants of landscape dominance that verify our theory: over $14$ out of $18$ systems ($78\%$), on average, confirm that the dominating models can enable much better tuning performance than their dominated counterpart in the pair, majority of which show statistical significance. The results are even more obvious on the percentage of pairs in which the model belongs to the dominating group wins: up to $87.5\%$ for a system (under ``no aggregation''). \revision{\mytag{R1C3} Among the variants of realizing landscape dominance, we also note that the ``no aggregation'' yields a higher percentage of DG win, implying that using a particular combination of the global and local features aligns with our theory best.} All the above demonstrate the validity and robustness of our theory on the relative model usefulness for tuning based on landscape features. In conclusion, we say:


\begin{tcbitemize}[%
    raster columns=1, 
    raster rows=1
    ]
  \tcbitem[myhbox={}{Finding \thefindingcount}]   \textit{In the majority of the cases ($78\%$), a model-emulated landscape with smaller deviation of the overall structure to the real landscape and less severe local optima can generally reflect a more useful model for tuning. Notably, capturing landscape dominance on a particular combination of the global and local features can be the most promising option.}
\end{tcbitemize}
\addtocounter{findingcount}{1}


\input{Tables/rq_landscape_dominance}

\subsection{RQ2: Correlating Accuracy and Landscape Features for Model Usefulness}
\label{sec:rq2}

\input{Tables/rq1_overview}

\subsubsection{Method}

To answer \textbf{RQ2}, we train all $10$ models over the $18$ systems with $30$ repeats, after which we measure their accuracy (MAPE and $\mu$RD) and the landscape features using the samples in Table~\ref{tb:systems}. Following our theory of model usefulness and \textbf{RQ1}, we use the $\Delta g$ and $l$ to represent the landscape features. A smaller $\Delta g$, $l$, MAPE, and $\mu$RD is preferred\footnote{All landscape features and accuracy metrics are computed using the same points in the testing dataset.}.




We then analyze the monotonic relationships between accuracy metrics and $\Delta g$/$l$ of the models using Spearman correlation. Therefore, considering the 10 models and 30 repeated runs, for each accuracy-landscape feature pair under a system, we have $10 \times 30=300$ points in the correlation analysis. When they have a strong positive correlation, we say that the information expressed by the landscape features for model usefulness is largely consistent with that of the accuracy, i.e., a better accuracy can also reflect a more preferred landscape feature.

In addition to the results for every accuracy-landscape feature pair under each system, we also summarize the results from two aspects:

\begin{itemize}
    \item \textbf{Per accuracy-landscape feature pair basis:} This means that, for each accuracy-landscape feature pair, we have $18$ cases since there are $18$ systems.
    \item \textbf{Per system basis:} For each system, we have $2$ accuracy metrics $\times$ $8$ landscape features $=16$ cases.
\end{itemize}

Note that we only consider correlation higher than the moderate level while $p<0.05$ as a significant correlation; all other cases (weak and negligible, or with $p>0.05$) have limited practical implications. A significant positive correlation means that the landscape features are more preferred when the accuracy becomes better, and hence the accuracy metrics can serve as their ``delegate''. Otherwise, either negative or insignificant correlations suggest that the landscape features indeed provide additional information that cannot be covered by the accuracy metrics.


\subsubsection{Results}

\begin{figure}[t!]
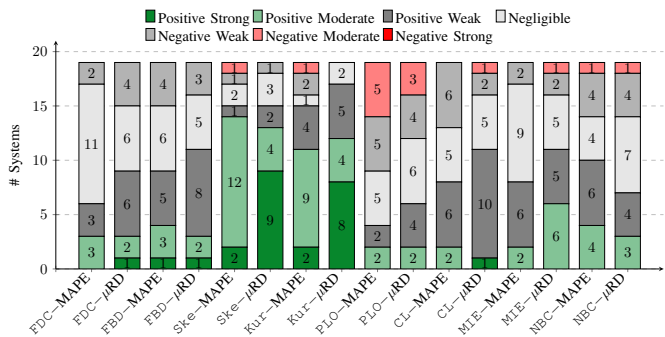

\centering
\includestandalone[width=0.8\columnwidth]{Figures/rq1_res_summary_metrics}
\caption{Per accuracy-landscape feature pair correlation.}
\label{fig:rq1-sum_from_metrics}
\end{figure}

\begin{figure}[t!]
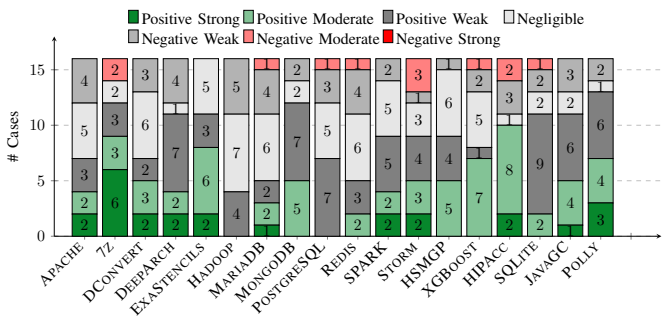

\centering
\includestandalone[width=0.8\columnwidth]{Figures/rq1_res_summary_system}
\caption{Per system correlation.}
\label{fig:rq1-sum_from_sytem}
\end{figure}


As shown in Table~\ref{tb: rq1:correlations}, a clear pattern we can observe is that, although in certain cases the landscape features have strongly and positively correlated with the accuracy metrics, they mostly have insignificant or negative correlation: only $29\%$ cases show moderate or higher positive correlations. Indeed, some systems exhibit more positive correlations than the others, e.g., \textsc{ExaStencils} shows positive correlations to the accuracy mainly for the global features while those on the local features are rather weak/negligible and with $p>0.05$ in general, despite being positive; in contrast, \textsc{Batik} has all the cases being negative or insignificant correlations. This is attributed to the real landscape formed by different systems: some systems are more difficult to tune than others due to the limited structural guidance and/or the presence of complex local optima.

We can see clearer results when investigating the per accuracy-landscape feature across the system, as shown in Figure~\ref{fig:rq1-sum_from_metrics}: the majority of the landscape features reveal insignificant correlations (on average, only $5.3$ out of $18$ systems show at least moderately positive correlation). The only exceptions are for \texttt{Ske} and \texttt{Kur} such that the positive correlations are more common (with up to $14$/$18$ systems), which is also the main contributor to the significant positive correlation in Table~\ref{tb: rq1:correlations}. This makes sense, since they are global features that reflect on the distribution of the performance (fitness) values, which is directly relevant to the way the accuracy metrics are calculated. However, we note that even for \texttt{Ske} and \texttt{Kur}, the correlations are rarely strong.

To better understand whether the correlations differ on the systems, in Figure~\ref{fig:rq1-sum_from_sytem}, we note that there indeed exist discrepancies; for example, three systems, i.e., \textsc{7z}, \textsc{ExaStencils}, and \textsc{HIPAcc}, have $50\%$ or more accuracy-landscape pairs with at least moderately positive correlation. However, overall, there are only $4.7$ out of $16$ pairs that have a positive correlation of at least the moderate level, demonstrating the generally insignificant correlation per system. Therefore, we say that:

\begin{tcbitemize}[%
    raster columns=1, 
    raster rows=1
    ]
  \tcbitem[myhbox={}{Finding \thefindingcount}]   \textit{The correlations between accuracy and landscape feature for surrogate model are not positively significant in general (for over $70\%$ cases), especially for local features.}
\end{tcbitemize}
\addtocounter{findingcount}{1}

\subsection{RQ3: Fidelity of the Model Emulated Landscape}

\subsubsection{Method}

For \textbf{RQ3}, we focus on the deviation between model-emulated landscape features ($v_{model}$) and those of the real system ($v_{system}$). We report on the relative deviation as $v_{model} - v_{system}$, covering $300$ data points (identical to that in \textbf{RQ2}), which is also verified by Wilcoxon rank-sum test for confirming a statistically significant difference. Note that $v_{model}$ and $v_{system}$ are computed using the same points in the testing dataset from Table~\ref{tb:systems}. To understand the direction of deviation, we do not invert the values of the landscape features. As such, depending on the landscape feature, a positive or negative value of the relative deviation has different meanings.


\subsubsection{Results}


As shown in Table~\ref{tb: rq2-overview}, a general observation is that the landscapes emulated by the models mostly exhibit a significant difference compared with the real one: there are $134$ out of $144$ cases in which more than $50\%$ models have $p<0.05$ difference ($100\%$ for many cases). According to our theory of landscape dominance in Section~\ref{Sec:Theory}, such a deviation is not ideal for global features but could be preferred for the local ones. For the global features, we see that the models tend to build a landscape with higher global landscape values than having smaller ones, except for \texttt{Ske}. This means that the model-emulated landscapes are generally more optimistic: they tend to create stronger guidance with a more dense and isolated optimal performance, but assuming that better performance is common, which could cause misleading issues in the tuning. For the local features, models tend to make the tuning more difficult on \texttt{PLO} and \texttt{MIE} ($|-\Delta| < |+\Delta|$), but they tend to be beneficial on \texttt{CL} ($|-\Delta| < |+\Delta|$) and \texttt{NBC} ($|-\Delta| > |+\Delta|$). This suggests that while models often emulate a landscape deviated from the real ones on the local features, they might be preferred for tuning since the issues of local optima can be relaxed, making the task easier for a tuner.

\input{Tables/rq2_overview}

\input{Tables/best_performing_model_in_landscape}

For particular systems, we see that the model emulated landscape shares similar characteristics with certain ones, but for others, their results are highly deviated, even for those systems of the same type, e.g., \textsc{SQLite} and \textsc{MongoDB}. This makes sense, since the systems are highly complex---any of their alternative internal designs might dramatically alter the performance behaviors, hence the landscapes to be modeled.


Table~\ref{tab:system_rows_models} shows the model over a particular landscape feature and system with the smallest $\Delta g$ or $l$. We see that, even when referring to a single landscape feature, it is difficult to conclude which model is generally the most preferred, as this differs depending on the systems. However, surprisingly, the classic machine learning models are generally more suitable for tuning in terms of the single landscape metrics over the deep learning ones ($119$ cases versus $25$ cases), despite a vast amount of studies have shown that deep learning-based models provide ``better'' accuracy \cite{DaL,DeepPerf,HINNPerf}. This finding again aligns with the prior work \cite{accuracy_can_lie}, i.e., the accuracy can lie.

Overall, we conclude that:

\begin{tcbitemize}[%
    raster columns=1, 
    raster rows=1
    ]
  \tcbitem[myhbox={}{Finding \thefindingcount}]  \textit{The existing surrogate models are still far from being able to correctly emulate the real system landscapes in $93\%$ cases, but they might be helpful to relax the issues of local optima for a tuner.}
\end{tcbitemize}
\addtocounter{findingcount}{1}

\begin{tcbitemize}[%
    raster columns=1, 
    raster rows=1
    ]
  \tcbitem[myhbox={}{Finding \thefindingcount}]  \textit{No single model has the generally smallest $\Delta g$ or $l$ for all cases, but machine learning models often tend to be more fit than the deep learning counterpart in terms of the preferred landscape properties---119 cases versus 25 cases.}
\end{tcbitemize}
\addtocounter{findingcount}{1}




\subsection{RQ4: Implications of Individual Option to the Model Emulated Landscape}
\label{sec: rq4}

\subsubsection{Method}


Here, we are interested in understanding how domain knowledge of the systems can influence the features of the emulated landscape. Therefore, in \textbf{RQ4}, we perform the following for each system and model:

\begin{enumerate}
    \item Remove a configuration option $O$ from the dataset.
    \item For configurations that only differ on $O$, we keep one of the most frequently occurring performance values as a sample while discarding the others (or randomly choose one among the most common performance values). \mytag{R1C5}\revision{Note that we can do this because the dataset is sampled from the system, which is naturally discrete, i.e., we can count what value of the performance is the most common among the affected configurations. Yet, this only occurs in a small number of cases.}
    \item Train the model with training size from Table~\ref{tb:systems}.
    \item Compute the landscape feature of the landscape built by the model without $O$ and that of a model trained on all options using the testing size from Table~\ref{tb:systems}. 
    \item Move to the next option until every option has been ruled out once.
\end{enumerate}



\begin{figure}[t!]
\centering
\includegraphics[width=0.4\columnwidth]{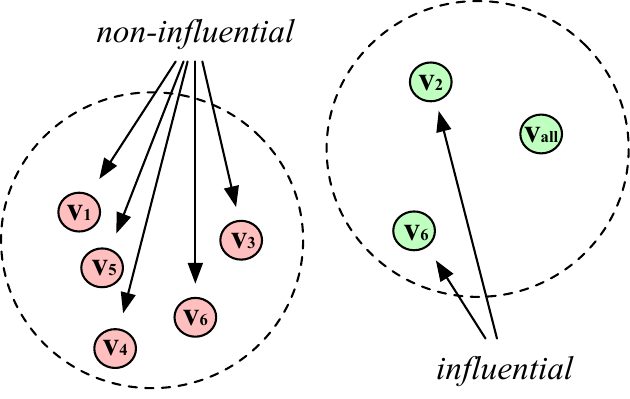}
\caption{Illustration of the clustering results to distinguish influential and non-influential options for a landscape feature.}
\label{fig:exp-cluster}
\end{figure}

To systematically analyze the impact of a configuration option on each landscape feature, we use a clustering-based approach for a system and a given landscape feature: 



\begin{enumerate}

\item \textbf{Build a model-feature matrix:} We train each model on the system with one of the options removed in turn:
\input{Tables/matrix}
\noindent whereby $\mathcal{M}_n$ is the $n$th model. $v_{n,k}$ is the landscape feature value for the $n$th model when the $k$th option is removed; $\mathbf{v_k}$ is the vector of landscape feature values from all models trained without using the $k$th option. Similarly, $v_{n,all}$ is the landscape feature value for the $n$th model trained using all options; $\mathbf{v_{all}}$ is a vector of the landscape feature values for all models with all options.

\item \textbf{Cluster option vector:} We apply $k$-Mean to cluster the option vector $\mathbf{V}=\{\mathbf{v_1},\mathbf{v_2},...,\mathbf{v_k},\mathbf{v_{all}}\}$, where each element is represented by the landscape feature values of all models. We are interested in dividing a $\mathbf{v_k}$ based on whether it is sufficiently distant from $\mathbf{v_{all}}$. As such, we set $k=2$, i.e., only two clusters.

\item \textbf{Identify influential options:} For a $\mathbf{v_k}$ that is assigned into the same cluster as $\mathbf{v_{all}}$, we consider that the corresponding $k$th option, if removed, is influential to the model under the given landscape feature and system; otherwise, the said option can be safely removed without significantly affecting the landscape feature. For example, in Figure~\ref{fig:exp-cluster}, $\mathbf{v_2}$ and $\mathbf{v_6}$ are placed in the same cluster as $\mathbf{v_{all}}$, hence we say the second and the sixth options are both generally influential to the corresponding landscape feature while the others are not.

\end{enumerate}

\input{Tables/option_catgory}

The above helps us to quickly identify the influential options on the landscape features while being robust to the confounding factor of overlapping groups~\cite{DBLP:journals/tse/Tantithamthavorn19} when using a multi-comparison statistical testing without degrading the statistical power.

To aid the interpretation of the results, we categorize the types of options as shown in Table~\ref{tb:option} by manually inspecting the related configuration code and documentation of the systems, following a similar approach used by Liang et al.~\cite{DBLP:conf/icse/LiangHC25}. Broadly, the functional options include those that control the core features and utility of the system: for example, in \textsc{MariaDB}, ``\texttt{binaryLog}'' is a utility option that controls certain granularity of the logging functionality. In contrast, the resource type involves options that tend to significantly influence the software/hardware provisioning, i.e., CPU, memory, storage and queue, e.g., the ``\texttt{mapreduce\_reduce\_merge\_inmem\_threshold}'' in \textsc{Hadoop} determines the memory allocation of the merge operations; the ``\texttt{maxClients}'' in \textsc{Apache} controls the maximum number of client connections before the requests need to be queued.

\input{Tables/RQ3_knob_influence}



\subsubsection{Results}



\begin{figure}[t!]
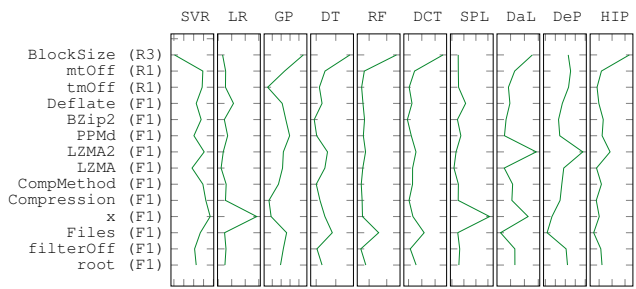
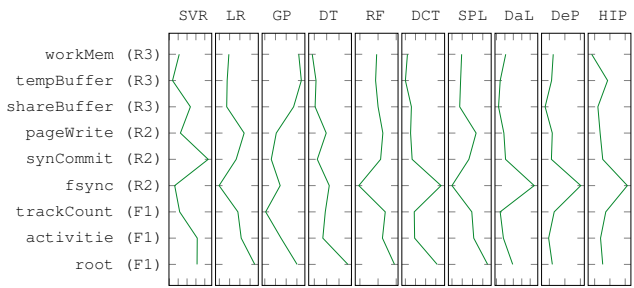

\centering
\subfloat[\texttt{FDC} on \textsc{7z}]{
\includestandalone[width=0.47\columnwidth]{Figures/rq4-exp1}
}
\hfill
\subfloat[\texttt{CL} on \textsc{PorstgreSQL}]{
\includestandalone[width=0.47\columnwidth]{Figures/rq4-exp2}
}
\caption{Extent of option influence to the models' landscape features in terms of the deviation between $v_{n,k}$ to $v_{n,all}$ for two exampled cases: \texttt{FDC} on \textsc{7z} and \texttt{CL} on \textsc{PostgreSQL}.}
\label{fig:rq4-details}
\end{figure}

From Table~\ref{tb: knob}, we note that most of the performance-sensitive options are related to core functionalities, followed by CPU and memory-related options. However, within each of the categories, around one fourth of the functional options are influential to the model emulated landscapes, which is similar to those of R1 and R2 ($20\%$ - $26\%$). This is because neither functional aspects nor CPU/storage-related aspects can dramatically influence the performance, and hence the impact on the landscape is limited. In contrast, for R3 and R4, which denote memory and queue, almost half of the options can pose significant shifts to the model landscape if removed. This makes sense, since both categories denote options that control the ``space'' of the system; therefore, if information of such a space is omitted, the configuration landscape learned can be severely affected. \mytag{R2C4}\revision{Importantly, this also helps to explain some findings from the other RQs: for example, the results show that the influence of memory/queue-related options on landscape features is approximately twice more influential than that of CPU/utility-related options---a root cause that, on \textbf{RQ2}, the model's landscape features on systems have most memory-related options are rather fluctuated (e.g., \textsc{Hadoop}), but are more stable on systems dominated by CPU/utility-related options.}


From the examples illustrated in Figure~\ref{fig:rq4-details}, we see that some options are highly influential to the landscape feature for some models but not always, e.g., ``\texttt{BlockSize}'' in Figure~\ref{fig:rq4-details}a. These are the same observations across different systems and models. Also, depending on the feature and system, options belonging to R3 might not always lead to the biggest influence, e.g., ``\texttt{workMen}'' versus ``\texttt{fsync}'' in Figure~\ref{fig:rq4-details}b, despite the fact that generally R3 options are more influential.



We therefore conclude that:


\begin{tcbitemize}[%
    raster columns=1, 
    raster rows=1
    ]
  \tcbitem[myhbox={}{Finding \thefindingcount}]  \textit{Options that control either memory or queue are generally easier to be influential ($\approx 2\times$) for the model emulated landscape compared with the others, but the extent of influence can be case-dependent.}
\end{tcbitemize}
\addtocounter{findingcount}{1}

\subsection{\mytag{R1C4}\revision{Discussion: On the Validity of Neighborhood}}

\revision{In this work, we use Hamming distance as the main distance metric when computing the neighborhood for \texttt{PLO} and \texttt{CL}, since it is the most widely used one~\cite{10.1145/3803859}. However, admittedly, it might not fit well with the granularity required for numeric options. As such, we examine two additional treatments to the numeric options (the way the other options are handled remains unchanged):}

\begin{itemize}
    \item \revision{\textbf{Numeric Step Neighborhood:} For a given numeric option $i$, we normalize its range in the datasets using max-min scaling. Suppose that $\alpha_i$ and $\beta_i$ are the normalized values of $i$ from two configurations, we then compute the distance $d_{i,\{\alpha,\beta\}}$ between them on $i$ using a common step $s=0.05$ \cite{neighborhood5, 10375909}:} 
    \begin{equation}
d_{i,\{\alpha,\beta\}} =
\begin{cases}
1, & |\alpha_i - \beta_i| > s\\
0, & \text{otherwise}
\end{cases}
\label{eq:stagnation-update}
\end{equation}

    \item \revision{\textbf{Adjacent Bin Neighborhood:} For a given numeric option $i$, we divide the values into $10$ bins, which is a common value \cite{Implications,tsai2025anollm}. This is achieved by computing the boundary $b={i_{max}-i_{min} \over 10}$ using the maximum ($i_{max}$) and minimum ($i_{min}$) values, then the bins are $\{b_1=[i_{min},i_{min}+b],b_2=[i_{min}+b,i_{min}+2b],\dots\}$. Suppose that $b_{i,\alpha}$ and $b_{i,\beta}$ are the bin indices that the values of two configurations belong on $i$, we then compute their distance $d_{i,\{\alpha,\beta\}}$ for $i$ as $d_{i,\{\alpha,\beta\}}=|b_{i,\alpha}-b_{i,\beta}|$.}

\end{itemize}

\revision{Finally, we compute the neighboring using Hamming distance based on the above; A configuration is a neighbor of another if their overall distance, considering both the numeric and other options, is 1. We compare the consistency between the results under the above two ablations to the original version (without specific treatments to the numeric options) for \texttt{PLO} and \texttt{CL}, using two metrics for each system:}

\begin{itemize}
    \item \revision{\textbf{Spearman's Rank Correlation ($\rho$):} Compute the nonlinearly monotonic consistency between two random variables within a range $[-1,1]$. A larger, positive score indicates a stronger correlation that one value increases also means an increment of the other value.}
    \item \revision{\textbf{Pairwise Flip Rate (PFR):} Quantify the proportion of relative order inversions within a range $[0,1]$. A smaller score reflects higher consistency.}
\end{itemize}

\input{Tables/neighour}

\revision{The results by averaging across all systems are shown in Table~\ref{tab:ablation-mean-stability}. Clearly, we confirm a drastically high positive (average) $\rho$ correlation between either ablation and the original results---a strong indication of the consistency ($\rho>0.69$ is a strong correlation~\cite{DBLP:journals/infsof/SamoladasAS10,DBLP:journals/tse/WattanakriengkraiWKTTIM23}). Further, the (average) PFR score is rather low, hence there are few discrepancies between them. {The above demonstrates that, even with diverse treatments of the numeric consistency, the results would not significantly affect the conclusions.}}

\revision{To further verify the implication, we have also re-computed the results of \texttt{PLO} and \texttt{CL} values using the above two ablations for \textbf{RQ1} and we confirm that the overall conclusions, especially with respect to the number of cases that confirm/violate our theory, are largely consistent\footnote{Detailed results can be found at: \url{https://github.com/ideas-labo/model4tune/tree/main/supp/neighbourhood.pdf}.}.}

%% file: Tables/tuners.tex
\begin{table}[t!]
\centering
\footnotesize
\caption{Specification of the tuners considered from~\cite{accuracy_can_lie}.}
\label{tb:tuners}
\setlength{\tabcolsep}{0.8mm}
\begin{adjustbox}{width=.8\columnwidth,center}
\begin{tabular}{lll||lll}
\toprule
\textbf{Tuner} & \textbf{Type} & \textbf{Domain} &\textbf{Tuner} & \textbf{Type} & \textbf{Domain}  \\ \midrule

\texttt{BOCA} \cite{BOCA} & Sequential & Compilers & \texttt{ConEx} \cite{conex} & Batch & Distributed systems \\
\rowcolor{gree!15}\texttt{ATconf} \cite{atconf} & Sequential & Big data systems &  \texttt{BestConfig} \cite{bestconfig} & Batch & Database systems \\
\texttt{FLASH} \cite{flash} & Sequential & Configurable systems &  \texttt{Irace} \cite{lopez2016irace} & Batch & Parameter optimization \\
\rowcolor{gree!15}\texttt{OtterTune} \cite{Ottertune} & Sequential & Database systems & \texttt{GGA}\cite{gga} & Batch & Parameter optimization \\
\texttt{ResTune} \cite{restune} & Sequential & Database systems & \texttt{ParamILS} \cite{paramils} & Batch & Parameter optimization \\
\rowcolor{gree!15}\texttt{ROBOTune} \cite{ROBOTune} & Sequential & Distributed systems & \texttt{Random} & Batch & General \\
\texttt{SMAC} \cite{SMAC} & Sequential & Parameter optimization & \texttt{GA} \cite{k2vtune} & Batch & General \\
\rowcolor{gree!15}\texttt{Tuneful} \cite{tuneful} & Sequential & Data analytics & \texttt{SWAY} \cite{Sway} & Batch & Parameter optimization \\

\bottomrule

%
\end{tabular}
\end{adjustbox}
\label{tb:tuners}
\end{table}

%% file: Tables/budget.tex
\begin{table}[t!]
\centering
\footnotesize
\caption{The budgets used for each system from~\cite{accuracy_can_lie}.}
\label{tb:budgets}
\begin{adjustbox}{width=.6\columnwidth,center}
\begin{tabular}{ll||ll||ll}
\toprule
\textbf{System} & \textbf{Budget} &\textbf{System} & \textbf{Budget} &\textbf{System} & \textbf{Budget} \\ \midrule
\textsc{Apache}&271&\textsc{7z}&382&\textsc{DConvert}&335\\
\rowcolor{gree!15}\textsc{DeepArch}&207&\textsc{ExaStencils}&416&\textsc{Hadoop}&297\\
\textsc{MariaDB}&226&\textsc{MongoDB}&278&\textsc{PostgreSQL}&298\\
\rowcolor{gree!15}\textsc{Redis}&298&\textsc{Spark}&326&\textsc{Storm}&263\\
\textsc{HSMGP}&218&\textsc{XGBoost}&278&\textsc{HIPAcc}&371\\
\rowcolor{gree!15}\textsc{SQLite}&206&\textsc{JavaGC}&289&\textsc{Polly}&285\\
\bottomrule

%
\end{tabular}
\end{adjustbox}
\label{tb:budget}
\end{table}

%% file: Tables/rq1-example.tex
\begin{table}[t!]
\centering
\footnotesize
\caption{Exampled model pairs in the dominating (DG) and dominated (DD) groups on a given system via landscape dominance.}
\label{tb:rq1-exp}
\begin{tabular}{lcccc}
\toprule
\multirow{2}{*}{\textbf{Tuner}} & \multicolumn{2}{c}{\textbf{Objectives}} & \multicolumn{2}{c}{\textbf{Pairs of Models}}  \\

\cmidrule{2-5}

& \textbf{$\Delta g$} & \textbf{$l$} &\textbf{DG group}  &\textbf{DD group}  \\ \midrule
\texttt{GA} & \texttt{FDC} & \texttt{CL} & $\mathcal{A}$ & $\mathcal{B}$\\
\texttt{GA} & \texttt{FDC} & \texttt{CL} & $\mathcal{B}$ & $\mathcal{C}$\\

$\cdots$ & $\cdots$ & $\cdots$ & $\cdots$ & $\cdots$\\

\texttt{GA} & \texttt{Kur} & \texttt{PLO} & $\mathcal{C}$ & $\mathcal{B}$\\

$\cdots$ & $\cdots$ & $\cdots$ & $\cdots$ & $\cdots$\\

\texttt{SMAC} & \texttt{FDC}& \texttt{CL}& $\mathcal{A}$ & $\mathcal{B}$\\

$\cdots$ & $\cdots$ & $\cdots$ & $\cdots$ & $\cdots$\\

\bottomrule

%
\end{tabular}
\label{tb:rq1-example}
\end{table}

%% file: Tables/rq_landscape_dominance.tex
\begin{table}[t!]
\centering
\caption{Comparing the performance of dominating (DG) and dominated (DD) groups. $^\dagger$ and $^*$ denote that the performance difference exhibits $p<0.001$ and $0.001 \leq p < 0.05$, respectively. \colorbox{gree!30}{Green rows} and \colorbox{red!30}{red rows} denote the cases that confirm and violate our theory of model usefulness, respectively.}
\label{tab:domination_performance_merged}
\begin{adjustbox}{max width=\textwidth,center}
\begin{tabular}{lrrr|rrr|rrr}
\toprule
 & \multicolumn{3}{c|}{\textbf{No aggregation}} & \multicolumn{3}{c|}{\textbf{Majority vote}} & \multicolumn{3}{c}{\textbf{Weighted sum}} \\ \cmidrule{2-10}
\textbf{System} & $\Delta p$ & DG Lose (\%) & DG Win (\%)& $\Delta p$ & DG Lose (\%)& DG Win (\%)& $\Delta p$ & DG Lose (\%)& DG Win (\%)\\
\midrule
\textsc{Apache} & \cellcolor{gree!30}$-$0.0191 & \cellcolor{gree!30}39.1 & \cellcolor{gree!30}60.9 & \cellcolor{gree!30}$-$0.0028 & \cellcolor{gree!30}49.1 & \cellcolor{gree!30}50.9 & \cellcolor{gree!30}$-$0.0255 & \cellcolor{gree!30}43.6 & \cellcolor{gree!30}56.4 \\
\textsc{7z} & \cellcolor{gree!30}$-$173.1200$^\dagger$ & \cellcolor{gree!30}22.7 & \cellcolor{gree!30}77.3 & \cellcolor{gree!30}$-$40.3500$^*$ & \cellcolor{gree!30}42.9 & \cellcolor{gree!30}57.1 & \cellcolor{gree!30}$-$167.5533$^\dagger$ & \cellcolor{gree!30}31.7 & \cellcolor{gree!30}68.3 \\
\textsc{DConvert} & \cellcolor{gree!30}$-$0.0016$^*$ & \cellcolor{gree!30}45.3 & \cellcolor{gree!30}54.7 & \cellcolor{gree!30}$-$0.0003 & \cellcolor{gree!30}47.3 & \cellcolor{gree!30}51.5 & \cellcolor{gree!30}$-$0.0030$^*$ & \cellcolor{gree!30}41.1 & \cellcolor{gree!30}57.8 \\
\textsc{DeepArch} & \cellcolor{gree!30}$-$0.0079$^\dagger$ & \cellcolor{gree!30}18.0 & \cellcolor{gree!30}82.0 & \cellcolor{gree!30}$-$0.0011$^\dagger$ & \cellcolor{gree!30}29.1 & \cellcolor{gree!30}67.4 & \cellcolor{gree!30}$-$0.0011$^\dagger$ & \cellcolor{gree!30}30.8 & \cellcolor{gree!30}65.8 \\
\textsc{ExaStencils} & \cellcolor{gree!30}$-$192.1372$^\dagger$ & \cellcolor{gree!30}17.2 & \cellcolor{gree!30}82.8 & \cellcolor{gree!30}$-$23.7633$^*$ & \cellcolor{gree!30}42.2 & \cellcolor{gree!30}57.8 & \cellcolor{gree!30}$-$121.1267$^\dagger$ & \cellcolor{gree!30}25.8 & \cellcolor{gree!30}74.2 \\
\textsc{Hadoop} & \cellcolor{gree!30}$-$0.5338$^*$ & \cellcolor{gree!30}42.2 & \cellcolor{gree!30}57.8 & \cellcolor{gree!30}$-$0.7500 & \cellcolor{gree!30}45.8 & \cellcolor{gree!30}54.2 & \cellcolor{gree!30}$-$1.1167 & \cellcolor{gree!30}45.8 & \cellcolor{gree!30}54.2 \\
\textsc{MariaDB} & \cellcolor{gree!30}$-$0.0779 & \cellcolor{gree!30}39.1 & \cellcolor{gree!30}60.9 & \cellcolor{gree!30}$-$0.0453 & \cellcolor{gree!30}45.4 & \cellcolor{gree!30}54.6 & \cellcolor{gree!30}$-$0.0436 & \cellcolor{gree!30}45.6 & \cellcolor{gree!30}54.4 \\
\textsc{MongoDB} & \cellcolor{gree!30}$-$417.9015$^\dagger$ & \cellcolor{gree!30}25.8 & \cellcolor{gree!30}74.2 & \cellcolor{gree!30}$-$8.8733$^*$ & \cellcolor{gree!30}38.1 & \cellcolor{gree!30}51.2 & \cellcolor{gree!30}$-$27.8900$^\dagger$ & \cellcolor{gree!30}32.8 & \cellcolor{gree!30}56.1 \\
\textsc{PostgreSQL} & \cellcolor{red!30}2.3402$^*$ & \cellcolor{red!30}59.4 & \cellcolor{red!30}40.6 & \cellcolor{gree!30}$-$1.0633 & \cellcolor{gree!30}47.0 & \cellcolor{gree!30}53.0 & \cellcolor{red!30}5.7733$^\dagger$ & \cellcolor{red!30}63.6 & \cellcolor{red!30}36.4 \\
\textsc{Redis} & 0.0441 & 50.0 & 50.0 & \cellcolor{red!30}97.2126 & \cellcolor{red!30}53.8 & \cellcolor{red!30}46.2 & \cellcolor{gree!30}$-$102.6061 & \cellcolor{gree!30}41.2 & \cellcolor{gree!30}58.8 \\
\textsc{Spark} & \cellcolor{gree!30}$-$3.1976$^\dagger$ & \cellcolor{gree!30}35.2 & \cellcolor{gree!30}64.8 & \cellcolor{gree!30}$-$6.3667$^*$ & \cellcolor{gree!30}36.7 & \cellcolor{gree!30}62.5 & \cellcolor{gree!30}$-$5.7333$^*$ & \cellcolor{gree!30}35.8 & \cellcolor{gree!30}63.3 \\
\textsc{Storm} & \cellcolor{gree!30}0.0000 & \cellcolor{gree!30}24.2 & \cellcolor{gree!30}63.3 & \cellcolor{red!30}0.0000 & \cellcolor{red!30}47.0 & \cellcolor{red!30}39.2 & \cellcolor{gree!30}0.0000 & \cellcolor{gree!30}38.3 & \cellcolor{gree!30}47.2 \\
\textsc{HSMGP} & \cellcolor{red!30}0.8928$^\dagger$ & \cellcolor{red!30}80.5 & \cellcolor{red!30}19.5 & \cellcolor{red!30}0.2686$^\dagger$ & \cellcolor{red!30}59.6 & \cellcolor{red!30}39.2 & \cellcolor{red!30}0.0315 & \cellcolor{red!30}54.4 & \cellcolor{red!30}44.4 \\
\textsc{XGBoost} & \cellcolor{gree!30}$-$0.9543$^\dagger$ & \cellcolor{gree!30}25.0 & \cellcolor{gree!30}75.0 & \cellcolor{gree!30}$-$0.1782$^\dagger$ & \cellcolor{gree!30}39.6 & \cellcolor{gree!30}59.8 & \cellcolor{gree!30}$-$0.3178$^\dagger$ & \cellcolor{gree!30}35.6 & \cellcolor{gree!30}63.9 \\
\textsc{HIPAcc} & \cellcolor{red!30}0.0153$^\dagger$ & \cellcolor{red!30}62.5 & \cellcolor{red!30}37.5 & \cellcolor{red!30}0.0101$^*$ & \cellcolor{red!30}59.1 & \cellcolor{red!30}40.9 & \cellcolor{red!30}0.0066$^*$ & \cellcolor{red!30}57.5 & \cellcolor{red!30}42.5 \\
\textsc{SQLite} & \cellcolor{gree!30}$-$0.0173$^\dagger$ & \cellcolor{gree!30}34.4 & \cellcolor{gree!30}65.6 & \cellcolor{gree!30}$-$0.0179$^*$ & \cellcolor{gree!30}43.0 & \cellcolor{gree!30}57.0 & \cellcolor{gree!30}$-$0.0024$^*$ & \cellcolor{gree!30}48.1 & \cellcolor{gree!30}51.9 \\
\textsc{JavaGC} & \cellcolor{gree!30}$-$3.0529$^\dagger$ & \cellcolor{gree!30}12.5 & \cellcolor{gree!30}87.5 & \cellcolor{gree!30}$-$0.4750$^*$ & \cellcolor{gree!30}41.6 & \cellcolor{gree!30}58.4 & \cellcolor{gree!30}$-$0.4083$^*$ & \cellcolor{gree!30}45.3 & \cellcolor{gree!30}54.7 \\
\textsc{Polly} & \cellcolor{gree!30}$-$0.0836$^\dagger$ & \cellcolor{gree!30}14.8 & \cellcolor{gree!30}85.2 & \cellcolor{gree!30}$-$0.0212$^\dagger$ & \cellcolor{gree!30}31.7 & \cellcolor{gree!30}68.3 & \cellcolor{gree!30}$-$0.0411$^\dagger$ & \cellcolor{gree!30}27.2 & \cellcolor{gree!30}72.8 \\
\bottomrule
\end{tabular}
\end{adjustbox}
\end{table}

%% file: Tables/rq1_overview.tex
\begin{table*}[t!]
\caption{Spearman correlations between landscape features and accuracy metrics. $^\dagger$ and $^\star$ denote $p<0.001$ and $0.001 \leq p<0.05$, respectively. \colorbox{gree!30}{Green bars} and \colorbox{red!30}{red bars} indicate positive and negative correlation, respectively. Their darkness of color reflect the strength of correlations.}
\label{tb: rq1:correlations}
\centering
\setlength{\tabcolsep}{0.5mm}
\footnotesize
\begin{adjustbox}{width=\textwidth,center}
\begin{tabular}{lr|r|r|r|r|r|r|r||r|r|r|r|r|r|r|r}
\toprule
\textbf{System}&\rotatebox{0}{\textbf{\texttt{FDC}$-$MAPE}}&\textbf{\texttt{FDC}$-\mu$RD}&\textbf{\texttt{FBD}$-$MAPE}&\textbf{\texttt{FBD}$-\mu$RD}&\textbf{\texttt{Ske}$-$MAPE}&\textbf{\texttt{Ske}$-\mu$RD}&\textbf{\texttt{Kur}$-$MAPE}&\textbf{\texttt{Kur}$-\mu$RD}&\textbf{\texttt{PLO}$-$MAPE}&\textbf{\texttt{PLO}$-\mu$RD}&\textbf{\texttt{CL}$-$MAPE}&\textbf{\texttt{CL}$-\mu$RD}&\textbf{\texttt{MIE}$-$MAPE}&\textbf{\texttt{MIE}$-\mu$RD}&\textbf{\texttt{NBC}$-$MAPE}&\textbf{\texttt{NBC}$-\mu$RD}\\
\midrule
\textsc{Apache}&$-$\percent[n]{0.17}{1}{0.17$^\star$}{40}&$-$\percent[n]{0.15}{1}{0.15$^\star$}{40}&\phantom{$-$}\percent[p]{0.05}{1}{0.05$^{\,\,\,}$}{20}&\phantom{$-$}\percent[p]{0.08}{1}{0.08$^{\,\,\,}$}{20}&\phantom{$-$}\percent[p]{0.57}{1}{0.57$^\dagger$}{60}&\phantom{$-$}\percent[p]{0.73}{1}{0.73$^\dagger$}{80}&\phantom{$-$}\percent[p]{0.61}{1}{0.61$^\dagger$}{60}&\phantom{$-$}\percent[p]{0.72}{1}{0.72$^\dagger$}{80}&\phantom{$-$}\percent[p]{0.08}{1}{0.08$^{\,\,\,}$}{20}&\phantom{$-$}\percent[p]{0.34}{1}{0.34$^\dagger$}{40}&\phantom{$-$}\percent[p]{0.20}{1}{0.20$^\dagger$}{40}&\phantom{$-$}\percent[p]{0.28}{1}{0.28$^\dagger$}{40}&$-$\percent[n]{0.21}{1}{0.21$^\dagger$}{40}&$-$\percent[n]{0.07}{1}{0.07$^{\,\,\,}$}{20}&$-$\percent[n]{0.01}{1}{0.01$^{\,\,\,}$}{20}&$-$\percent[n]{0.32}{1}{0.32$^\dagger$}{40}\\
\textsc{7z}&\phantom{$-$}\percent[p]{0.24}{1}{0.24$^\dagger$}{40}&\phantom{$-$}\percent[p]{0.20}{1}{0.20$^\dagger$}{40}&\phantom{$-$}\percent[p]{0.74}{1}{0.74$^\dagger$}{80}&\phantom{$-$}\percent[p]{0.74}{1}{0.74$^\dagger$}{80}&\phantom{$-$}\percent[p]{0.83}{1}{0.83$^\dagger$}{80}&\phantom{$-$}\percent[p]{0.89}{1}{0.89$^\dagger$}{80}&\phantom{$-$}\percent[p]{0.86}{1}{0.86$^\dagger$}{80}&\phantom{$-$}\percent[p]{0.92}{1}{0.92$^\dagger$}{100}&$-$\percent[n]{0.45}{1}{0.45$^\dagger$}{60}&$-$\percent[n]{0.40}{1}{0.40$^\dagger$}{60}&\phantom{$-$}\percent[p]{0.39}{1}{0.39$^\dagger$}{60}&\phantom{$-$}\percent[p]{0.34}{1}{0.34$^\dagger$}{40}&$-$\percent[n]{0.01}{1}{0.01$^{\,\,\,}$}{20}&\phantom{$-$}\percent[p]{0.05}{1}{0.05$^{\,\,\,}$}{20}&\phantom{$-$}\percent[p]{0.47}{1}{0.47$^\dagger$}{60}&\phantom{$-$}\percent[p]{0.41}{1}{0.41$^\dagger$}{60}\\
\textsc{DConvert}&$-$\percent[n]{0.02}{1}{0.02$^{\,\,\,}$}{20}&$-$\percent[n]{0.21}{1}{0.21$^\dagger$}{40}&$-$\percent[n]{0.13}{1}{0.13$^\star$}{40}&\phantom{$-$}\percent[p]{0.01}{1}{0.01$^{\,\,\,}$}{20}&\phantom{$-$}\percent[p]{0.52}{1}{0.52$^\dagger$}{60}&\phantom{$-$}\percent[p]{0.77}{1}{0.77$^\dagger$}{80}&\phantom{$-$}\percent[p]{0.47}{1}{0.47$^\dagger$}{60}&\phantom{$-$}\percent[p]{0.79}{1}{0.79$^\dagger$}{80}&\phantom{$-$}\percent[p]{0.08}{1}{0.08$^{\,\,\,}$}{20}&$-$\percent[n]{0.08}{1}{0.08$^{\,\,\,}$}{20}&$-$\percent[n]{0.06}{1}{0.06$^{\,\,\,}$}{20}&\phantom{$-$}\percent[p]{0.05}{1}{0.05$^{\,\,\,}$}{20}&\phantom{$-$}\percent[p]{0.30}{1}{0.30$^\dagger$}{40}&\phantom{$-$}\percent[p]{0.55}{1}{0.55$^\dagger$}{60}&$-$\percent[n]{0.10}{1}{0.10$^{\,\,\,}$}{40}&\phantom{$-$}\percent[p]{0.09}{1}{0.09$^{\,\,\,}$}{40}\\
\textsc{DeepArch}&\phantom{$-$}\percent[p]{0.08}{1}{0.08$^{\,\,\,}$}{20}&$-$\percent[n]{0.19}{1}{0.19$^\dagger$}{40}&\phantom{$-$}\percent[p]{0.37}{1}{0.37$^\dagger$}{40}&\phantom{$-$}\percent[p]{0.11}{1}{0.11$^\star$}{40}&\phantom{$-$}\percent[p]{0.70}{1}{0.70$^\dagger$}{80}&\phantom{$-$}\percent[p]{0.41}{1}{0.41$^\dagger$}{60}&\phantom{$-$}\percent[p]{0.71}{1}{0.71$^\dagger$}{80}&\phantom{$-$}\percent[p]{0.38}{1}{0.38$^\dagger$}{40}&$-$\percent[n]{0.32}{1}{0.32$^\dagger$}{40}&\phantom{$-$}\percent[p]{0.31}{1}{0.31$^\dagger$}{40}&$-$\percent[n]{0.21}{1}{0.21$^\dagger$}{40}&\phantom{$-$}\percent[p]{0.24}{1}{0.24$^\dagger$}{40}&\phantom{$-$}\percent[p]{0.36}{1}{0.36$^\dagger$}{40}&\phantom{$-$}\percent[p]{0.42}{1}{0.42$^\dagger$}{60}&\phantom{$-$}\percent[p]{0.31}{1}{0.31$^\dagger$}{40}&$-$\percent[n]{0.35}{1}{0.35$^\dagger$}{40}\\
\textsc{ExaStencils}&\phantom{$-$}\percent[p]{0.57}{1}{0.57$^\dagger$}{60}&\phantom{$-$}\percent[p]{0.66}{1}{0.66$^\dagger$}{60}&\phantom{$-$}\percent[p]{0.55}{1}{0.55$^\dagger$}{60}&\phantom{$-$}\percent[p]{0.59}{1}{0.59$^\dagger$}{60}&\phantom{$-$}\percent[p]{0.61}{1}{0.61$^\dagger$}{60}&\phantom{$-$}\percent[p]{0.77}{1}{0.77$^\dagger$}{80}&\phantom{$-$}\percent[p]{0.64}{1}{0.64$^\dagger$}{60}&\phantom{$-$}\percent[p]{0.81}{1}{0.81$^\dagger$}{80}&$-$\percent[n]{0.06}{1}{0.06$^{\,\,\,}$}{20}&\phantom{$-$}\percent[p]{0.01}{1}{0.01$^{\,\,\,}$}{20}&\phantom{$-$}\percent[p]{0.30}{1}{0.30$^\dagger$}{40}&\phantom{$-$}\percent[p]{0.15}{1}{0.15$^\star$}{40}&\phantom{$-$}\percent[p]{0.07}{1}{0.07$^{\,\,\,}$}{20}&\phantom{$-$}\percent[p]{0.22}{1}{0.22$^\dagger$}{40}&\phantom{$-$}\percent[p]{0.05}{1}{0.05$^{\,\,\,}$}{20}&\phantom{$-$}\percent[p]{0.05}{1}{0.05$^{\,\,\,}$}{20}\\
\textsc{Hadoop}&$-$\percent[n]{0.03}{1}{0.03$^{\,\,\,}$}{20}&$-$\percent[n]{0.03}{1}{0.03$^{\,\,\,}$}{20}&$-$\percent[n]{0.16}{1}{0.16$^\star$}{40}&\phantom{$-$}\percent[p]{0.11}{1}{0.11$^\star$}{40}&$-$\percent[n]{0.10}{1}{0.10$^{\,\,\,}$}{40}&\phantom{$-$}\percent[p]{0.06}{1}{0.06$^{\,\,\,}$}{20}&$-$\percent[n]{0.24}{1}{0.24$^\dagger$}{40}&\phantom{$-$}\percent[p]{0.02}{1}{0.02$^{\,\,\,}$}{20}&\phantom{$-$}\percent[p]{0.27}{1}{0.27$^\dagger$}{40}&\phantom{$-$}\percent[p]{0.05}{1}{0.05$^{\,\,\,}$}{20}&$-$\percent[n]{0.24}{1}{0.24$^\dagger$}{40}&\phantom{$-$}\percent[p]{0.21}{1}{0.21$^\dagger$}{40}&\phantom{$-$}\percent[p]{0.04}{1}{0.04$^{\,\,\,}$}{20}&\phantom{$-$}\percent[p]{0.21}{1}{0.21$^\dagger$}{40}&$-$\percent[n]{0.39}{1}{0.39$^\dagger$}{40}&\phantom{$-$}\percent[p]{0.05}{1}{0.05$^{\,\,\,}$}{20}\\
\textsc{MariaDB}&$-$\percent[n]{0.05}{1}{0.05$^{\,\,\,}$}{20}&\phantom{$-$}\percent[p]{0.03}{1}{0.03$^{\,\,\,}$}{20}&$-$\percent[n]{0.08}{1}{0.08$^{\,\,\,}$}{20}&$-$\percent[n]{0.21}{1}{0.21$^\dagger$}{40}&\phantom{$-$}\percent[p]{0.56}{1}{0.56$^\dagger$}{60}&\phantom{$-$}\percent[p]{0.76}{1}{0.76$^\dagger$}{80}&\phantom{$-$}\percent[p]{0.36}{1}{0.36$^\dagger$}{40}&\phantom{$-$}\percent[p]{0.63}{1}{0.63$^\dagger$}{60}&$-$\percent[n]{0.41}{1}{0.41$^\dagger$}{60}&$-$\percent[n]{0.35}{1}{0.35$^\dagger$}{40}&$-$\percent[n]{0.32}{1}{0.32$^\dagger$}{40}&$-$\percent[n]{0.34}{1}{0.34$^\dagger$}{40}&$-$\percent[n]{0.04}{1}{0.04$^{\,\,\,}$}{20}&\phantom{$-$}\percent[p]{0.18}{1}{0.18$^\dagger$}{40}&$-$\percent[n]{0.07}{1}{0.07$^{\,\,\,}$}{20}&\phantom{$-$}\percent[p]{0.07}{1}{0.07$^{\,\,\,}$}{20}\\
\textsc{MongoDB}&\phantom{$-$}\percent[p]{0.27}{1}{0.27$^\dagger$}{40}&\phantom{$-$}\percent[p]{0.38}{1}{0.38$^\dagger$}{40}&\phantom{$-$}\percent[p]{0.10}{1}{0.10$^{\,\,\,}$}{40}&\phantom{$-$}\percent[p]{0.27}{1}{0.27$^\dagger$}{40}&\phantom{$-$}\percent[p]{0.47}{1}{0.47$^\dagger$}{60}&\phantom{$-$}\percent[p]{0.56}{1}{0.56$^\dagger$}{60}&\phantom{$-$}\percent[p]{0.38}{1}{0.38$^\dagger$}{40}&\phantom{$-$}\percent[p]{0.44}{1}{0.44$^\dagger$}{60}&\phantom{$-$}\percent[p]{0.47}{1}{0.47$^\dagger$}{60}&\phantom{$-$}\percent[p]{0.41}{1}{0.41$^\dagger$}{60}&\phantom{$-$}\percent[p]{0.13}{1}{0.13$^\star$}{40}&\phantom{$-$}\percent[p]{0.14}{1}{0.14$^\star$}{40}&\phantom{$-$}\percent[p]{0.04}{1}{0.04$^{\,\,\,}$}{20}&\phantom{$-$}\percent[p]{0.06}{1}{0.06$^{\,\,\,}$}{20}&$-$\percent[n]{0.32}{1}{0.32$^\dagger$}{40}&$-$\percent[n]{0.24}{1}{0.24$^\dagger$}{40}\\
\textsc{PostgreSQL}&\phantom{$-$}\percent[p]{0.08}{1}{0.08$^{\,\,\,}$}{20}&\phantom{$-$}\percent[p]{0.11}{1}{0.11$^\star$}{40}&\phantom{$-$}\percent[p]{0.03}{1}{0.03$^{\,\,\,}$}{20}&\phantom{$-$}\percent[p]{0.01}{1}{0.01$^{\,\,\,}$}{20}&\phantom{$-$}\percent[p]{0.13}{1}{0.13$^\star$}{40}&\phantom{$-$}\percent[p]{0.19}{1}{0.19$^\dagger$}{40}&\phantom{$-$}\percent[p]{0.07}{1}{0.07$^{\,\,\,}$}{20}&\phantom{$-$}\percent[p]{0.26}{1}{0.26$^\dagger$}{40}&$-$\percent[n]{0.18}{1}{0.18$^\star$}{40}&\phantom{$-$}\percent[p]{0.26}{1}{0.26$^\dagger$}{40}&\phantom{$-$}\percent[p]{0.03}{1}{0.03$^{\,\,\,}$}{20}&\phantom{$-$}\percent[p]{0.24}{1}{0.24$^\dagger$}{40}&$-$\percent[n]{0.15}{1}{0.15$^\star$}{40}&$-$\percent[n]{0.46}{1}{0.46$^\dagger$}{60}&\phantom{$-$}\percent[p]{0.19}{1}{0.19$^\dagger$}{40}&$-$\percent[n]{0.19}{1}{0.19$^\dagger$}{40}\\
\textsc{Redis}&$-$\percent[n]{0.04}{1}{0.04$^{\,\,\,}$}{20}&\phantom{$-$}\percent[p]{0.24}{1}{0.24$^\dagger$}{40}&$-$\percent[n]{0.13}{1}{0.13$^\star$}{40}&$-$\percent[n]{0.04}{1}{0.04$^{\,\,\,}$}{20}&$-$\percent[n]{0.01}{1}{0.01$^{\,\,\,}$}{20}&$-$\percent[n]{0.08}{1}{0.08$^{\,\,\,}$}{20}&$-$\percent[n]{0.18}{1}{0.18$^\dagger$}{40}&\phantom{$-$}\percent[p]{0.27}{1}{0.27$^\dagger$}{40}&\phantom{$-$}\percent[p]{0.45}{1}{0.45$^\dagger$}{60}&$-$\percent[n]{0.35}{1}{0.35$^\dagger$}{40}&$-$\percent[n]{0.22}{1}{0.22$^\dagger$}{40}&\phantom{$-$}\percent[p]{0.14}{1}{0.14$^\star$}{40}&$-$\percent[n]{0.06}{1}{0.06$^{\,\,\,}$}{20}&\phantom{$-$}\percent[p]{0.08}{1}{0.08$^{\,\,\,}$}{20}&$-$\percent[n]{0.58}{1}{0.58$^\dagger$}{60}&\phantom{$-$}\percent[p]{0.49}{1}{0.49$^\dagger$}{60}\\
\textsc{Spark}&$-$\percent[n]{0.05}{1}{0.05$^{\,\,\,}$}{20}&$-$\percent[n]{0.03}{1}{0.03$^{\,\,\,}$}{20}&\phantom{$-$}\percent[p]{0.16}{1}{0.16$^\star$}{40}&\phantom{$-$}\percent[p]{0.20}{1}{0.20$^\dagger$}{40}&\phantom{$-$}\percent[p]{0.54}{1}{0.54$^\dagger$}{60}&\phantom{$-$}\percent[p]{0.71}{1}{0.71$^\dagger$}{80}&\phantom{$-$}\percent[p]{0.61}{1}{0.61$^\dagger$}{60}&\phantom{$-$}\percent[p]{0.83}{1}{0.83$^\dagger$}{80}&$-$\percent[n]{0.28}{1}{0.28$^\dagger$}{40}&$-$\percent[n]{0.36}{1}{0.36$^\dagger$}{40}&\phantom{$-$}\percent[p]{0.06}{1}{0.06$^{\,\,\,}$}{20}&\phantom{$-$}\percent[p]{0.17}{1}{0.17$^\star$}{40}&\phantom{$-$}\percent[p]{0.08}{1}{0.08$^{\,\,\,}$}{20}&\phantom{$-$}\percent[p]{0.03}{1}{0.03$^{\,\,\,}$}{20}&\phantom{$-$}\percent[p]{0.17}{1}{0.17$^\star$}{40}&\phantom{$-$}\percent[p]{0.12}{1}{0.12$^\star$}{40}\\
\textsc{Storm}&$-$\percent[n]{0.05}{1}{0.05$^{\,\,\,}$}{20}&\phantom{$-$}\percent[p]{0.02}{1}{0.02$^{\,\,\,}$}{20}&\phantom{$-$}\percent[p]{0.18}{1}{0.18$^\star$}{40}&\phantom{$-$}\percent[p]{0.10}{1}{0.10$^{\,\,\,}$}{40}&\phantom{$-$}\percent[p]{0.57}{1}{0.57$^\dagger$}{60}&\phantom{$-$}\percent[p]{0.78}{1}{0.78$^\dagger$}{80}&\phantom{$-$}\percent[p]{0.52}{1}{0.52$^\dagger$}{60}&\phantom{$-$}\percent[p]{0.77}{1}{0.77$^\dagger$}{80}&$-$\percent[n]{0.62}{1}{0.62$^\dagger$}{60}&$-$\percent[n]{0.66}{1}{0.66$^\dagger$}{60}&$-$\percent[n]{0.22}{1}{0.22$^\dagger$}{40}&$-$\percent[n]{0.42}{1}{0.42$^\dagger$}{60}&\phantom{$-$}\percent[p]{0.04}{1}{0.04$^{\,\,\,}$}{20}&\phantom{$-$}\percent[p]{0.29}{1}{0.29$^\dagger$}{40}&\phantom{$-$}\percent[p]{0.39}{1}{0.39$^\dagger$}{60}&\phantom{$-$}\percent[p]{0.38}{1}{0.38$^\dagger$}{40}\\
\textsc{HSMGP}&$-$\percent[n]{0.02}{1}{0.02$^{\,\,\,}$}{20}&\phantom{$-$}\percent[p]{0.18}{1}{0.18$^\star$}{40}&$-$\percent[n]{0.02}{1}{0.02$^{\,\,\,}$}{20}&$-$\percent[n]{0.17}{1}{0.17$^\star$}{40}&\phantom{$-$}\percent[p]{0.47}{1}{0.47$^\dagger$}{60}&\phantom{$-$}\percent[p]{0.42}{1}{0.42$^\dagger$}{60}&\phantom{$-$}\percent[p]{0.47}{1}{0.47$^\dagger$}{60}&\phantom{$-$}\percent[p]{0.42}{1}{0.42$^\dagger$}{60}&$-$\percent[n]{0.03}{1}{0.03$^{\,\,\,}$}{20}&\phantom{$-$}\percent[p]{0.28}{1}{0.28$^\dagger$}{40}&$-$\percent[n]{0.01}{1}{0.01$^{\,\,\,}$}{20}&$-$\percent[n]{0.03}{1}{0.03$^{\,\,\,}$}{20}&\phantom{$-$}\percent[p]{0.25}{1}{0.25$^\dagger$}{40}&\phantom{$-$}\percent[p]{0.68}{1}{0.68$^\dagger$}{60}&\phantom{$-$}\percent[p]{0.14}{1}{0.14$^\star$}{40}&$-$\percent[n]{0.03}{1}{0.03$^{\,\,\,}$}{20}\\
\textsc{XGBoost}&\phantom{$-$}\percent[p]{0.12}{1}{0.12$^\star$}{40}&\phantom{$-$}\percent[p]{0.01}{1}{0.01$^{\,\,\,}$}{20}&\phantom{$-$}\percent[p]{0.03}{1}{0.03$^{\,\,\,}$}{20}&$-$\percent[n]{0.22}{1}{0.22$^\dagger$}{40}&\phantom{$-$}\percent[p]{0.56}{1}{0.56$^\dagger$}{60}&\phantom{$-$}\percent[p]{0.63}{1}{0.63$^\dagger$}{60}&\phantom{$-$}\percent[p]{0.53}{1}{0.53$^\dagger$}{60}&\phantom{$-$}\percent[p]{0.60}{1}{0.60$^\dagger$}{60}&$-$\percent[n]{0.41}{1}{0.41$^\dagger$}{60}&\phantom{$-$}\percent[p]{0.06}{1}{0.06$^{\,\,\,}$}{20}&$-$\percent[n]{0.02}{1}{0.02$^{\,\,\,}$}{20}&$-$\percent[n]{0.15}{1}{0.15$^\star$}{40}&\phantom{$-$}\percent[p]{0.50}{1}{0.50$^\dagger$}{60}&\phantom{$-$}\percent[p]{0.51}{1}{0.51$^\dagger$}{60}&\phantom{$-$}\percent[p]{0.42}{1}{0.42$^\dagger$}{60}&$-$\percent[n]{0.04}{1}{0.04$^{\,\,\,}$}{20}\\
\textsc{HIPAcc}&$-$\percent[n]{0.18}{1}{0.18$^\dagger$}{40}&$-$\percent[n]{0.24}{1}{0.24$^\dagger$}{40}&\phantom{$-$}\percent[p]{0.60}{1}{0.60$^\dagger$}{60}&\phantom{$-$}\percent[p]{0.62}{1}{0.62$^\dagger$}{60}&\phantom{$-$}\percent[p]{0.65}{1}{0.65$^\dagger$}{60}&\phantom{$-$}\percent[p]{0.73}{1}{0.73$^\dagger$}{80}&\phantom{$-$}\percent[p]{0.62}{1}{0.62$^\dagger$}{60}&\phantom{$-$}\percent[p]{0.70}{1}{0.70$^\dagger$}{80}&$-$\percent[n]{0.49}{1}{0.49$^\dagger$}{60}&$-$\percent[n]{0.48}{1}{0.48$^\dagger$}{60}&$-$\percent[n]{0.11}{1}{0.11$^{\,\,\,}$}{40}&$-$\percent[n]{0.06}{1}{0.06$^{\,\,\,}$}{20}&\phantom{$-$}\percent[p]{0.56}{1}{0.56$^\dagger$}{60}&\phantom{$-$}\percent[p]{0.62}{1}{0.62$^\dagger$}{60}&\phantom{$-$}\percent[p]{0.57}{1}{0.57$^\dagger$}{60}&\phantom{$-$}\percent[p]{0.58}{1}{0.58$^\dagger$}{60}\\
\textsc{SQLite}&\phantom{$-$}\percent[p]{0.07}{1}{0.07$^{\,\,\,}$}{20}&\phantom{$-$}\percent[p]{0.17}{1}{0.17$^\star$}{40}&\phantom{$-$}\percent[p]{0.22}{1}{0.22$^\dagger$}{40}&\phantom{$-$}\percent[p]{0.15}{1}{0.15$^\star$}{40}&$-$\percent[n]{0.06}{1}{0.06$^{\,\,\,}$}{20}&$-$\percent[n]{0.27}{1}{0.27$^\dagger$}{40}&\phantom{$-$}\percent[p]{0.28}{1}{0.28$^\dagger$}{40}&\phantom{$-$}\percent[p]{0.36}{1}{0.36$^\dagger$}{40}&\phantom{$-$}\percent[p]{0.24}{1}{0.24$^\dagger$}{40}&\phantom{$-$}\percent[p]{0.54}{1}{0.54$^\dagger$}{60}&\phantom{$-$}\percent[p]{0.09}{1}{0.09$^{\,\,\,}$}{40}&\phantom{$-$}\percent[p]{0.17}{1}{0.17$^\star$}{40}&\phantom{$-$}\percent[p]{0.38}{1}{0.38$^\dagger$}{40}&\phantom{$-$}\percent[p]{0.47}{1}{0.47$^\dagger$}{60}&$-$\percent[n]{0.14}{1}{0.14$^\star$}{40}&$-$\percent[n]{0.45}{1}{0.45$^\dagger$}{60}\\
\textsc{JavaGC}&\phantom{$-$}\percent[p]{0.45}{1}{0.45$^\dagger$}{60}&\phantom{$-$}\percent[p]{0.42}{1}{0.42$^\dagger$}{60}&$-$\percent[n]{0.29}{1}{0.29$^\star$}{40}&\phantom{$-$}\percent[p]{0.29}{1}{0.29$^\star$}{40}&\phantom{$-$}\percent[p]{0.67}{1}{0.67$^\dagger$}{60}&\phantom{$-$}\percent[p]{0.27}{1}{0.27$^\star$}{40}&\phantom{$-$}\percent[p]{0.30}{1}{0.30$^\star$}{40}&\phantom{$-$}\percent[p]{0.34}{1}{0.34$^\star$}{40}&$-$\percent[n]{0.35}{1}{0.35$^\star$}{40}&\phantom{$-$}\percent[p]{0.03}{1}{0.03$^{\,\,\,}$}{20}&\phantom{$-$}\percent[p]{0.44}{1}{0.44$^\dagger$}{60}&\phantom{$-$}\percent[p]{0.72}{1}{0.72$^\dagger$}{80}&\phantom{$-$}\percent[p]{0.37}{1}{0.37$^\star$}{40}&$-$\percent[n]{0.18}{1}{0.18$^{\,\,\,}$}{40}&\phantom{$-$}\percent[p]{0.38}{1}{0.38$^\star$}{40}&\phantom{$-$}\percent[p]{0.01}{1}{0.01$^{\,\,\,}$}{20}\\
\textsc{Polly}&\phantom{$-$}\percent[p]{0.53}{1}{0.53$^\dagger$}{60}&\phantom{$-$}\percent[p]{0.70}{1}{0.70$^\dagger$}{80}&\phantom{$-$}\percent[p]{0.40}{1}{0.40$^\dagger$}{60}&\phantom{$-$}\percent[p]{0.37}{1}{0.37$^\dagger$}{40}&\phantom{$-$}\percent[p]{0.44}{1}{0.44$^\dagger$}{60}&\phantom{$-$}\percent[p]{0.73}{1}{0.73$^\dagger$}{80}&\phantom{$-$}\percent[p]{0.50}{1}{0.50$^\dagger$}{60}&\phantom{$-$}\percent[p]{0.77}{1}{0.77$^\dagger$}{80}&$-$\percent[n]{0.15}{1}{0.15$^\star$}{40}&$-$\percent[n]{0.22}{1}{0.22$^\dagger$}{40}&\phantom{$-$}\percent[p]{0.19}{1}{0.19$^\dagger$}{40}&\phantom{$-$}\percent[p]{0.06}{1}{0.06$^{\,\,\,}$}{20}&\phantom{$-$}\percent[p]{0.25}{1}{0.25$^\dagger$}{40}&\phantom{$-$}\percent[p]{0.32}{1}{0.32$^\dagger$}{40}&\phantom{$-$}\percent[p]{0.16}{1}{0.16$^\star$}{40}&\phantom{$-$}\percent[p]{0.23}{1}{0.23$^\dagger$}{40}\\
\bottomrule
\end{tabular}
\end{adjustbox}
\end{table*}

%% file: Figures/rq1_res_summary_metrics.tex
\begin{tikzpicture}
\begin{axis}[
    ybar stacked,
    ymajorgrids=true,
    grid style = {dashed},
    width=16cm,
    height=7cm,
    ymax=21,
    ymin=0,
    xmin=0,
    xmax=17,
    bar width=0.6cm,
    nodes near coords, 
    enlarge x limits=0.05,
    axis x line = bottom,
    axis y line = left,
    legend style={draw=none,at={(0.5,1.15)},
      anchor=north,legend columns=4},
    ylabel={\# Systems},
    legend cell align={left},
    symbolic x coords={0,1,2,3,4,5,6,7,8,9,10,11,12,13,14,15,16,17},
    xticklabels={, \texttt{FDC}$-$MAPE, \texttt{FDC}$-$$\mu$RD, \texttt{FBD}$-$MAPE, \texttt{FBD}$-$$\mu$RD, \texttt{Ske}$-$MAPE, \texttt{Ske}$-$$\mu$RD, \texttt{Kur}$-$MAPE, \texttt{Kur}$-$$\mu$RD, \texttt{PLO}$-$MAPE, \texttt{PLO}$-$$\mu$RD, \texttt{CL}$-$MAPE, \texttt{CL}$-$$\mu$RD, \texttt{MIE}$-$MAPE, \texttt{MIE}$-$$\mu$RD, \texttt{NBC}$-$MAPE, \texttt{NBC}$-$$\mu$RD},
    xticklabel style={rotate=45, anchor=east},
    xtick=data,
    bar shift=0pt,
]

\addplot[ybar, fill=gree, draw=black, line width=0.5pt] coordinates {
    (0,0) (1,0) (2,1) (3,1) (4,1) (5,2) (6,9) (7,2) (8,8) (9,0) (10,0) (11,0) (12,1) (13,0) (14,0) (15,0) (16,0)
};

\addplot[ybar, fill=gree!50, draw=black, line width=0.5pt] coordinates {
    (0,0) (1,3) (2,2) (3,3) (4,2) (5,12) (6,4) (7,9) (8,4) (9,2) (10,2) (11,2) (12,0) (13,2) (14,6) (15,4) (16,3)
};

\addplot[ybar, fill=black!50, draw=black, line width=0.5pt] coordinates {
    (0,0) (1,3) (2,6) (3,5) (4,8) (5,1) (6,2) (7,4) (8,5) (9,2) (10,4) (11,6) (12,10) (13,6) (14,5) (15,6) (16,4)
};

\addplot[ybar, fill=black!10, draw=black, line width=0.5pt] coordinates {
    (0,0) (1,11) (2,6) (3,6) (4,5) (5,2) (6,3) (7,1) (8,2) (9,5) (10,6) (11,5) (12,5) (13,9) (14,5) (15,4) (16,7)
};

\addplot[ybar, fill=black!30, draw=black, line width=0.5pt] coordinates {
    (0,0) (1,2) (2,4) (3,4) (4,3) (5,1) (6,1) (7,2) (8,0) (9,5) (10,4) (11,6) (12,2) (13,2) (14,2) (15,4) (16,4)
};

\addplot[ybar, fill=red!50, draw=black, line width=0.5pt] coordinates {
    (0,0) (1,0) (2,0) (3,0) (4,0) (5,1) (6,0) (7,1) (8,0) (9,5) (10,3) (11,0) (12,1) (13,0) (14,1) (15,1) (16,1)
};

\addplot[ybar, fill=red, draw=black, line width=0.5pt] coordinates {
    (0,0) (1,0) (2,0) (3,0) (4,0) (5,0) (6,0) (7,0) (8,0) (9,0) (10,0) (11,0) (12,0) (13,0) (14,0) (15,0) (16,0)
};

\addlegendentry{Positive Strong }
\addlegendentry{Positive Moderate}
\addlegendentry{Positive Weak}
\addlegendentry{Negligible }
\addlegendentry{Negative Weak }
\addlegendentry{Negative Moderate}
\addlegendentry{Negative Strong}

\end{axis}
\end{tikzpicture}

%% file: Figures/rq1_res_summary_system.tex
\begin{tikzpicture}
\begin{axis}[
    ybar stacked,
    ymajorgrids=true,
    grid style = {dashed},
    width=15cm,
    height=6cm,
    ymax=18,
    ymin=0,
    xmin=0,
    xmax=20,
    bar width=0.55cm,
    nodes near coords, 
    enlarge x limits=0.05,
    axis x line = bottom,
    axis y line = left,
    legend style={draw=none,at={(0.5,1.15)},
      anchor=north,legend columns=4},
    ylabel={\# Cases},
    legend cell align={left},
    xtick={1,...,19},
    xticklabels={\textsc{Apache},\textsc{7z},\textsc{DConvert},\textsc{DeepArch},\textsc{ExaStencils},\textsc{Hadoop},\textsc{MariaDB},\textsc{MongoDB},\textsc{PostgreSQL},\textsc{Redis},\textsc{SPARK},\textsc{Storm},\textsc{HSMGP},\textsc{XGBoost},\textsc{HIPAcc},\textsc{SQLite},\textsc{JavaGC},\textsc{Polly}},
    xticklabel style={rotate=45, anchor=east},
]

\addplot[ybar, fill=gree, draw=black] coordinates {
    (1,2) (2,6) (3,2) (4,2) (5,2) (6,0) (7,1) (8,0) (9,0) (10,0) (11,2) (12,2) (13,0) (14,0) (15,2) (16,0) (17,1) (18,3)
};

\addplot[ybar, fill=gree!50, draw=black] coordinates {
    (1,2) (2,3) (3,3) (4,2) (5,6) (6,0) (7,2) (8,5) (9,0) (10,2) (11,2) (12,3) (13,5) (14,7) (15,8) (16,2) (17,4) (18,4)
};

\addplot[ybar, fill=black!50, draw=black] coordinates {
    (1,3) (2,3) (3,2) (4,7) (5,3) (6,4) (7,2) (8,7) (9,7) (10,3) (11,5) (12,4) (13,4) (14,1) (15,0) (16,9) (17,6) (18,6)
};

\addplot[ybar, fill=black!10, draw=black] coordinates {
    (1,5) (2,2) (3,6) (4,1) (5,5) (6,7) (7,6) (8,2) (9,5) (10,6) (11,5) (12,3) (13,6) (14,5) (15,1) (16,2) (17,2) (18,1)
};

\addplot[ybar, fill=black!30, draw=black] coordinates {
    (1,4) (2,0) (3,3) (4,4) (5,0) (6,5) (7,4) (8,2) (9,3) (10,4) (11,2) (12,1) (13,1) (14,2) (15,3) (16,2) (17,3) (18,2)
};

\addplot[ybar, fill=red!50, draw=black] coordinates {
    (1,0) (2,2) (3,0) (4,0) (5,0) (6,0) (7,1) (8,0) (9,1) (10,1) (11,0) (12,3) (13,0) (14,1) (15,2) (16,1) (17,0) (18,0)
};

\addplot[ybar, fill=red, draw=black] coordinates {
    (1,0) (2,0) (3,0) (4,0) (5,0) (6,0) (7,0) (8,0) (9,0) (10,0) (11,0) (12,0) (13,0) (14,0) (15,0) (16,0) (17,0) (18,0)
};

\addlegendentry{Positive Strong }
\addlegendentry{Positive Moderate}
\addlegendentry{Positive Weak}
\addlegendentry{Negligible }
\addlegendentry{Negative Weak }
\addlegendentry{Negative Moderate}
\addlegendentry{Negative Strong}

\end{axis}
\end{tikzpicture}

%% file: Tables/rq2_overview.tex
\begin{table*}[t!]
\caption{Relative deviation between the landscape emulated by a surrogate model and that of the real systems. ``\textbf{$+\Delta$}'' and ``\textbf{$-\Delta$}'' denote the average relative deviation when it is $\geq 0$ and $\leq 0$, respectively, within the $300$ data points; N/A means no applicable cases under the condition. ``\textbf{SS\%}'' means the percentage of models such that the relative deviation has statistical significance ($p<0.05$) across the $10$ models. \colorbox{red!30}{Red cells} highlight the cases where SS\%$>$50. $\rightarrow$ 0, $\uparrow$, and $\downarrow$ represents a value that is close to 0, bigger, and smaller, respectively. The ``better'' means a surrogate model resembles the real landscape in a more preferred way on the landscape feature, as derived from our theory of landscape dominance on model usefulness.}
\label{tb: rq2-overview}
\centering
\setlength{\tabcolsep}{1mm}
\begin{adjustbox}{width=\textwidth,center}
\begin{tabular}{lrrr|rrr|rrr|rrr||rrr|rrr|rrr|rrr}
\toprule
\multirow{2}{*}{\textbf{System}} & \multicolumn{3}{c|}{\textbf{\texttt{FDC}} ($\rightarrow$ 0 is better)} & \multicolumn{3}{c|}{\textbf{\texttt{FBD}} ($\rightarrow$ 0 is better)} & \multicolumn{3}{c|}{\textbf{\texttt{Ske}} ($\rightarrow$ 0 is better)} & \multicolumn{3}{c||}{\textbf{\texttt{Kur}} ($\rightarrow$ 0 is better)} & \multicolumn{3}{c|}{\textbf{\texttt{PLO}} ($\downarrow$ is better)} & \multicolumn{3}{c|}{\textbf{\texttt{CL}} ($\uparrow$ is better)} & \multicolumn{3}{c|}{\textbf{\texttt{MIE}} ($\downarrow$ is better)} & \multicolumn{3}{c}{\textbf{\texttt{NBC}} ($\downarrow$ is better)} \\ 
\cmidrule{2-25}
 & $+$$\Delta$  & $-$$\Delta$ & SS\% & $+$$\Delta$  & $-$$\Delta$ & SS\% & $+$$\Delta$  & $-$$\Delta$ & SS\% & $+$$\Delta$  & $-$$\Delta$ & SS\% & $+$$\Delta$  & $-$$\Delta$ & SS\% & $+$$\Delta$  & $-$$\Delta$ & SS\% & $+$$\Delta$  & $-$$\Delta$ & SS\% & $+$$\Delta$  & $-$$\Delta$ & SS\% \\ 
\midrule
\textsc{Apache} & \cellcolor{red!30}{0.205} & \cellcolor{red!30}{N/A} & \cellcolor{red!30}{90.0} & \cellcolor{red!30}{4.533} & \cellcolor{red!30}{N/A} & \cellcolor{red!30}{100.0} & \cellcolor{red!30}{0.003} & \cellcolor{red!30}{$-$0.001} & \cellcolor{red!30}{70.0} & \cellcolor{red!30}{0.144} & \cellcolor{red!30}{$-$0.003} & \cellcolor{red!30}{80.0} & \cellcolor{red!30}{0.051} & \cellcolor{red!30}{$-$0.055} & \cellcolor{red!30}{90.0} & \cellcolor{red!30}{1.615} & \cellcolor{red!30}{N/A} & \cellcolor{red!30}{100.0} & \cellcolor{red!30}{N/A} & \cellcolor{red!30}{$-$0.028} & \cellcolor{red!30}{80.0} & \cellcolor{red!30}{0.010} & \cellcolor{red!30}{$-$0.030} & \cellcolor{red!30}{60.0} \\
\textsc{7z} & \cellcolor{red!30}{0.133} & \cellcolor{red!30}{N/A} & \cellcolor{red!30}{100.0} & \cellcolor{red!30}{3.987} & \cellcolor{red!30}{N/A} & \cellcolor{red!30}{100.0} & \cellcolor{red!30}{0.068} & \cellcolor{red!30}{$-$2.499} & \cellcolor{red!30}{60.0} & \cellcolor{red!30}{48.902} & \cellcolor{red!30}{$-$8.920} & \cellcolor{red!30}{80.0} & \cellcolor{red!30}{0.008} & \cellcolor{red!30}{$-$0.006} & \cellcolor{red!30}{100.0} & \cellcolor{red!30}{0.787} & \cellcolor{red!30}{$-$0.658} & \cellcolor{red!30}{80.0} & \cellcolor{red!30}{0.028} & \cellcolor{red!30}{$-$0.022} & \cellcolor{red!30}{80.0} & \cellcolor{red!30}{0.032} & \cellcolor{red!30}{$-$0.241} & \cellcolor{red!30}{100.0} \\
\textsc{DConvert} & \cellcolor{red!30}{0.116} & \cellcolor{red!30}{$-$0.050} & \cellcolor{red!30}{80.0} & \cellcolor{red!30}{3.340} & \cellcolor{red!30}{N/A} & \cellcolor{red!30}{100.0} & \cellcolor{red!30}{0.012} & \cellcolor{red!30}{$-$0.988} & \cellcolor{red!30}{90.0} & \cellcolor{red!30}{0.042} & \cellcolor{red!30}{$-$1.091} & \cellcolor{red!30}{70.0} & \cellcolor{red!30}{0.033} & \cellcolor{red!30}{$-$0.049} & \cellcolor{red!30}{100.0} & \cellcolor{red!30}{0.492} & \cellcolor{red!30}{$-$0.020} & \cellcolor{red!30}{90.0} & \cellcolor{red!30}{0.025} & \cellcolor{red!30}{$-$0.024} & \cellcolor{red!30}{90.0} & \cellcolor{red!30}{0.029} & \cellcolor{red!30}{$-$0.188} & \cellcolor{red!30}{100.0} \\
\textsc{DeepArch} & \cellcolor{red!30}{0.079} & \cellcolor{red!30}{$-$0.045} & \cellcolor{red!30}{70.0} & \cellcolor{red!30}{2.894} & \cellcolor{red!30}{N/A} & \cellcolor{red!30}{100.0} & 0.013  & $-$0.445 & 30.0 & \cellcolor{red!30}{0.060} & \cellcolor{red!30}{$-$0.472} & \cellcolor{red!30}{70.0} & \cellcolor{red!30}{0.152} & \cellcolor{red!30}{N/A} & \cellcolor{red!30}{100.0} & 0.355  & $-$0.317 & 50.0 & \cellcolor{red!30}{0.010} & \cellcolor{red!30}{$-$0.010} & \cellcolor{red!30}{90.0} & \cellcolor{red!30}{N/A} & \cellcolor{red!30}{$-$0.054} & \cellcolor{red!30}{100.0} \\
\textsc{ExaStencils} & \cellcolor{red!30}{0.103} & \cellcolor{red!30}{N/A} & \cellcolor{red!30}{90.0} & \cellcolor{red!30}{4.384} & \cellcolor{red!30}{N/A} & \cellcolor{red!30}{100.0} & \cellcolor{red!30}{0.010} & \cellcolor{red!30}{$-$1.054} & \cellcolor{red!30}{70.0} & \cellcolor{red!30}{271.778} & \cellcolor{red!30}{$-$0.753} & \cellcolor{red!30}{90.0} & \cellcolor{red!30}{0.009} & \cellcolor{red!30}{$-$0.000} & \cellcolor{red!30}{100.0} & \cellcolor{red!30}{1.041} & \cellcolor{red!30}{$-$1.266} & \cellcolor{red!30}{80.0} & \cellcolor{red!30}{0.015} & \cellcolor{red!30}{$-$0.026} & \cellcolor{red!30}{100.0} & \cellcolor{red!30}{0.020} & \cellcolor{red!30}{$-$0.360} & \cellcolor{red!30}{100.0} \\
\textsc{Hadoop} & \cellcolor{red!30}{0.208} & \cellcolor{red!30}{N/A} & \cellcolor{red!30}{100.0} & \cellcolor{red!30}{6.970} & \cellcolor{red!30}{N/A} & \cellcolor{red!30}{100.0} & \cellcolor{red!30}{0.227} & \cellcolor{red!30}{$-$0.282} & \cellcolor{red!30}{80.0} & \cellcolor{red!30}{1.009} & \cellcolor{red!30}{N/A} & \cellcolor{red!30}{90.0} & \cellcolor{red!30}{0.000} & \cellcolor{red!30}{N/A} & \cellcolor{red!30}{100.0} & 1.135  & $-$0.104 & 40.0 & \cellcolor{red!30}{N/A} & \cellcolor{red!30}{$-$0.021} & \cellcolor{red!30}{100.0} & \cellcolor{red!30}{0.017} & \cellcolor{red!30}{$-$0.082} & \cellcolor{red!30}{90.0} \\
\textsc{MariaDB} & 0.071  & $-$0.022 & 30.0 & \cellcolor{red!30}{3.440} & \cellcolor{red!30}{N/A} & \cellcolor{red!30}{100.0} & \cellcolor{red!30}{N/A} & \cellcolor{red!30}{$-$0.199} & \cellcolor{red!30}{80.0} & \cellcolor{red!30}{1.173} & \cellcolor{red!30}{$-$0.185} & \cellcolor{red!30}{70.0} & \cellcolor{red!30}{0.063} & \cellcolor{red!30}{$-$0.027} & \cellcolor{red!30}{90.0} & \cellcolor{red!30}{1.547} & \cellcolor{red!30}{$-$0.278} & \cellcolor{red!30}{100.0} & \cellcolor{red!30}{0.023} & \cellcolor{red!30}{$-$0.012} & \cellcolor{red!30}{70.0} & \cellcolor{red!30}{0.021} & \cellcolor{red!30}{$-$0.003} & \cellcolor{red!30}{100.0} \\
\textsc{MongoDB} & \cellcolor{red!30}{0.031} & \cellcolor{red!30}{$-$0.076} & \cellcolor{red!30}{80.0} & \cellcolor{red!30}{3.253} & \cellcolor{red!30}{N/A} & \cellcolor{red!30}{100.0} & \cellcolor{red!30}{0.040} & \cellcolor{red!30}{$-$0.173} & \cellcolor{red!30}{60.0} & \cellcolor{red!30}{60.338} & \cellcolor{red!30}{$-$0.016} & \cellcolor{red!30}{100.0} & \cellcolor{red!30}{0.065} & \cellcolor{red!30}{$-$0.022} & \cellcolor{red!30}{90.0} & \cellcolor{red!30}{0.500} & \cellcolor{red!30}{$-$1.664} & \cellcolor{red!30}{100.0} & \cellcolor{red!30}{N/A} & \cellcolor{red!30}{$-$0.052} & \cellcolor{red!30}{100.0} & \cellcolor{red!30}{0.011} & \cellcolor{red!30}{$-$0.152} & \cellcolor{red!30}{70.0} \\
\textsc{PostgreSQL} & 0.063  & $-$0.084 & 30.0 & \cellcolor{red!30}{3.740} & \cellcolor{red!30}{N/A} & \cellcolor{red!30}{100.0} & \cellcolor{red!30}{1.773} & \cellcolor{red!30}{$-$0.426} & \cellcolor{red!30}{100.0} & \cellcolor{red!30}{4.253} & \cellcolor{red!30}{$-$0.133} & \cellcolor{red!30}{90.0} & \cellcolor{red!30}{0.035} & \cellcolor{red!30}{$-$0.028} & \cellcolor{red!30}{100.0} & \cellcolor{red!30}{0.496} & \cellcolor{red!30}{$-$1.525} & \cellcolor{red!30}{100.0} & \cellcolor{red!30}{0.037} & \cellcolor{red!30}{$-$0.032} & \cellcolor{red!30}{100.0} & \cellcolor{red!30}{0.004} & \cellcolor{red!30}{$-$0.057} & \cellcolor{red!30}{70.0} \\
\textsc{Redis} & \cellcolor{red!30}{0.127} & \cellcolor{red!30}{$-$0.006} & \cellcolor{red!30}{60.0} & \cellcolor{red!30}{7.296} & \cellcolor{red!30}{N/A} & \cellcolor{red!30}{100.0} & 0.145  & $-$0.145 & 50.0 & \cellcolor{red!30}{19.413} & \cellcolor{red!30}{$-$1.039} & \cellcolor{red!30}{100.0} & \cellcolor{red!30}{0.002} & \cellcolor{red!30}{N/A} & \cellcolor{red!30}{100.0} & \cellcolor{red!30}{0.863} & \cellcolor{red!30}{$-$0.065} & \cellcolor{red!30}{70.0} & \cellcolor{red!30}{0.019} & \cellcolor{red!30}{$-$0.012} & \cellcolor{red!30}{90.0} & \cellcolor{red!30}{0.010} & \cellcolor{red!30}{$-$0.031} & \cellcolor{red!30}{100.0} \\
\textsc{Spark} & \cellcolor{red!30}{0.315} & \cellcolor{red!30}{$-$0.016} & \cellcolor{red!30}{60.0} & \cellcolor{red!30}{8.920} & \cellcolor{red!30}{N/A} & \cellcolor{red!30}{100.0} & \cellcolor{red!30}{0.357} & \cellcolor{red!30}{$-$0.018} & \cellcolor{red!30}{60.0} & \cellcolor{red!30}{1.549} & \cellcolor{red!30}{$-$0.047} & \cellcolor{red!30}{90.0} & \cellcolor{red!30}{0.003} & \cellcolor{red!30}{$-$0.001} & \cellcolor{red!30}{90.0} & \cellcolor{red!30}{0.966} & \cellcolor{red!30}{N/A} & \cellcolor{red!30}{90.0} & \cellcolor{red!30}{0.011} & \cellcolor{red!30}{$-$0.002} & \cellcolor{red!30}{70.0} & \cellcolor{red!30}{0.027} & \cellcolor{red!30}{$-$0.136} & \cellcolor{red!30}{100.0} \\
\textsc{Storm} & 0.052  & $-$0.005 & 20.0 & \cellcolor{red!30}{2.700} & \cellcolor{red!30}{N/A} & \cellcolor{red!30}{100.0} & \cellcolor{red!30}{N/A} & \cellcolor{red!30}{$-$0.485} & \cellcolor{red!30}{80.0} & \cellcolor{red!30}{0.003} & \cellcolor{red!30}{$-$0.893} & \cellcolor{red!30}{60.0} & \cellcolor{red!30}{0.044} & \cellcolor{red!30}{$-$0.004} & \cellcolor{red!30}{90.0} & \cellcolor{red!30}{1.135} & \cellcolor{red!30}{$-$0.157} & \cellcolor{red!30}{60.0} & \cellcolor{red!30}{0.087} & \cellcolor{red!30}{$-$0.008} & \cellcolor{red!30}{80.0} & \cellcolor{red!30}{N/A} & \cellcolor{red!30}{$-$0.000} & \cellcolor{red!30}{100.0} \\
\textsc{HSMGP} & \cellcolor{red!30}{0.018} & \cellcolor{red!30}{$-$0.047} & \cellcolor{red!30}{80.0} & \cellcolor{red!30}{2.632} & \cellcolor{red!30}{N/A} & \cellcolor{red!30}{100.0} & N/A  & $-$0.270 & 30.0 & N/A  & $-$1.285 & 30.0 & \cellcolor{red!30}{0.187} & \cellcolor{red!30}{N/A} & \cellcolor{red!30}{100.0} & \cellcolor{red!30}{0.181} & \cellcolor{red!30}{N/A} & \cellcolor{red!30}{60.0} & \cellcolor{red!30}{0.182} & \cellcolor{red!30}{N/A} & \cellcolor{red!30}{100.0} & \cellcolor{red!30}{N/A} & \cellcolor{red!30}{$-$0.251} & \cellcolor{red!30}{100.0} \\
\textsc{XGBoost} & \cellcolor{red!30}{0.855} & \cellcolor{red!30}{N/A} & \cellcolor{red!30}{100.0} & \cellcolor{red!30}{3.953} & \cellcolor{red!30}{N/A} & \cellcolor{red!30}{100.0} & \cellcolor{red!30}{0.018} & \cellcolor{red!30}{$-$0.590} & \cellcolor{red!30}{60.0} & \cellcolor{red!30}{N/A} & \cellcolor{red!30}{$-$0.651} & \cellcolor{red!30}{70.0} & \cellcolor{red!30}{0.045} & \cellcolor{red!30}{$-$0.008} & \cellcolor{red!30}{100.0} & \cellcolor{red!30}{0.882} & \cellcolor{red!30}{$-$0.141} & \cellcolor{red!30}{90.0} & \cellcolor{red!30}{0.029} & \cellcolor{red!30}{$-$0.015} & \cellcolor{red!30}{70.0} & \cellcolor{red!30}{0.004} & \cellcolor{red!30}{$-$0.029} & \cellcolor{red!30}{100.0} \\
\textsc{HIPAcc} & \cellcolor{red!30}{0.082} & \cellcolor{red!30}{$-$0.028} & \cellcolor{red!30}{90.0} & \cellcolor{red!30}{3.459} & \cellcolor{red!30}{N/A} & \cellcolor{red!30}{100.0} & \cellcolor{red!30}{0.509} & \cellcolor{red!30}{$-$0.428} & \cellcolor{red!30}{90.0} & \cellcolor{red!30}{17.966} & \cellcolor{red!30}{$-$1.152} & \cellcolor{red!30}{100.0} & \cellcolor{red!30}{0.284} & \cellcolor{red!30}{N/A} & \cellcolor{red!30}{100.0} & \cellcolor{red!30}{N/A} & \cellcolor{red!30}{$-$0.936} & \cellcolor{red!30}{100.0} & \cellcolor{red!30}{0.053} & \cellcolor{red!30}{$-$0.007} & \cellcolor{red!30}{100.0} & \cellcolor{red!30}{N/A} & \cellcolor{red!30}{$-$0.103} & \cellcolor{red!30}{100.0} \\
\textsc{SQLite} & \cellcolor{red!30}{0.116} & \cellcolor{red!30}{$-$0.116} & \cellcolor{red!30}{70.0} & \cellcolor{red!30}{11.613} & \cellcolor{red!30}{N/A} & \cellcolor{red!30}{100.0} & \cellcolor{red!30}{0.240} & \cellcolor{red!30}{$-$2.951} & \cellcolor{red!30}{90.0} & \cellcolor{red!30}{27.620} & \cellcolor{red!30}{N/A} & \cellcolor{red!30}{90.0} & \cellcolor{red!30}{0.293} & \cellcolor{red!30}{N/A} & \cellcolor{red!30}{100.0} & 0.321  & $-$0.420 & 50.0 & \cellcolor{red!30}{0.003} & \cellcolor{red!30}{$-$0.009} & \cellcolor{red!30}{80.0} & \cellcolor{red!30}{0.053} & \cellcolor{red!30}{$-$0.002} & \cellcolor{red!30}{90.0} \\
\textsc{JavaGC} & \cellcolor{red!30}{0.169} & \cellcolor{red!30}{N/A} & \cellcolor{red!30}{70.0} & \cellcolor{red!30}{19.631} & \cellcolor{red!30}{N/A} & \cellcolor{red!30}{80.0} & \cellcolor{red!30}{N/A} & \cellcolor{red!30}{$-$3.938} & \cellcolor{red!30}{70.0} & \cellcolor{red!30}{N/A} & \cellcolor{red!30}{$-$88.941} & \cellcolor{red!30}{70.0} & \cellcolor{red!30}{0.282} & \cellcolor{red!30}{$-$0.035} & \cellcolor{red!30}{80.0} & \cellcolor{red!30}{0.988} & \cellcolor{red!30}{$-$0.366} & \cellcolor{red!30}{60.0} & \cellcolor{red!30}{0.031} & \cellcolor{red!30}{$-$0.051} & \cellcolor{red!30}{80.0} & \cellcolor{red!30}{0.020} & \cellcolor{red!30}{$-$0.418} & \cellcolor{red!30}{80.0} \\
\textsc{Polly} & \cellcolor{red!30}{0.039} & \cellcolor{red!30}{$-$0.017} & \cellcolor{red!30}{90.0} & \cellcolor{red!30}{5.540} & \cellcolor{red!30}{N/A} & \cellcolor{red!30}{100.0} & \cellcolor{red!30}{2.193} & \cellcolor{red!30}{$-$0.008} & \cellcolor{red!30}{90.0} & \cellcolor{red!30}{0.022} & \cellcolor{red!30}{$-$3.595} & \cellcolor{red!30}{70.0} & \cellcolor{red!30}{0.249} & \cellcolor{red!30}{$-$0.033} & \cellcolor{red!30}{100.0} & \cellcolor{red!30}{0.566} & \cellcolor{red!30}{$-$0.026} & \cellcolor{red!30}{90.0} & \cellcolor{red!30}{0.005} & \cellcolor{red!30}{$-$0.023} & \cellcolor{red!30}{60.0} & \cellcolor{red!30}{0.014} & \cellcolor{red!30}{$-$0.078} & \cellcolor{red!30}{100.0} \\
\midrule
\textbf{Overall (Median)} & \cellcolor{red!30}{0.109} & \cellcolor{red!30}{$-$0.036} & \cellcolor{red!30}{80.0} & \cellcolor{red!30}{3.970} & \cellcolor{red!30}{N/A} & \cellcolor{red!30}{100.0} & \cellcolor{red!30}{0.107} & \cellcolor{red!30}{$-$0.427} & \cellcolor{red!30}{70.0} & \cellcolor{red!30}{1.549} & \cellcolor{red!30}{$-$0.823} & \cellcolor{red!30}{80.0} & \cellcolor{red!30}{0.048} & \cellcolor{red!30}{$-$0.024} & \cellcolor{red!30}{100.0} & \cellcolor{red!30}{0.863} & \cellcolor{red!30}{$-$0.317} & \cellcolor{red!30}{85.0} & \cellcolor{red!30}{0.025} & \cellcolor{red!30}{$-$0.021} & \cellcolor{red!30}{85.0} & \cellcolor{red!30}{0.019} & \cellcolor{red!30}{$-$0.080} & \cellcolor{red!30}{100.0} \\
\bottomrule
\end{tabular}
\end{adjustbox}
\end{table*}

%% file: Tables/best_performing_model_in_landscape.tex
\begin{table}[t!]
\centering
\caption{The model that emulates a landscape with the smallest $\Delta g$ for global feature or $l$ for local features on a system. A \colorbox{gree!30}{Green cell} denotes a classic machine learning model or otherwise it is a deep learning model.}
\begin{adjustbox}{width=0.5\textwidth,center}
\begin{tabular}{lllll|llll}
\toprule
\textbf{System} & \textbf{\texttt{FDC}} & \textbf{\texttt{FBD}} & \textbf{\texttt{Ske}} & \textbf{\texttt{Kur}} & \textbf{\texttt{PLO}} & \textbf{\texttt{CL}} & \textbf{\texttt{MIE}} & \textbf{\texttt{NBC}} \\ \midrule
\textsc{Apache} & \cellcolor{gree!30}\texttt{DT} & \cellcolor{gree!30}\texttt{DCT} & \texttt{HIP} & \cellcolor{gree!30}\texttt{LR} & \texttt{HIP} & \texttt{DaL} & \cellcolor{gree!30}\texttt{SVR} & \cellcolor{gree!30}\texttt{DCT} \\
\textsc{7z} & \cellcolor{gree!30}\texttt{DCT} & \cellcolor{gree!30}\texttt{DCT} & \cellcolor{gree!30}\texttt{RF} & \cellcolor{gree!30}\texttt{SPL} & \cellcolor{gree!30}\texttt{RF} & \cellcolor{gree!30}\texttt{DCT} & \cellcolor{gree!30}\texttt{GP} & \cellcolor{gree!30}\texttt{DCT} \\
\textsc{DConvert} & \cellcolor{gree!30}\texttt{SVR} & \cellcolor{gree!30}\texttt{DCT} & \texttt{DeP} & \cellcolor{gree!30}\texttt{SPL} & \texttt{DaL} & \cellcolor{gree!30}\texttt{DCT} & \texttt{DaL} & \cellcolor{gree!30}\texttt{DCT} \\
\textsc{DeepArch} & \cellcolor{gree!30}\texttt{DT} & \cellcolor{gree!30}\texttt{DCT} & \cellcolor{gree!30}\texttt{RF} & \texttt{DeP} & \cellcolor{gree!30}\texttt{RF} & \cellcolor{gree!30}\texttt{SPL} & \cellcolor{gree!30}\texttt{RF} & \cellcolor{gree!30}\texttt{DCT} \\
\textsc{ExaStencils} & \cellcolor{gree!30}\texttt{DT} & \cellcolor{gree!30}\texttt{DCT} & \cellcolor{gree!30}\texttt{RF} & \cellcolor{gree!30}\texttt{SVR} & \cellcolor{gree!30}\texttt{RF} & \cellcolor{gree!30}\texttt{DCT} & \cellcolor{gree!30}\texttt{DCT} & \cellcolor{gree!30}\texttt{DCT} \\
\textsc{Hadoop} & \cellcolor{gree!30}\texttt{RF} & \cellcolor{gree!30}\texttt{DCT} & \cellcolor{gree!30}\texttt{DT} & \cellcolor{gree!30}\texttt{GP} & \cellcolor{gree!30}\texttt{DT} & \cellcolor{gree!30}\texttt{DCT} & \texttt{DaL} & \cellcolor{gree!30}\texttt{DCT} \\
\textsc{MariaDB} & \cellcolor{gree!30}\texttt{SPL} & \cellcolor{gree!30}\texttt{DCT} & \texttt{DeP} & \cellcolor{gree!30}\texttt{SPL} & \texttt{DeP} & \cellcolor{gree!30}\texttt{LR} & \texttt{DeP} & \cellcolor{gree!30}\texttt{DT} \\
\textsc{MongoDB} & \cellcolor{gree!30}\texttt{DCT} & \cellcolor{gree!30}\texttt{DCT} & \cellcolor{gree!30}\texttt{DCT} & \cellcolor{gree!30}\texttt{SPL} & \cellcolor{gree!30}\texttt{LR} & \cellcolor{gree!30}\texttt{DCT} & \cellcolor{gree!30}\texttt{GP} & \cellcolor{gree!30}\texttt{DCT} \\
\textsc{PostgreSQL} & \texttt{HIP} & \cellcolor{gree!30}\texttt{DCT} & \cellcolor{gree!30}\texttt{SVR} & \cellcolor{gree!30}\texttt{GP} & \cellcolor{gree!30}\texttt{SVR} & \cellcolor{gree!30}\texttt{SPL} & \cellcolor{gree!30}\texttt{RF} & \cellcolor{gree!30}\texttt{DT} \\
\textsc{Redis} & \cellcolor{gree!30}\texttt{RF} & \texttt{DaL} & \cellcolor{gree!30}\texttt{DT} & \cellcolor{gree!30}\texttt{LR} & \cellcolor{gree!30}\texttt{SPL} & \cellcolor{gree!30}\texttt{DT} & \cellcolor{gree!30}\texttt{GP} & \cellcolor{gree!30}\texttt{DCT} \\
\textsc{Spark} & \cellcolor{gree!30}\texttt{SVR} & \cellcolor{gree!30}\texttt{DCT} & \cellcolor{gree!30}\texttt{DCT} & \cellcolor{gree!30}\texttt{GP} & \cellcolor{gree!30}\texttt{DCT} & \cellcolor{gree!30}\texttt{DCT} & \cellcolor{gree!30}\texttt{DCT} & \cellcolor{gree!30}\texttt{DCT} \\
\textsc{Storm} & \texttt{DeP} & \cellcolor{gree!30}\texttt{DCT} & \cellcolor{gree!30}\texttt{DT} & \cellcolor{gree!30}\texttt{LR} & \cellcolor{gree!30}\texttt{DT} & \cellcolor{gree!30}\texttt{SVR} & \texttt{DeP} & \cellcolor{gree!30}\texttt{DCT} \\
\textsc{HSMGP} & \cellcolor{gree!30}\texttt{SVR} & \texttt{DaL} & \texttt{DaL} & \cellcolor{gree!30}\texttt{GP} & \texttt{DaL} & \cellcolor{gree!30}\texttt{DCT} & \cellcolor{gree!30}\texttt{SVR} & \cellcolor{gree!30}\texttt{DCT} \\
\textsc{XGBoost} & \texttt{HIP} & \cellcolor{gree!30}\texttt{DCT} & \texttt{DaL} & \cellcolor{gree!30}\texttt{GP} & \texttt{DeP} & \cellcolor{gree!30}\texttt{DCT} & \cellcolor{gree!30}\texttt{DCT} & \cellcolor{gree!30}\texttt{DCT} \\
\textsc{HIPAcc} & \cellcolor{gree!30}\texttt{GP} & \cellcolor{gree!30}\texttt{DCT} & \texttt{DaL} & \cellcolor{gree!30}\texttt{SPL} & \cellcolor{gree!30}\texttt{DT} & \cellcolor{gree!30}\texttt{LR} & \texttt{HIP} & \cellcolor{gree!30}\texttt{DCT} \\
\textsc{SQLite} & \cellcolor{gree!30}\texttt{SVR} & \cellcolor{gree!30}\texttt{RF} & \texttt{HIP} & \cellcolor{gree!30}\texttt{SVR} & \cellcolor{gree!30}\texttt{DT} & \cellcolor{gree!30}\texttt{GP} & \texttt{HIP} & \cellcolor{gree!30}\texttt{DCT} \\
\textsc{JavaGC} & \cellcolor{gree!30}\texttt{DCT} & \cellcolor{gree!30}\texttt{GP} & \cellcolor{gree!30}\texttt{DCT} & \cellcolor{gree!30}\texttt{RF} & \cellcolor{gree!30}\texttt{DCT} & \cellcolor{gree!30}\texttt{DCT} & \cellcolor{gree!30}\texttt{DCT} & \cellcolor{gree!30}\texttt{DCT} \\
\textsc{Polly} & \cellcolor{gree!30}\texttt{RF} & \cellcolor{gree!30}\texttt{DCT} & \cellcolor{gree!30}\texttt{DT} & \cellcolor{gree!30}\texttt{SPL} & \cellcolor{gree!30}\texttt{DT} & \cellcolor{gree!30}\texttt{DCT} & \cellcolor{gree!30}\texttt{DCT} & \cellcolor{gree!30}\texttt{DCT} \\

\bottomrule
\end{tabular}
\end{adjustbox}
\label{tab:system_rows_models}
\end{table}

%% file: Tables/matrix.tex
\begin{equation}
\label{eq:matrix2}
 \begin{blockarray}{c ccccc}
& \mathbf{v_1}  & \mathbf{v_2} & \cdots & \mathbf{v_k} & \mathbf{v_{all}}  \\
\cmidrule{2-6}
\begin{block}{c [ccccc]}
\mathcal{M}_1 &  v_{1,1} &  v_{1,2} & \cdots &  v_{1,k} &  v_{1,all}  \\
\mathcal{M}_2 &  v_{2,1} & v_{2,2} & \cdots & v_{2,k} & v_{2,all}  \\
\vdots & \vdots & \vdots & \ddots & \vdots & \vdots  \\
\mathcal{M}_{n} &  v_{n,1} & v_{n,2} & \cdots & v_{n,k} & v_{n,all}  \\
\end{block}
\noalign{\vspace{-100.5ex}}
\\
\end{blockarray}
\end{equation}

%% file: Tables/option_catgory.tex
\begin{table}[t!]
\centering
\caption{Option categories and descriptions.}
\label{tb:option}
\begin{adjustbox}{width=\textwidth,center}
\begin{tabular}{llll}
\toprule
\textbf{Type} & \textbf{Category} & \textbf{Descriptions} & \textbf{Symbol} \\ \midrule
Core &Functional &  Governing the system's core logic, directly impacting execution behavior. & F1 \\
\rowcolor{gree!15}Utility &Functional &  Providing utilization functionalities applicable across multiple modules. & F2 \\
CPU & Resource & Controlling CPU usage, parallel processing, or core multi-threading. & R1 \\
\rowcolor{gree!15}Storage & Resource  & Managing persistent data storage. & R2 \\
Memory  & Resource &  Handling cache, memory pools, and buffer. & R3 \\
\rowcolor{gree!15}Queue  & Resource &  Regulating scheduling, timeouts, and delays. & R4 \\ \bottomrule
\end{tabular}
\end{adjustbox}
\end{table}

%% file: Tables/RQ3_knob_influence.tex
\begin{table}[t!]
\centering
\footnotesize
\caption{The number of influential and non-influential options with respect to option types and landscape features (totaling all systems). $\mathcal{N}$ and $\mathcal{I}$ count the non-influential and influential options, respectively.}

\setlength{\tabcolsep}{1mm}
\begin{adjustbox}{width=.5\linewidth,center}
\begin{tabular}{lcc|cc|cc|cc|cc|cc}
\toprule
\multirow{2}{*}{\textbf{Feature}} & \multicolumn{2}{c|}{\textbf{F1}} & \multicolumn{2}{c|}{\textbf{F2}} & \multicolumn{2}{c|}{\textbf{R1}} & \multicolumn{2}{c|}{\textbf{R2}} & \multicolumn{2}{c|}{\textbf{R3}} & \multicolumn{2}{c}{\textbf{R4}} \\ \cmidrule{2-13}
 & $\mathcal{N}$ & $\mathcal{I}$ & $\mathcal{N}$ & $\mathcal{I}$ & $\mathcal{N}$ & $\mathcal{I}$ & $\mathcal{N}$ & $\mathcal{I}$ & $\mathcal{N}$ & $\mathcal{I}$ & $\mathcal{N}$ & $\mathcal{I}$ \\ \midrule
\texttt{FDC} & 107 & 46 & 5 & 4 & 35 & 6 & 11 & 0 & 28 & 16 & 1 & 2 \\
\texttt{FBD} & 112 & 41 & 7 & 2 & 31 & 10 & 5 & 6 & 23 & 21 & 3 & 0 \\
\texttt{Ske} & 131 & 22 & 7 & 2 & 34 & 7 & 9 & 2 & 27 & 17 & 2 & 1 \\
\texttt{Kur} & 121 & 32 & 7 & 2 & 33 & 8 & 8 & 3 & 26 & 18 & 2 & 1 \\
\texttt{PLO} & 119 & 34 & 6 & 3 & 31 & 10 & 10 & 1 & 29 & 15 & 0 & 3 \\
\texttt{CL} & 110 & 43 & 7 & 2 & 30 & 11 & 9 & 2 & 22 & 22 & 1 & 2 \\
\texttt{MIE} & 107 & 46 & 7 & 2 & 32 & 9 & 7 & 4 & 27 & 17 & 2 & 1 \\
\texttt{NBC} & 118 & 35 & 7 & 2 & 36 & 5 & 10 & 1 & 26 & 18 & 3 & 0 \\
\midrule
{\textbf{Average $\mathbf{\#}$}} & 116 & 37 & 7 & 2 & 33 & 8 & 9 & 2 & 26 & 18 & 2 & 1 \\
{\textbf{Average $\mathbf{\%}$}} & 76\% & 24\% & 74\% & 26\% & 80\% & 20\% & 78\% & 22\% & 59\% & 41\% & 58\% & 42\% \\
\bottomrule
\end{tabular}
\end{adjustbox}
\label{tb: knob}
\end{table}

%% file: Figures/rq4-exp1.tex
    \begin{tikzpicture}[]
    \begin{axis}[
    width=2.5cm,height=7cm,
    table/col sep=comma,
        ytick={1,2,3,4,5,6,7,8,9,10,11,12,13,14,15},xticklabels={,,},  
        yticklabels={\texttt{root (F1)},\texttt{filterOff (F1)},\texttt{Files (F1)},\texttt{x (F1)},\texttt{Compression (F1)},\texttt{CompMethod (F1)},\texttt{LZMA (F1)},\texttt{LZMA2 (F1)},\texttt{PPMd (F1)},\texttt{BZip2 (F1)},\texttt{Deflate (F1)},\texttt{tmOff (R1)},\texttt{mtOff (R1)},\texttt{BlockSize (R3)}},
       legend columns=2, 
       legend style={font=\tiny,draw=none,row sep=-0.05cm,fill=none,at={(1,0.35)}},
    max space between ticks=20,
    ]

\addplot[mark=none,draw=gree] 
	coordinates {
                (0.9255793591316208,1) (0.9093762754993632,2) (0.967162184678716,3) (1.0669621696716547,4) (1.0228193139440882,5) (0.993065037099815,6) (0.8836007011615508,7) (1.0062297533188609,8) (0.9025996584123456,9) (0.9755784160444224,10) (0.9259408010952014,11) (0.9859291690044764,12) (0.992656629782522,13) (0.7063661722285657,14)
    };

    \end{axis}
    \node[align=left,anchor=west,black] (source) at (0.08,5.8){ \texttt{SVR}};
   \node[align=left,anchor=west,black] (source) at (1.1,5.8){ \texttt{LR}};
 \node[align=left,anchor=west,black] (source) at (2.1,5.8){ \texttt{GP}};
  \node[align=left,anchor=west,black] (source) at (3.1,5.8){ \texttt{DT}};
   \node[align=left,anchor=west,black] (source) at (4.1,5.8){ \texttt{RF}};
    \node[align=left,anchor=west,black] (source) at (5.08,5.8){ \texttt{DCT}};
     \node[align=left,anchor=west,black] (source) at (6.08,5.8){ \texttt{SPL}};
      \node[align=left,anchor=west,black] (source) at (7.08,5.8){ \texttt{DaL}};
       \node[align=left,anchor=west,black] (source) at (8.08,5.8){ \texttt{DeP}};
        \node[align=left,anchor=west,black] (source) at (9.1,5.8){ \texttt{HIP}};

     \begin{axis}[
    width=2.5cm,height=7cm,
    table/col sep=comma,
    xshift=1cm,
        ytick={1,2,3,4,5,6,7,8,9,10,11,12,13,14},xticklabels={,,},  
        yticklabels={,,},
       legend columns=2, 
       legend style={font=\tiny,draw=none,row sep=-0.05cm,fill=none,at={(1,0.35)}},
    max space between ticks=20,
    ]

\addplot[mark=none,draw=gree]  	coordinates {         (0.0500577879573427,1) (0.0578880893050603,2) (0.0475884475158299,3) (0.2850487811034482,4) (0.0540707606798816,5) (0.0591592668389373,6) (0.0218032708082763,7) (0.0376238204191152,8) (0.0701580529036461,9) (0.0457306672949432,10) (0.1132895612442515,11) (0.0539922208750704,12) (0.0553347568749785,13) (0.0327475581882477,14)};

    \end{axis}

     \begin{axis}[
    width=2.5cm,height=7cm,
    table/col sep=comma,
    xshift=2cm,
        ytick={1,2,3,4,5,6,7,8,9,10,11,12,13,14},xticklabels={,,},  
        yticklabels={,,},
       legend columns=2, 
       legend style={font=\tiny,draw=none,row sep=-0.05cm,fill=none,at={(1,0.35)}},
    max space between ticks=20,
    ]

\addplot[mark=none,draw=gree]  	    coordinates {
        (0.0985372672268974,1) (0.1048541148704179,2) (0.1122663627383966,3) (0.0773563582781014,4) (0.0725388782743892,5) (0.0939174881322086,6) (0.1031248625774275,7) (0.1049750505285123,8) (0.1190887198003537,9) (0.1105018208195078,10) (0.1023367897384686,11) (0.0696556655230462,12) (0.1077481380794777,13) (0.1499366430857846,14)
    };

    \end{axis}

     \begin{axis}[
    width=2.5cm,height=7cm,
    table/col sep=comma,
    xshift=3cm,
        ytick={1,2,3,4,5,6,7,8,9,10,11,12,13,14},xticklabels={,,},  
        yticklabels={,,},
       legend columns=2, 
       legend style={font=\tiny,draw=none,row sep=-0.05cm,fill=none,at={(1,0.35)}},
    max space between ticks=20,
    ]

\addplot[mark=none,draw=gree]  	    coordinates {
        (0.0649854173028018,1) (0.0520224023034537,2) (0.0911101915995923,3) (0.0722838530990738,4) (0.059722600919377,5) (0.0502038315635786,6) (0.0721598696181865,7) (0.07883861304931,8) (0.0512718609664361,9) (0.0450800592193691,10) (0.0649405097760266,11) (0.0581529954200242,12) (0.0725278690911122,13) (0.1359705869363906,14)
    };

    \end{axis}

     \begin{axis}[
    width=2.5cm,height=7cm,
    table/col sep=comma,
    xshift=4cm,
        ytick={1,2,3,4,5,6,7,8,9,10,11,12,13,14},xticklabels={,,},  
        yticklabels={,,},
       legend columns=2, 
       legend style={font=\tiny,draw=none,row sep=-0.05cm,fill=none,at={(1,0.35)}},
    max space between ticks=20,
    ]

\addplot[mark=none,draw=gree]  	    coordinates {
        (0.0811432611481268,1) (0.0528916384312898,2) (0.154091615899361,3) (0.0622931782354223,4) (0.0602708215218373,5) (0.0552873132087752,6) (0.0639964491130473,7) (0.0797771483306279,8) (0.06712527559995,9) (0.0739458748692618,10) (0.0651177824626557,11) (0.0574444715913975,12) (0.0727332691276766,13) (0.2552578708644438,14)
    };

    \end{axis}

     \begin{axis}[
    width=2.5cm,height=7cm,
    table/col sep=comma,
    xshift=5cm,
        ytick={1,2,3,4,5,6,7,8,9,10,11,12,13,14},xticklabels={,,},  
        yticklabels={,,},
       legend columns=2, 
       legend style={font=\tiny,draw=none,row sep=-0.05cm,fill=none,at={(1,0.35)}},
    max space between ticks=20,
    ]

\addplot[mark=none,draw=gree]  	    coordinates {
        (0.0785467124875922,1) (0.056590569151189,2) (0.1088243898601552,3) (0.0665626633325788,4) (0.0519223300769833,5) (0.0650556714239086,6) (0.0653286528090493,7) (0.0780338182182687,8) (0.0603354614589626,9) (0.045721406872701,10) (0.0629471614439132,11) (0.0523980475555877,12) (0.0726543830968493,13) (0.1799571985378036,14)
    };

    \end{axis}

     \begin{axis}[
    width=2.5cm,height=7cm,
    table/col sep=comma,
    xshift=6cm,
        ytick={1,2,3,4,5,6,7,8,9,10,11,12,13,14},xticklabels={,,},  
        yticklabels={,,},
       legend columns=2, 
       legend style={font=\tiny,draw=none,row sep=-0.05cm,fill=none,at={(1,0.35)}},
    max space between ticks=20,
    ]

\addplot[mark=none,draw=gree]  	    coordinates {
        (0.0580798508030623,1) (0.0651554693117844,2) (0.0487147479286183,3) (0.3141676495833386,4) (0.0525066124902066,5) (0.0540463748967307,6) (0.0173350915032737,7) (0.0346580361561933,8) (0.0745681650397392,9) (0.046159738392834,10) (0.1142427881596083,11) (0.0555954888557629,12) (0.053432349907038,13) (0.0525863179295587,14)
    };

    \end{axis}

     \begin{axis}[
    width=2.5cm,height=7cm,
    table/col sep=comma,
    xshift=7cm,
        ytick={1,2,3,4,5,6,7,8,9,10,11,12,13,14},xticklabels={,,},  
        yticklabels={,,},
       legend columns=2, 
       legend style={font=\tiny,draw=none,row sep=-0.05cm,fill=none,at={(1,0.35)}},
    max space between ticks=20,
    ]

\addplot[mark=none,draw=gree]      coordinates {
        (0.1156683048829221,1) (0.115447718240386,2) (0.0638857188584496,3) (0.1632460859196049,4) (0.1047690919227482,5) (0.1074672639490249,6) (0.0772645176572565,7) (0.1924652707870188,8) (0.0780055509783516,9) (0.0837457737565248,10) (0.0975747416390028,11) (0.0923659510761082,12) (0.114797143362433,13) (0.178682041432354,14)
    };

    \end{axis}

     \begin{axis}[
    width=2.5cm,height=7cm,
    table/col sep=comma,
    xshift=8cm,
        ytick={1,2,3,4,5,6,7,8,9,10,11,12,13,14},xticklabels={,,},  
        yticklabels={,,},
       legend columns=2, 
       legend style={font=\tiny,draw=none,row sep=-0.05cm,fill=none,at={(1,0.35)}},
    max space between ticks=20,
    ]

\addplot[mark=none,draw=gree]      coordinates {
        (0.1609492708711433,1) (0.153882413677678,2) (0.058754304088009,3) (0.083057368812052,4) (0.1236754877039159,5) (0.1324085957281298,6) (0.1421508187069643,7) (0.2375601026861755,8) (0.1221142003550012,9) (0.116040428800254,10) (0.135077083222928,11) (0.1662656260897352,12) (0.1765306541044614,13) (0.1647131726957549,14)
    };

    \end{axis}

     \begin{axis}[
    width=2.5cm,height=7cm,
    table/col sep=comma,
    xshift=9cm,
        ytick={1,2,3,4,5,6,7,8,9,10,11,12,13,14},xticklabels={,,},  
        yticklabels={,,},
       legend columns=2, 
       legend style={font=\tiny,draw=none,row sep=-0.05cm,fill=none,at={(1,0.35)}},
    max space between ticks=20,
    ]

\addplot[mark=none,draw=gree]  	   coordinates {
        (0.1222530165189617,1) (0.1130407556259435,2) (0.0581228607267299,3) (0.1011110076527325,4) (0.077944864324463,5) (0.1177406150531699,6) (0.079826367122936,7) (0.1846057344305106,8) (0.1232149987464203,9) (0.1318855000284516,10) (0.0979306122304761,11) (0.0840649321552457,12) (0.1177518888501483,13) (0.3333511560656249,14)
    };

    \end{axis}
    
    \end{tikzpicture}

%% file: Figures/rq4-exp2.tex
    \begin{tikzpicture}[]
    \begin{axis}[
    width=2.5cm,height=7cm,
    table/col sep=comma,
        ytick={1,2,3,4,5,6,7,8,9,10,11,12,13,14,15},xticklabels={,,},  
        yticklabels={\texttt{root (F1)},\texttt{activitie (F1)},\texttt{trackCount (F1)},\texttt{fsync (R2)},\texttt{synCommit (R2)},\texttt{pageWrite (R2)},\texttt{shareBuffer (R3)},\texttt{tempBuffer (R3)},\texttt{workMem (R3)}},
       legend columns=2, 
       legend style={font=\tiny,draw=none,row sep=-0.05cm,fill=none,at={(1,0.35)}},
    max space between ticks=20,
    ]

\addplot[mark=none,draw=gree] 
    coordinates {
        (1.62583626020368,1) (1.6301478610504403,2) (1.2337864953014477,3) (1.1212434719285225,4) (1.876842149384839,5) (1.2505042975141154,6) (1.4754401218523197,7) (1.0723274671358651,8) (1.2235745767591806,9)
    };

    \end{axis}
    \node[align=left,anchor=west,black] (source) at (0.08,5.8){ \texttt{SVR}};
   \node[align=left,anchor=west,black] (source) at (1.1,5.8){ \texttt{LR}};
 \node[align=left,anchor=west,black] (source) at (2.1,5.8){ \texttt{GP}};
  \node[align=left,anchor=west,black] (source) at (3.1,5.8){ \texttt{DT}};
   \node[align=left,anchor=west,black] (source) at (4.1,5.8){ \texttt{RF}};
    \node[align=left,anchor=west,black] (source) at (5.08,5.8){ \texttt{DCT}};
     \node[align=left,anchor=west,black] (source) at (6.08,5.8){ \texttt{SPL}};
      \node[align=left,anchor=west,black] (source) at (7.08,5.8){ \texttt{DaL}};
       \node[align=left,anchor=west,black] (source) at (8.08,5.8){ \texttt{DeP}};
        \node[align=left,anchor=west,black] (source) at (9.1,5.8){ \texttt{HIP}};

     \begin{axis}[
    width=2.5cm,height=7cm,
    table/col sep=comma,
    xshift=1cm,
        ytick={1,2,3,4,5,6,7,8,9,10,11,12,13,14},xticklabels={,,},  
        yticklabels={,,},
       legend columns=2, 
       legend style={font=\tiny,draw=none,row sep=-0.05cm,fill=none,at={(1,0.35)}},
    max space between ticks=20,
    ]

\addplot[mark=none,draw=gree]  	    coordinates {
        (0.4911337068340424,1) (0.2268196529224169,2) (0.1617994788623016,3) (-0.2049191397310825,4) (0.1282734356221371,5) (0.2839706567754927,6) (-0.0566784288461925,7) (-0.0474889768658925,8) (-0.0143008843898759,9)
    };

    \end{axis}

     \begin{axis}[
    width=2.5cm,height=7cm,
    table/col sep=comma,
    xshift=2cm,
        ytick={1,2,3,4,5,6,7,8,9,10,11,12,13,14},xticklabels={,,},  
        yticklabels={,,},
       legend columns=2, 
       legend style={font=\tiny,draw=none,row sep=-0.05cm,fill=none,at={(1,0.35)}},
    max space between ticks=20,
    ]

\addplot[mark=none,draw=gree]  	       coordinates {
        (-2.399123411198476,1) (-2.4742045533931107,2) (-2.545825863562188,3) (-2.478091404554487,4) (-2.5196781951566,5) (-2.4963344931835145,6) (-2.415346098400927,7) (-2.379295968909644,8) (-2.391437398902854,9)
    };

    \end{axis}

     \begin{axis}[
    width=2.5cm,height=7cm,
    table/col sep=comma,
    xshift=3cm,
        ytick={1,2,3,4,5,6,7,8,9,10,11,12,13,14},xticklabels={,,},  
        yticklabels={,,},
       legend columns=2, 
       legend style={font=\tiny,draw=none,row sep=-0.05cm,fill=none,at={(1,0.35)}},
    max space between ticks=20,
    ]

\addplot[mark=none,draw=gree]  	       coordinates {
        (-2.027866834840201,1) (-2.2311689702077797,2) (-2.2112412468787626,3) (-2.1789108228764245,4) (-2.275909545371694,5) (-2.2036183422028914,6) (-2.2936946655471955,7) (-2.2862768498656054,8) (-2.316791803722704,9)
    };

    \end{axis}

     \begin{axis}[
    width=2.5cm,height=7cm,
    table/col sep=comma,
    xshift=4cm,
        ytick={1,2,3,4,5,6,7,8,9,10,11,12,13,14},xticklabels={,,},  
        yticklabels={,,},
       legend columns=2, 
       legend style={font=\tiny,draw=none,row sep=-0.05cm,fill=none,at={(1,0.35)}},
    max space between ticks=20,
    ]

\addplot[mark=none,draw=gree]  	       coordinates {
        (-1.047790634296874,1) (-1.3791764322906812,2) (-1.302107155692803,3) (-2.029595090713649,4) (-1.4299381993978275,5) (-1.3679545750311353,6) (-1.499577324493395,7) (-1.5677818889824224,8) (-1.5385756301200428,9)
    };

    \end{axis}

     \begin{axis}[
    width=2.5cm,height=7cm,
    table/col sep=comma,
    xshift=5cm,
        ytick={1,2,3,4,5,6,7,8,9,10,11,12,13,14},xticklabels={,,},  
        yticklabels={,,},
       legend columns=2, 
       legend style={font=\tiny,draw=none,row sep=-0.05cm,fill=none,at={(1,0.35)}},
    max space between ticks=20,
    ]

\addplot[mark=none,draw=gree]  	        coordinates {
        (-0.9714023329199196,1) (-1.417295975222604,2) (-1.4246550560546505,3) (-0.8948061137270309,4) (-1.4667298350703732,5) (-1.497286479318118,6) (-1.481142681172798,7) (-1.599832327270899,8) (-1.5594864095321668,9)
    };

    \end{axis}

     \begin{axis}[
    width=2.5cm,height=7cm,
    table/col sep=comma,
    xshift=6cm,
        ytick={1,2,3,4,5,6,7,8,9,10,11,12,13,14},xticklabels={,,},  
        yticklabels={,,},
       legend columns=2, 
       legend style={font=\tiny,draw=none,row sep=-0.05cm,fill=none,at={(1,0.35)}},
    max space between ticks=20,
    ]

\addplot[mark=none,draw=gree]  	     coordinates {
        (0.49453482503711,1) (0.2241659476334661,2) (0.1862007012277851,3) (-0.2000808536831431,4) (0.1293099335438574,5) (0.2700445021359225,6) (-0.0495593883844867,7) (-0.0320228644114894,8) (-0.0116948399638121,9)
    };

    \end{axis}

     \begin{axis}[
    width=2.5cm,height=7cm,
    table/col sep=comma,
    xshift=7cm,
        ytick={1,2,3,4,5,6,7,8,9,10,11,12,13,14},xticklabels={,,},  
        yticklabels={,,},
       legend columns=2, 
       legend style={font=\tiny,draw=none,row sep=-0.05cm,fill=none,at={(1,0.35)}},
    max space between ticks=20,
    ]

\addplot[mark=none,draw=gree]       coordinates {
        (-2.0149064963168986,1) (-2.235552081283646,2) (-2.3120137956539155,3) (-1.5053210039360267,4) (-2.184598809813383,5) (-2.2243265821627896,6) (-2.3507006015788416,7) (-2.311559706051826,8) (-2.180692735032169,9)
    };

    \end{axis}

     \begin{axis}[
    width=2.5cm,height=7cm,
    table/col sep=comma,
    xshift=8cm,
        ytick={1,2,3,4,5,6,7,8,9,10,11,12,13,14},xticklabels={,,},  
        yticklabels={,,},
       legend columns=2, 
       legend style={font=\tiny,draw=none,row sep=-0.05cm,fill=none,at={(1,0.35)}},
    max space between ticks=20,
    ]

\addplot[mark=none,draw=gree]         coordinates {
        (-1.808877114711694,1) (-2.0475185627830337,2) (-1.6887136956152868,3) (0.0249613501937608,4) (-1.8685687148623,5) (-1.8124667021471128,6) (-2.274200700168466,7) (-1.858678563211759,8) (-1.7510054783593163,9)
    };

    \end{axis}

     \begin{axis}[
    width=2.5cm,height=7cm,
    table/col sep=comma,
    xshift=9cm,
        ytick={1,2,3,4,5,6,7,8,9,10,11,12,13,14},xticklabels={,,},  
        yticklabels={,,},
       legend columns=2, 
       legend style={font=\tiny,draw=none,row sep=-0.05cm,fill=none,at={(1,0.35)}},
    max space between ticks=20,
    ]

\addplot[mark=none,draw=gree]  	      coordinates {
        (-1.235554319894394,1) (-1.3067404706036685,2) (-1.135488238973415,3) (-0.4299027010227472,4) (-1.2349539860603516,5) (-1.3203181691370052,6) (-1.3939900451477842,7) (-1.0762752365716604,8) (-1.596238366983217,9)
    };

    \end{axis}
    
    \end{tikzpicture}

%% file: Tables/neighour.tex
\begin{table}[t]
\centering
\adjustbox{max width=\textwidth}{
\begin{tabular}{llll}
\toprule
\textbf{Feature} & \textbf{Comparison} & \textbf{$\rho$} & \textbf{PFR}\\
\midrule
\texttt{CL} & No treatment vs. Numeric step & 0.8478 & 0.0942 \\
\texttt{CL} & No treatment vs. Adjacent bin & 0.8067 & 0.1064 \\
\texttt{PLO} & No treatment vs. Numeric step & 0.9630 & 0.0331  \\
\texttt{PLO} & No treatment vs. Adjacent bin & 0.9601 & 0.0332  \\

\bottomrule
\end{tabular}
}
\caption{Consistency scores by averaging over all systems (No treatment means no special treatment applies to the numeric options).}
\label{tab:ablation-mean-stability}
\end{table}

%% file: assessment.tex

\section{Insights from the Findings for Configuration Tuning}
\label{Sec:Insights}

The findings from our empirical study are not only confirmatory but also exploratory, both of which could influence future research directions of the field. Specifically, the most impactful insight is perhaps the confirmation that:

\begin{tcolorbox}[enhanced,boxrule=0.2mm, title=\textbf{\textit{Finding 1 $\rightarrow$ Insight 1}},
colframe=black,colback=white,colbacktitle=white,
coltitle=black,attach boxed title to top center=
{yshift=-0.25mm-\tcboxedtitleheight/2,yshifttext=2mm-\tcboxedtitleheight/2},
boxed title style={boxrule=-0.2mm,
frame code={ \path[tcb fill frame] ([xshift=-3mm]frame.west)
-- (frame.north west) -- (frame.north east) -- ([xshift=3mm]frame.east)
-- (frame.south east) -- (frame.south west) -- cycle; },
interior code={ \path[tcb fill interior] ([xshift=-2.1mm]interior.west)
-- (interior.north west) -- (interior.north east)
-- ([xshift=2.1mm]interior.east) -- (interior.south east) -- (interior.south west)
-- cycle;} }]
Jointly considering the deviation to the true overall structure (global feature) and severity of local optima (local feature) from the a particular combination of the features from the fitness landscape can serve as a good proxy to estimate and evaluate the model usefulness for configuration tuning.
\end{tcolorbox}

We have also revealed an extended explanation to the observations of ``accuracy can lie'' in prior work \cite{accuracy_can_lie}, that is:

\begin{tcolorbox}[enhanced,boxrule=0.2mm, title=\textbf{\textit{Finding 2 $\rightarrow$ Insight 2}},
colframe=black,colback=white,colbacktitle=white,
coltitle=black,attach boxed title to top center=
{yshift=-0.25mm-\tcboxedtitleheight/2,yshifttext=2mm-\tcboxedtitleheight/2},
boxed title style={boxrule=-0.2mm,
frame code={ \path[tcb fill frame] ([xshift=-3mm]frame.west)
-- (frame.north west) -- (frame.north east) -- ([xshift=3mm]frame.east)
-- (frame.south east) -- (frame.south west) -- cycle; },
interior code={ \path[tcb fill interior] ([xshift=-2.1mm]interior.west)
-- (interior.north west) -- (interior.north east)
-- ([xshift=2.1mm]interior.east) -- (interior.south east) -- (interior.south west)
-- cycle;} }]
The accuracy metrics and landscape features provide generally diverse aspects of information regarding the model usefulness for configuration tuning, hence they should be jointly considered.
\end{tcolorbox}

Finally, the results from diverse cases, including different systems and types of options, have implied that:

\begin{tcolorbox}[enhanced,boxrule=0.2mm, title=\textbf{\textit{Findings 2-5 $\rightarrow$ Insight 3}},
colframe=black,colback=white,colbacktitle=white,
coltitle=black,attach boxed title to top center=
{yshift=-0.25mm-\tcboxedtitleheight/2,yshifttext=2mm-\tcboxedtitleheight/2},
boxed title style={boxrule=-0.2mm,
frame code={ \path[tcb fill frame] ([xshift=-3mm]frame.west)
-- (frame.north west) -- (frame.north east) -- ([xshift=3mm]frame.east)
-- (frame.south east) -- (frame.south west) -- cycle; },
interior code={ \path[tcb fill interior] ([xshift=-2.1mm]interior.west)
-- (interior.north west) -- (interior.north east)
-- ([xshift=2.1mm]interior.east) -- (interior.south east) -- (interior.south west)
-- cycle;} }]
It is unlikely to have a generally most useful model for configuration tuning, but rather,  it depends on the actual systems and the option types thereof.
\end{tcolorbox}

%% file: approach.tex
\section{Fine-grained Prediction of Model Usefulness for Configuration Tuning via \approach}
\label{sec:approach}

\subsection{Motivation and Problem Formulation}

While \textbf{RQ1} has confirmed our theory of landscape dominance on model usefulness for tuning, it remains coarse-grained:

\begin{itemize}
    \item We do not know which exact pair of the global and local features is the most effective.
    \item For two non-dominated models (e.g., $\mathcal{A}$ has better $\Delta g$ but with less preferred $l$ than $\mathcal{B}$), it is unclear which one is more useful.
    \item There is a lack of tool support to estimate the model usefulness without having to profile the models paired with different tuners. 
\end{itemize}

Therefore, what is really helpful for the practitioners of configuration tuning is a fine-grained approach that can predict the relative model usefulness for a target system with a set of possible tuners, hence a more appropriate one can be chosen without the need to conduct expensive tuning and measurements. To that end, we formalize a learning to rank problem as:
\begin{equation}
\min \sum^n_{i=1}\mathcal{L}(\mathbf{F_i},r_i)    
\end{equation}
whereby the goal is to minimize the loss $\mathcal{L}$ using a vector of features $\mathbf{F_i}$ with respect to the relative ranking of the $i$th model-tuner pair ($r_i$) across a total of $n$ pairs.

Drawing on our insights from Section~\ref{Sec:Insights}, in what follows, we delineate \approach, an approach that predictably ranks the more useful models and tuners for tuning a target system.


\subsection{\approach~Designs}

\subsubsection{Features for Prediction}

\approach~considers three categories of features in the learning to rank model usefulness:

\input{Tables/tuner-features}
\input{Tables/pred-samples}

\begin{itemize}
    \item \textbf{Landscape Features $\mathbf{F_l}$:} Drawing on \textbf{Insight 1}, we consider the relative deviation between the model emulated landscape and that of the real system for one global feature and one local feature, i.e., $v_{model} - v_{system}$. For the former, we found that taking the direction of deviation into account enables better prediction accuracy than purely the absolute deviation. For the latter, a relative deviation can still serve the same purpose as directly using the local feature of the model-emulated landscape, but with better numeric stability. For example, we still prefer a smaller value for \texttt{PLO} regardless if the relative deviation is used. Another advantage is that this could incorporate knowledge of the real system (in terms of its landscape), which we have found to be beneficial. \approach~uses \texttt{Kur} with \texttt{MIE} for sequential model-based tuners and \texttt{Ske} with \texttt{PLO} for batch model-based tuners as the respective global and local features for all systems by default, as our experiments suggest that they provide the best prediction in general for their tuner types\footnote{We found that all combinations of the global and local features perform much better than the random ranking baseline, but \texttt{Kur} with \texttt{MIE} and \texttt{Ske} with \texttt{PLO} lead to particularly promising results.}.

      \item \textbf{Accuracy Metrics $\mathbf{F_a}$:} Here, we consider both the {MAPE} and $\mu$RD as the features for accuracy since they provide different aspects of information compared with the landscape features according to \textbf{Insight 2}.

     \item \textbf{Tuner Characteristics $\mathbf{F_t}$:} The model usefulness is inevitably tuner dependent; thus, we further classify the characteristics of tuners based on those studied in this work, as shown in Table~\ref{tb:tuner-f}. Note that we use one-hot encoding for this due to its highly categorical nature, but the values can be easily extended and retrain \approach. For example, \texttt{BOCA} \cite{BOCA} is a sequential model-based tuner that uses Gini score for option reduction; Expected Improvement (EI) as the acquisition function to be searched by selective exploration, in which only the important options are explored fully. Therefore, following the order in Table~\ref{tb:tuner-f}, it is represented as $\{\langle0,1,0,0\rangle\,\langle1,0,0,0\rangle\,\langle0,0,0,1\rangle\}$.
     
\end{itemize}

\begin{figure}[t!]
\centering
\includegraphics[width=0.6\columnwidth]{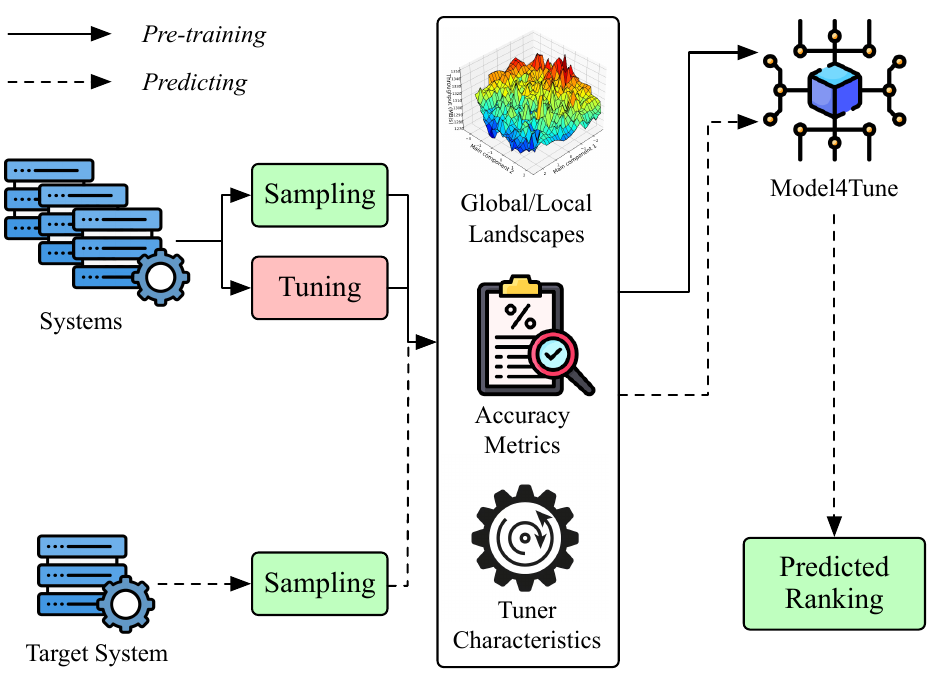}
\caption{\approach~workflow.}
\label{fig:m4t-workflow}
\end{figure}

\subsubsection{Training Procedure}

As shown in Figure~\ref{fig:m4t-workflow}, \approach~uses \texttt{LambdaRank} \cite{DBLP:conf/nips/KeMFWCMYL17} as the underlying learning to rank algorithm, which is pre-trained following the steps below:

\begin{enumerate}[label=(\roman*)]
    \item Pick a system.
    \item Sample a set of configuration-performance measurements (\textbf{\textit{sampling}}) and run different tuner-model pairs to collect their tuning performance, e.g., runtime, with repeated runs, i.e., \textbf{\textit{tuning}}, in order to obtain their relative ranks $\mathbf{y}$ for training (the same as \textbf{RQ1}). 
    \item Select a subset of configuration-performance measurements to train different surrogate models, then test their landscape features ($\mathbf{F_l}$) and accuracy metrics ($\mathbf{F_a}$) using the remaining measurements. The results are paired with the tuner's characteristics ($\mathbf{F_t}$) and the ranked performance of the corresponding tuner-model pair under the system (on average across the runs). Since the sequential and batch model-based tuners are rather different in terms of how the surrogate model is leveraged and the system measurements, we independently train two separate prediction models for them. The above is important, as \textbf{Insight 3} suggests that the most useful model for tuning is system and/or options-dependent.
    \item Repeat (i) until data for all available systems have been collected.
    \item Merge all data into a dataset to train \texttt{LambdaRank} as the example shown in Table~\ref{tb:train-example}.
\end{enumerate}

Note that since \approach~works on generic features, it naturally works as a cross-system prediction model. Upon prediction, one merely needs to collect some samples of configuration-performance measurements from the target system to train the models to be chosen from and compute their landscape/accuracy features (\textbf{\textit{sampling}}), but there is no need to profile the tuners with the models via \textbf{\textit{tuning}}; \approach~would then output the ranks of different model-tuner pairs.

\subsection{Evaluation}

\subsubsection{Setup}
To evaluate \approach, we compare it against a random guess that randomly ranks the model-tuner pairs, \mytag{R2C5}\revision{i.e., it produces a list of ranking for the pair considered in random order,} under two common metrics \cite{DBLP:journals/tois/JarvelinK02,liu2009learning}:
\begin{itemize}
    \item \textbf{Normalized Discounted Cumulative Gain (NDCG)} is a normalized metrics that assess the relative correct order of the top $k$ items in the ranking returned.
    \item \textbf{Average Precision (AP)} evaluates the ability of returning correct ranks in the top $k$ results while placing more relevant items at the top. Since AP requires a notion of ``relevance'', we say that only the true top $50\%$ of the model-tuner pairs are relevant.
\end{itemize}

Since we consider $10$ models and $16$ tuners (as in \textbf{RQ1}), there are $10 \times 16=160$ model-tuner pairs to rank for each system. We perform the actual tuning when pairing each model for every tuner on a system under the same procedure in \textbf{RQ1}, which serves as the ground truth of ranking data. The $k=1$, $k=10$, $k=20$, and full ranks are reported for both metrics. 

We conduct leave-one-out evaluation across the systems: all ranking data of $n-1$ systems are used to pre-train \approach~and it is tested over all the ranking data from the remaining one system. For each system (the training and testing ones), we build and test each surrogate model the same way as \textbf{RQ1-3}, which gives us the values of $\mathbf{F_l}$ and $\mathbf{F_a}$ used in training and prediction. All the experiments are repeated $30$ times, and we use Wilcoxon rank-sum test to verify the statistical significance. In total, we have $18$ systems $\times$ $2$ metrics $\times$ $4$ values of $k=144$ cases. \mytag{R1C6}\revision{Note that the 30 repeats use different random seeds that change the sampled configurations used to compute the accuracy of the model and landscape features, which are important features for \approach~to learn the ranking. Further, the stochastic behaviors in the learning-to-rank process can also differ.}

\subsubsection{Results}

\input{Tables/predict_perf_model}
\input{Tables/predict_perf_free}

As can be seen from Tables~\ref{tab: SMBT_prediction} and~\ref{tab: BMBT_prediction}, \approach~performs significantly better than a random guess, showing clear advances in $82\%$ ($118$ out of $144$ cases) and  $79\%$ ($114$ out of $144$ cases) for sequential and batch model-based tuners, respectively, with the majority of them exhibiting statistical significance. In particular, we see that with NDCG@1 and AP@1, \approach~achieves perfect prediction for some systems, meaning that it ranks the best model-tuner pairs correctly. As for the improvement over random ranking, the magnitude is also remarkable: up to $152.3\%$ (\texttt{PostgreSQL}) for NDCG@1 and $201.7\%$ (\texttt{XGBoost}) for AP@10 on sequential model-based tuners and up to $141.9\%$ (\texttt{Polly}) for NDCG@1 and $243.7\%$ (\texttt{SQLite}) for AP@10 on batch model-based tuners. The above suggests the practical effectiveness of \approach~in assisting practitioners for model-based configuration tuning.

\subsection{How \approach~can Help?}

We do not expect that \approach~would replace human expertise when designing configuration tuners, though, but what we anticipate is that it can serve as a powerful tool to assist practitioners in building better model-based tuners for an unforeseen system, saving the significant efforts that would otherwise be required for empirical verification. As such, we envisage several application scenarios of \approach~based on the extent of prior knowledge available:

\begin{itemize}
    \item \textbf{Understand the generally ``best'' model-tuner pair.} While in our results, we have not found a generally best model-tuner pair, in practice, it is possible that such a pair can be identified for a smaller set of systems. \approach~would be able to quickly provide an initial idea as to which pair is more promising across, e.g., a category of systems, hence providing insights on tuner designs for the target system when none of its data is available.

    \item \textbf{Foresee the promising model-tuner pairs for a target system}. When it is possible to measure configuration performance samples for a target system, \approach~can then directly predict the promising model-tuner pair(s). One can then choose to follow the same pairs, or at least the output serves as a foundation to reason about different design alternatives.

    \item \textbf{Predict the useful models under a tuner for a target system.} It is not uncommon that the tuner has already been decided for a target system. As such, with the measured data, \approach~can also hint at the more useful models for the selected tuner without the need to investigate all the alternatives.
\end{itemize}

\subsection{Overhead Saving}

\mytag{R2C6}\revision{The key overhead of profiling configurable systems lies in the measurements, for which the exact cost differs depending on systems, but is often expensive: taking \textsc{PostgreSQL} as an example, measuring the performance of one configuration can take around $30$ minutes. As such, for one model-tuner pair under a budget of 100, this is already taking $\approx 3,000$ minutes, let alone the time required to train the models. The above needs to be repeated for every model-tuner pair, hence the overhead increases linearly with more pairs to be examined. In general, the measurements constitute the majority of the overhead.}

\mytag{R2C6}\revision{\approach, as a predictive method, is designed to mitigate such an overhead: its pre-trained nature merely incurs a one-off cost, and it is the cross-system transferability that makes it cost-effective. Upon prediction, one would need to sample a small number of configurations for measurement, which is used to compute the landscape and train individual models. As such, unlike the traditional case where we need to do an exhaustive profiling, the same set of sampled configuration performance can be used for predicting the ranking of different model-tuner pairs, reducing a significant amount of overhead.}

\subsection{On The Best Model-Tuner Pair}

\mytag{R2C7}\revision{We have shown in our prior work~\cite{accuracy_can_lie} that the optimal model-tuner can differ from different systems, as part of the ground truth. As such, the predicted best model-tuner would also be expected to be different across systems. Indeed, if a system is changed to a different version, its configuration landscape might also be changed, in which case it can be treated as a new system, and the proposed \approach~can be reused again. The resulting best model-tuner might also be changed.}

%% file: Tables/tuner-features.tex
\begin{table*}[t!]
\centering
\footnotesize
\caption{Characteristics of tuners as the features. Note that the values can be extended as needed since it is easy to retrain \approach.}
\begin{adjustbox}{width=\linewidth,center}
\begin{tabular}{lll}
\toprule

\textbf{Feature} & \textbf{Values} & \textbf{Description}   \\ \midrule
\rowcolor{lightgray}\multicolumn{3}{c}{\textbf{Sequential model-based}} \\\midrule


Reduction & $\{$Lasso, Gini, Multi-sensitivity analysis, None$\}$& Metrics/method to reduce the dimensionality of options. \\
Acquisition & $\{$EI, UCB, Hedge, Max Mean$\}$&Acquisition function. \\
Heuristic & $\{$Greedy, Gradient descent, Local search, Selective exploration$\}$&Heuristic family that searches on the acquisition. \\
\midrule
\rowcolor{lightgray} \multicolumn{3}{c}{\textbf{Batch model-based}}  \\\midrule
Domain &$\{$Database, Hyperparameter, Design models, General$\}$&Tuner with domain specific designs (common for batch model-based tuners). \\
Incremental&$\{$True, False$\}$&Whether configurations from the previous iteration are exploited.\\
Heuristic &$\{$Evolutionary, Local search, Sampling, Random$\}$&Heuristic family oof the search process.\\

\bottomrule

%
\end{tabular}
\end{adjustbox}
\label{tb:tuner-f}
\end{table*}

%% file: Tables/pred-samples.tex
\begin{table*}[t!]
\centering
\footnotesize
\caption{Exampled dataset of samples for training \approach. The best global and local landscape features for sequential model-based tuners are \texttt{Kur} and \texttt{MIE}; for batch model-based tuners, these are \texttt{Ske} and \texttt{PLO}.}
\label{tb:train-example}
\begin{adjustbox}{width=\textwidth,center}
\begin{tabular}{l||ll|ll|lll||l}
\toprule
&\multicolumn{2}{c|}{\textbf{$\mathbf{F_l}$}} & \multicolumn{2}{c|}{\textbf{$\mathbf{F_a}$}}  & \multicolumn{3}{c||}{\textbf{$\mathbf{F_t}$}} & \multirow{2}{*}{\textbf{$\mathbf{y}$}} \\

\cmidrule{2-8}

& \textbf{\texttt{Kur}/\texttt{Ske}} & \textbf{\texttt{MIE}/\texttt{PLO}} &\textbf{MAPE}  &\textbf{$\mu$RD} &\textbf{Reduction/Domain} &\textbf{Acquisition/Incremental} &\textbf{Heuristic}&   \\ \midrule

\texttt{BOCA} with \texttt{SVR} on \textsc{Apache}&0.1510/$-$0.0013&$-$0.0494/$-$0.0687&64.0293&135.5281&0100&1000&0001&1\\

\texttt{BOCA} with \texttt{LR} on \textsc{Apache}&0.2338/0.0001&0.0004/$-$0.0734&0.8130&122.0664&0100&1000&0001&5\\

$\cdots$&$\cdots$&$\cdots$&$\cdots$&$\cdots$&$\cdots$&$\cdots$&$\cdots$&$\cdots$\\

\texttt{GA} with \texttt{DaL} on \textsc{Polly}&0.0585/$-$0.0076&$-$0.0314/$-$0.0404&1.4244&6816.5666&0001&1&1000&2\\

$\cdots$&$\cdots$&$\cdots$&$\cdots$&$\cdots$&$\cdots$&$\cdots$&$\cdots$&$\cdots$\\

\bottomrule

%
\end{tabular}
\end{adjustbox}

\end{table*}

%% file: Tables/predict_perf_model.tex
\begin{table*}[t!]
\centering
\caption{Average ranking performance of \approach~over the random ranking for sequential model-based tuners. The $\alpha^{\pm\beta\%}$ in each cell means \approach~achieves $\alpha$ on a ranking metric which is $+\beta\%$ better than (or $-\beta\%$ worse than) the random ranking. $^\dagger$ and $^\star$ respectively denote $p < 0.001$ and $0.001 \leq p < 0.05$ when comparing \approach~and the random ranking. \colorbox{gree!30}{gree cells} and \colorbox{red!30}{red cells} highlight the cases where \approach~is better and worse, respectively. Results for other combinations of global and local features can be found here: \textcolor{blue}{\texttt{\href{https://github.com/ideas-labo/model4tune/blob/main/Results/Batch.pdf}{https://github.com/ideas-labo/model4tune/blob/main/Results/Batch.pdf}}}.}
\label{tab: SMBT_prediction}
\begin{adjustbox}{width=\textwidth,center}
\begin{tabular}{l|llll|llll}
\toprule
System& NDCG@1 & NDCG@10 & NDCG@20 & NDCG & AP@1 & AP@10 & AP@20 & AP \\ \hline
\textsc{Apache} & \cellcolor{red!30}0.3000\textsuperscript{-22.4\%}$^{\,\,\,}$ & \cellcolor{red!30}0.1670\textsuperscript{-64.3\%}$^\dagger$ & \cellcolor{red!30}0.1896\textsuperscript{-63.5\%}$^\dagger$ & \cellcolor{red!30}0.7151\textsuperscript{-12.5\%}$^\dagger$ & \cellcolor{red!30}0.0000\textsuperscript{-100.0\%}$^\dagger$ & \cellcolor{red!30}0.0000\textsuperscript{-100.0\%}$^\dagger$ & \cellcolor{red!30}0.0001\textsuperscript{-100.0\%}$^\dagger$ & \cellcolor{red!30}0.1722\textsuperscript{-34.5\%}$^\dagger$ \\
\textsc{7z} & \cellcolor{gree!30}0.6967\textsuperscript{+63.3\%}$^\dagger$ & \cellcolor{gree!30}0.7204\textsuperscript{+46.8\%}$^\dagger$ & \cellcolor{gree!30}0.6439\textsuperscript{+21.2\%}$^\dagger$ & \cellcolor{gree!30}0.8823\textsuperscript{+7.1\%}$^\dagger$ & \cellcolor{gree!30}0.6000\textsuperscript{+63.6\%}$^{\,\,\,}$ & \cellcolor{gree!30}0.5787\textsuperscript{+100.7\%}$^\dagger$ & \cellcolor{gree!30}0.4000\textsuperscript{+39.5\%}$^\dagger$ & \cellcolor{gree!30}0.2887\textsuperscript{+9.2\%}$^\dagger$ \\
\textsc{DConvert} & \cellcolor{gree!30}0.6700\textsuperscript{+47.8\%}$^\star$ & \cellcolor{gree!30}0.6820\textsuperscript{+41.7\%}$^\dagger$ & \cellcolor{gree!30}0.6934\textsuperscript{+35.2\%}$^\dagger$ & \cellcolor{gree!30}0.8860\textsuperscript{+8.4\%}$^\dagger$ & \cellcolor{gree!30}0.9333\textsuperscript{+86.7\%}$^\dagger$ & \cellcolor{gree!30}0.5416\textsuperscript{+69.0\%}$^\dagger$ & \cellcolor{gree!30}0.5144\textsuperscript{+78.1\%}$^\dagger$ & \cellcolor{gree!30}0.3256\textsuperscript{+25.7\%}$^\dagger$ \\
\textsc{DeepArch} & \cellcolor{gree!30}1.0000\textsuperscript{+101.3\%}$^\dagger$ & \cellcolor{gree!30}0.6927\textsuperscript{+37.0\%}$^\dagger$ & \cellcolor{gree!30}0.6751\textsuperscript{+27.5\%}$^\dagger$ & \cellcolor{gree!30}0.8900\textsuperscript{+8.0\%}$^\dagger$ & \cellcolor{gree!30}1.0000\textsuperscript{+57.9\%}$^\dagger$ & \cellcolor{gree!30}0.3828\textsuperscript{+17.6\%}$^{\,\,\,}$ & \cellcolor{gree!30}0.3510\textsuperscript{+20.9\%}$^\star$ & \cellcolor{gree!30}0.2819\textsuperscript{+8.0\%}$^\dagger$ \\
\textsc{ExaStencils} & \cellcolor{gree!30}1.0000\textsuperscript{+125.6\%}$^\dagger$ & \cellcolor{gree!30}0.7629\textsuperscript{+59.4\%}$^\dagger$ & \cellcolor{gree!30}0.7340\textsuperscript{+40.6\%}$^\dagger$ & \cellcolor{gree!30}0.9120\textsuperscript{+11.3\%}$^\dagger$ & \cellcolor{gree!30}1.0000\textsuperscript{+87.5\%}$^\dagger$ & \cellcolor{gree!30}0.7261\textsuperscript{+148.7\%}$^\dagger$ & \cellcolor{gree!30}0.6106\textsuperscript{+110.3\%}$^\dagger$ & \cellcolor{gree!30}0.3518\textsuperscript{+34.0\%}$^\dagger$ \\
\textsc{Hadoop} & \cellcolor{red!30}0.2767\textsuperscript{-32.5\%}$^{\,\,\,}$ & \cellcolor{red!30}0.2196\textsuperscript{-53.0\%}$^\dagger$ & \cellcolor{red!30}0.3888\textsuperscript{-24.4\%}$^\dagger$ & \cellcolor{red!30}0.7666\textsuperscript{-6.3\%}$^\dagger$ & \cellcolor{red!30}0.0000\textsuperscript{-100.0\%}$^\dagger$ & \cellcolor{red!30}0.0007\textsuperscript{-99.8\%}$^\dagger$ & \cellcolor{red!30}0.0795\textsuperscript{-71.2\%}$^\dagger$ & \cellcolor{red!30}0.2059\textsuperscript{-21.2\%}$^\dagger$ \\
\textsc{MariaDB} & \cellcolor{gree!30}0.6067\textsuperscript{+29.1\%}$^{\,\,\,}$ & \cellcolor{gree!30}0.6140\textsuperscript{+26.4\%}$^\dagger$ & \cellcolor{gree!30}0.6073\textsuperscript{+16.8\%}$^\dagger$ & \cellcolor{gree!30}0.8533\textsuperscript{+3.9\%}$^\dagger$ & \cellcolor{gree!30}1.0000\textsuperscript{+87.5\%}$^\dagger$ & \cellcolor{gree!30}0.6291\textsuperscript{+101.2\%}$^\dagger$ & \cellcolor{gree!30}0.4836\textsuperscript{+68.7\%}$^\dagger$ & \cellcolor{gree!30}0.3169\textsuperscript{+22.3\%}$^\dagger$ \\
\textsc{MongoDB} & \cellcolor{gree!30}0.8600\textsuperscript{+61.2\%}$^\dagger$ & \cellcolor{gree!30}0.7342\textsuperscript{+41.2\%}$^\dagger$ & \cellcolor{gree!30}0.6539\textsuperscript{+22.1\%}$^\dagger$ & \cellcolor{gree!30}0.8824\textsuperscript{+6.6\%}$^\dagger$ & \cellcolor{gree!30}1.0000\textsuperscript{+50.0\%}$^\dagger$ & \cellcolor{gree!30}0.8052\textsuperscript{+122.7\%}$^\dagger$ & \cellcolor{gree!30}0.5837\textsuperscript{+89.8\%}$^\dagger$ & \cellcolor{gree!30}0.3257\textsuperscript{+22.6\%}$^\dagger$ \\
\textsc{PostgreSQL} & \cellcolor{gree!30}0.9000\textsuperscript{+152.3\%}$^\dagger$ & \cellcolor{gree!30}0.8218\textsuperscript{+77.5\%}$^\dagger$ & \cellcolor{gree!30}0.7346\textsuperscript{+46.1\%}$^\dagger$ & \cellcolor{gree!30}0.9077\textsuperscript{+11.7\%}$^\dagger$ & \cellcolor{gree!30}1.0000\textsuperscript{+130.8\%}$^\dagger$ & \cellcolor{gree!30}0.6829\textsuperscript{+123.0\%}$^\dagger$ & \cellcolor{gree!30}0.4781\textsuperscript{+72.9\%}$^\dagger$ & \cellcolor{gree!30}0.2951\textsuperscript{+13.7\%}$^\dagger$ \\
\textsc{Redis} & \cellcolor{gree!30}0.7033\textsuperscript{+46.5\%}$^\dagger$ & \cellcolor{gree!30}0.6117\textsuperscript{+18.0\%}$^\star$ & \cellcolor{gree!30}0.6585\textsuperscript{+23.3\%}$^\dagger$ & \cellcolor{gree!30}0.8746\textsuperscript{+5.7\%}$^\dagger$ & \cellcolor{gree!30}0.8667\textsuperscript{+52.9\%}$^\star$ & \cellcolor{gree!30}0.4663\textsuperscript{+30.0\%}$^\star$ & \cellcolor{gree!30}0.4756\textsuperscript{+57.5\%}$^\dagger$ & \cellcolor{gree!30}0.3366\textsuperscript{+26.8\%}$^\dagger$ \\
\textsc{Spark} & \cellcolor{gree!30}1.0000\textsuperscript{+106.9\%}$^\dagger$ & \cellcolor{gree!30}0.8147\textsuperscript{+62.7\%}$^\dagger$ & \cellcolor{gree!30}0.8325\textsuperscript{+59.8\%}$^\dagger$ & \cellcolor{gree!30}0.9301\textsuperscript{+13.1\%}$^\dagger$ & \cellcolor{gree!30}1.0000\textsuperscript{+66.7\%}$^\dagger$ & \cellcolor{gree!30}0.5955\textsuperscript{+90.0\%}$^\dagger$ & \cellcolor{gree!30}0.5983\textsuperscript{+117.3\%}$^\dagger$ & \cellcolor{gree!30}0.3389\textsuperscript{+32.2\%}$^\dagger$ \\
\textsc{Storm} & \cellcolor{gree!30}0.8400\textsuperscript{+65.8\%}$^\dagger$ & \cellcolor{gree!30}0.9200\textsuperscript{+79.8\%}$^\dagger$ & \cellcolor{gree!30}0.8451\textsuperscript{+59.3\%}$^\dagger$ & \cellcolor{gree!30}0.9414\textsuperscript{+14.1\%}$^\dagger$ & \cellcolor{gree!30}1.0000\textsuperscript{+57.9\%}$^\dagger$ & \cellcolor{gree!30}1.0000\textsuperscript{+184.0\%}$^\dagger$ & \cellcolor{gree!30}0.8290\textsuperscript{+175.3\%}$^\dagger$ & \cellcolor{gree!30}0.3968\textsuperscript{+51.8\%}$^\dagger$ \\
\textsc{HSMGP} & \cellcolor{red!30}0.4000\textsuperscript{-20.0\%}$^{\,\,\,}$ & \cellcolor{gree!30}0.6961\textsuperscript{+37.9\%}$^\dagger$ & \cellcolor{gree!30}0.6535\textsuperscript{+24.5\%}$^\dagger$ & \cellcolor{gree!30}0.8704\textsuperscript{+5.5\%}$^\dagger$ & \cellcolor{red!30}0.3333\textsuperscript{-44.4\%}$^\star$ & \cellcolor{gree!30}0.5786\textsuperscript{+66.7\%}$^\dagger$ & \cellcolor{gree!30}0.4334\textsuperscript{+47.4\%}$^\dagger$ & \cellcolor{gree!30}0.3011\textsuperscript{+14.3\%}$^\dagger$ \\
\textsc{XGBoost} & \cellcolor{gree!30}0.8867\textsuperscript{+106.2\%}$^\dagger$ & \cellcolor{gree!30}0.9299\textsuperscript{+91.4\%}$^\dagger$ & \cellcolor{gree!30}0.8807\textsuperscript{+68.5\%}$^\dagger$ & \cellcolor{gree!30}0.9580\textsuperscript{+17.0\%}$^\dagger$ & \cellcolor{gree!30}1.0000\textsuperscript{+100.0\%}$^\dagger$ & \cellcolor{gree!30}0.9867\textsuperscript{+201.7\%}$^\dagger$ & \cellcolor{gree!30}0.8078\textsuperscript{+164.6\%}$^\dagger$ & \cellcolor{gree!30}0.4032\textsuperscript{+53.6\%}$^\dagger$ \\
\textsc{HIPAcc} & \cellcolor{red!30}0.4000\textsuperscript{-24.5\%}$^{\,\,\,}$ & \cellcolor{gree!30}0.5592\textsuperscript{+14.5\%}$^\star$ & \cellcolor{gree!30}0.5322\textsuperscript{+1.3\%}$^{\,\,\,}$ & \cellcolor{gree!30}0.8268\textsuperscript{+0.2\%}$^{\,\,\,}$ & \cellcolor{red!30}0.3333\textsuperscript{-41.2\%}$^{\,\,\,}$ & \cellcolor{gree!30}0.3632\textsuperscript{+15.8\%}$^{\,\,\,}$ & \cellcolor{gree!30}0.3076\textsuperscript{+5.9\%}$^{\,\,\,}$ & \cellcolor{gree!30}0.2653\textsuperscript{+1.4\%}$^{\,\,\,}$ \\
\textsc{SQLite} & \cellcolor{gree!30}0.4200\textsuperscript{+5.0\%}$^{\,\,\,}$ & \cellcolor{red!30}0.4333\textsuperscript{-13.6\%}$^\star$ & \cellcolor{gree!30}0.5464\textsuperscript{+1.7\%}$^{\,\,\,}$ & \cellcolor{red!30}0.8181\textsuperscript{-0.9\%}$^{\,\,\,}$ & \cellcolor{gree!30}0.8000\textsuperscript{+100.0\%}$^\star$ & \cellcolor{red!30}0.2751\textsuperscript{-16.0\%}$^{\,\,\,}$ & \cellcolor{gree!30}0.4085\textsuperscript{+33.4\%}$^\dagger$ & \cellcolor{gree!30}0.3012\textsuperscript{+12.8\%}$^\dagger$ \\
\textsc{JavaGC} & \cellcolor{gree!30}0.7017\textsuperscript{+43.2\%}$^\star$ & \cellcolor{gree!30}0.6137\textsuperscript{+24.5\%}$^\dagger$ & \cellcolor{gree!30}0.6168\textsuperscript{+18.4\%}$^\dagger$ & \cellcolor{gree!30}0.8674\textsuperscript{+5.5\%}$^\dagger$ & \cellcolor{gree!30}0.8333\textsuperscript{+56.2\%}$^\star$ & \cellcolor{gree!30}0.3906\textsuperscript{+25.3\%}$^\star$ & \cellcolor{gree!30}0.2949\textsuperscript{+6.5\%}$^{\,\,\,}$ & \cellcolor{gree!30}0.2697\textsuperscript{+4.6\%}$^\star$ \\
\textsc{Polly} & \cellcolor{red!30}0.3533\textsuperscript{-25.4\%}$^{\,\,\,}$ & \cellcolor{red!30}0.4903\textsuperscript{-3.1\%}$^{\,\,\,}$ & \cellcolor{gree!30}0.5499\textsuperscript{+1.8\%}$^{\,\,\,}$ & \cellcolor{gree!30}0.8305\textsuperscript{+0.3\%}$^{\,\,\,}$ & \cellcolor{red!30}0.4000\textsuperscript{-14.3\%}$^{\,\,\,}$ & \cellcolor{gree!30}0.3321\textsuperscript{+7.9\%}$^{\,\,\,}$ & \cellcolor{gree!30}0.3696\textsuperscript{+26.4\%}$^\dagger$ & \cellcolor{gree!30}0.2990\textsuperscript{+13.8\%}$^\dagger$ \\
\hline
\textbf{Average} & \cellcolor{gree!30}\textbf{0.6675}\textsuperscript{+45.3\%}$^\dagger$ & \cellcolor{gree!30}\textbf{0.6380}\textsuperscript{+29.5\%}$^\dagger$ & \cellcolor{gree!30}\textbf{0.6353}\textsuperscript{+21.1\%}$^\dagger$ & \cellcolor{gree!30}\textbf{0.8674}\textsuperscript{+5.5\%}$^\dagger$ & \cellcolor{gree!30}\textbf{0.7278}\textsuperscript{+39.9\%}$^\dagger$ & \cellcolor{gree!30}\textbf{0.5186}\textsuperscript{+62.2\%}$^\dagger$ & \cellcolor{gree!30}\textbf{0.4459}\textsuperscript{+53.4\%}$^\dagger$ & \cellcolor{gree!30}\textbf{0.3042}\textsuperscript{+16.2\%}$^\dagger$ \\
\hline
\end{tabular}
\end{adjustbox}
\end{table*}     

%% file: Tables/predict_perf_free.tex
\begin{table*}[t!]
\centering
\caption{Average ranking performance of \approach~over the random ranking for batch model-based tuners. The format is the same as Table~\ref{tab: SMBT_prediction}. Results for other combinations of global and local features can be found here: \textcolor{blue}{\texttt{\href{https://github.com/ideas-labo/model4tune/blob/main/Results/Batch.pdf}{https://github.com/ideas-labo/model4tune/blob/main/Results/Batch.pdf}}}.}
\label{tab: BMBT_prediction}
\begin{adjustbox}{width=\textwidth,center}
\begin{tabular}{l|llll|llll}
\toprule

System& NDCG@1 & NDCG@10 & NDCG@20 & NDCG & AP@1 & AP@10 & AP@20 & AP \\ \hline
\textsc{Apache} & \cellcolor{red!30}0.4000\textsuperscript{-14.3\%}$^{\,\,\,}$ & \cellcolor{red!30}0.4347\textsuperscript{-11.1\%}$^{\,\,\,}$ & \cellcolor{red!30}0.4013\textsuperscript{-22.7\%}$^\dagger$ & \cellcolor{red!30}0.7772\textsuperscript{-5.3\%}$^\dagger$ & \cellcolor{gree!30}0.6667\textsuperscript{+25.0\%}$^{\,\,\,}$ & \cellcolor{red!30}0.2959\textsuperscript{-2.4\%}$^{\,\,\,}$ & \cellcolor{red!30}0.1882\textsuperscript{-33.0\%}$^\dagger$ & \cellcolor{red!30}0.2142\textsuperscript{-17.3\%}$^\dagger$ \\
\textsc{7z} & \cellcolor{gree!30}0.9000\textsuperscript{+114.3\%}$^\dagger$ & \cellcolor{gree!30}0.7427\textsuperscript{+62.6\%}$^\dagger$ & \cellcolor{gree!30}0.7566\textsuperscript{+51.8\%}$^\dagger$ & \cellcolor{gree!30}0.9037\textsuperscript{+11.3\%}$^\dagger$ & \cellcolor{gree!30}1.0000\textsuperscript{+114.3\%}$^\dagger$ & \cellcolor{gree!30}0.6422\textsuperscript{+115.8\%}$^\dagger$ & \cellcolor{gree!30}0.5577\textsuperscript{+100.8\%}$^\dagger$ & \cellcolor{gree!30}0.3073\textsuperscript{+18.7\%}$^\dagger$ \\
\textsc{DConvert} & \cellcolor{gree!30}0.8767\textsuperscript{+102.3\%}$^\dagger$ & \cellcolor{gree!30}0.6128\textsuperscript{+29.4\%}$^\dagger$ & \cellcolor{gree!30}0.6292\textsuperscript{+23.9\%}$^\dagger$ & \cellcolor{gree!30}0.8698\textsuperscript{+6.6\%}$^\dagger$ & \cellcolor{gree!30}0.9333\textsuperscript{+86.7\%}$^\dagger$ & \cellcolor{gree!30}0.4967\textsuperscript{+76.3\%}$^\dagger$ & \cellcolor{gree!30}0.4358\textsuperscript{+65.5\%}$^\dagger$ & \cellcolor{gree!30}0.3021\textsuperscript{+18.8\%}$^\dagger$ \\
\textsc{DeepArch} & \cellcolor{red!30}0.3800\textsuperscript{-10.2\%}$^{\,\,\,}$ & \cellcolor{red!30}0.4307\textsuperscript{-13.6\%}$^\star$ & \cellcolor{red!30}0.4816\textsuperscript{-8.1\%}$^\star$ & \cellcolor{red!30}0.8147\textsuperscript{-0.9\%}$^{\,\,\,}$ & \cellcolor{red!30}0.0667\textsuperscript{-81.8\%}$^\star$ & \cellcolor{red!30}0.1500\textsuperscript{-53.1\%}$^\dagger$ & \cellcolor{red!30}0.1624\textsuperscript{-43.0\%}$^\dagger$ & \cellcolor{red!30}0.2496\textsuperscript{-4.6\%}$^{\,\,\,}$ \\
\textsc{ExaStencils} & \cellcolor{gree!30}0.7000\textsuperscript{+84.2\%}$^\dagger$ & \cellcolor{gree!30}0.5020\textsuperscript{+7.1\%}$^{\,\,\,}$ & \cellcolor{gree!30}0.6742\textsuperscript{+32.8\%}$^\dagger$ & \cellcolor{gree!30}0.8726\textsuperscript{+7.3\%}$^\dagger$ & \cellcolor{gree!30}1.0000\textsuperscript{+150.0\%}$^\dagger$ & \cellcolor{gree!30}0.4072\textsuperscript{+33.0\%}$^\star$ & \cellcolor{gree!30}0.5267\textsuperscript{+84.5\%}$^\dagger$ & \cellcolor{gree!30}0.3304\textsuperscript{+27.6\%}$^\dagger$ \\
\textsc{Hadoop} & \cellcolor{gree!30}0.9533\textsuperscript{+86.9\%}$^\dagger$ & \cellcolor{gree!30}0.8753\textsuperscript{+80.5\%}$^\dagger$ & \cellcolor{gree!30}0.9009\textsuperscript{+77.2\%}$^\dagger$ & \cellcolor{gree!30}0.9521\textsuperscript{+16.4\%}$^\dagger$ & \cellcolor{gree!30}1.0000\textsuperscript{+66.7\%}$^\dagger$ & \cellcolor{gree!30}0.9366\textsuperscript{+205.0\%}$^\dagger$ & \cellcolor{gree!30}0.9119\textsuperscript{+241.0\%}$^\dagger$ & \cellcolor{gree!30}0.4114\textsuperscript{+61.5\%}$^\dagger$ \\
\textsc{MariaDB} & \cellcolor{red!30}0.5000\textsuperscript{-5.7\%}$^{\,\,\,}$ & \cellcolor{gree!30}0.5230\textsuperscript{+2.2\%}$^{\,\,\,}$ & \cellcolor{gree!30}0.5499\textsuperscript{+2.6\%}$^{\,\,\,}$ & \cellcolor{gree!30}0.8327\textsuperscript{+0.4\%}$^{\,\,\,}$ & \cellcolor{gree!30}1.0000\textsuperscript{+57.9\%}$^\dagger$ & \cellcolor{red!30}0.3381\textsuperscript{-6.9\%}$^{\,\,\,}$ & \cellcolor{red!30}0.2831\textsuperscript{-9.8\%}$^{\,\,\,}$ & \cellcolor{red!30}0.2658\textsuperscript{-1.2\%}$^{\,\,\,}$ \\
\textsc{MongoDB} & \cellcolor{gree!30}1.0000\textsuperscript{+91.1\%}$^\dagger$ & \cellcolor{gree!30}0.9650\textsuperscript{+94.3\%}$^\dagger$ & \cellcolor{gree!30}0.8948\textsuperscript{+74.7\%}$^\dagger$ & \cellcolor{gree!30}0.9626\textsuperscript{+17.1\%}$^\dagger$ & \cellcolor{gree!30}1.0000\textsuperscript{+66.7\%}$^\dagger$ & \cellcolor{gree!30}0.9857\textsuperscript{+196.1\%}$^\dagger$ & \cellcolor{gree!30}0.8461\textsuperscript{+200.1\%}$^\dagger$ & \cellcolor{gree!30}0.4105\textsuperscript{+57.6\%}$^\dagger$ \\
\textsc{PostgreSQL} & \cellcolor{gree!30}0.5000\textsuperscript{+23.0\%}$^{\,\,\,}$ & \cellcolor{gree!30}0.5839\textsuperscript{+24.2\%}$^\dagger$ & \cellcolor{gree!30}0.5801\textsuperscript{+10.7\%}$^\star$ & \cellcolor{gree!30}0.8355\textsuperscript{+2.0\%}$^\star$ & \cellcolor{gree!30}1.0000\textsuperscript{+130.8\%}$^\dagger$ & \cellcolor{gree!30}0.6799\textsuperscript{+130.0\%}$^\dagger$ & \cellcolor{gree!30}0.5299\textsuperscript{+87.2\%}$^\dagger$ & \cellcolor{gree!30}0.2996\textsuperscript{+13.9\%}$^\dagger$ \\
\textsc{Redis} & \cellcolor{gree!30}0.9000\textsuperscript{+86.2\%}$^\dagger$ & \cellcolor{gree!30}0.8532\textsuperscript{+68.5\%}$^\dagger$ & \cellcolor{gree!30}0.8506\textsuperscript{+61.5\%}$^\dagger$ & \cellcolor{gree!30}0.9373\textsuperscript{+13.6\%}$^\dagger$ & \cellcolor{gree!30}1.0000\textsuperscript{+100.0\%}$^\dagger$ & \cellcolor{gree!30}0.7930\textsuperscript{+141.3\%}$^\dagger$ & \cellcolor{gree!30}0.6891\textsuperscript{+137.5\%}$^\dagger$ & \cellcolor{gree!30}0.3352\textsuperscript{+27.6\%}$^\dagger$ \\
\textsc{Spark} & \cellcolor{gree!30}0.6633\textsuperscript{+68.6\%}$^\dagger$ & \cellcolor{gree!30}0.5447\textsuperscript{+10.3\%}$^{\,\,\,}$ & \cellcolor{gree!30}0.5895\textsuperscript{+11.2\%}$^\star$ & \cellcolor{gree!30}0.8435\textsuperscript{+2.6\%}$^\star$ & \cellcolor{gree!30}0.6333\textsuperscript{+35.7\%}$^{\,\,\,}$ & \cellcolor{red!30}0.2687\textsuperscript{-21.7\%}$^{\,\,\,}$ & \cellcolor{red!30}0.2797\textsuperscript{-10.9\%}$^{\,\,\,}$ & \cellcolor{red!30}0.2518\textsuperscript{-6.5\%}$^{\,\,\,}$ \\
\textsc{Storm} & \cellcolor{gree!30}0.5633\textsuperscript{+44.4\%}$^\star$ & \cellcolor{gree!30}0.6238\textsuperscript{+32.7\%}$^\dagger$ & \cellcolor{gree!30}0.6346\textsuperscript{+22.6\%}$^\dagger$ & \cellcolor{gree!30}0.8601\textsuperscript{+5.2\%}$^\dagger$ & \cellcolor{gree!30}1.0000\textsuperscript{+172.7\%}$^\dagger$ & \cellcolor{gree!30}0.5301\textsuperscript{+83.1\%}$^\dagger$ & \cellcolor{gree!30}0.4491\textsuperscript{+61.6\%}$^\dagger$ & \cellcolor{gree!30}0.3042\textsuperscript{+17.3\%}$^\dagger$ \\
\textsc{HSMGP} & \cellcolor{red!30}0.4600\textsuperscript{-11.0\%}$^{\,\,\,}$ & \cellcolor{gree!30}0.5412\textsuperscript{+2.6\%}$^{\,\,\,}$ & \cellcolor{gree!30}0.6523\textsuperscript{+20.8\%}$^\dagger$ & \cellcolor{gree!30}0.8602\textsuperscript{+3.5\%}$^\dagger$ & \cellcolor{gree!30}0.8000\textsuperscript{+41.2\%}$^{\,\,\,}$ & \cellcolor{gree!30}0.4235\textsuperscript{+14.6\%}$^{\,\,\,}$ & \cellcolor{gree!30}0.4986\textsuperscript{+59.3\%}$^\dagger$ & \cellcolor{gree!30}0.3319\textsuperscript{+24.0\%}$^\dagger$ \\
\textsc{XGBoost} & \cellcolor{gree!30}0.4000\textsuperscript{+16.5\%}$^{\,\,\,}$ & \cellcolor{red!30}0.3867\textsuperscript{-18.9\%}$^\dagger$ & \cellcolor{red!30}0.4636\textsuperscript{-11.2\%}$^\dagger$ & \cellcolor{red!30}0.7986\textsuperscript{-2.3\%}$^\star$ & \cellcolor{red!30}0.0000\textsuperscript{-100.0\%}$^\dagger$ & \cellcolor{red!30}0.0041\textsuperscript{-98.6\%}$^\dagger$ & \cellcolor{red!30}0.0537\textsuperscript{-81.7\%}$^\dagger$ & \cellcolor{red!30}0.1985\textsuperscript{-25.0\%}$^\dagger$ \\
\textsc{HIPAcc} & \cellcolor{gree!30}0.8233\textsuperscript{+75.2\%}$^\dagger$ & \cellcolor{gree!30}0.9003\textsuperscript{+92.7\%}$^\dagger$ & \cellcolor{gree!30}0.8547\textsuperscript{+69.3\%}$^\dagger$ & \cellcolor{gree!30}0.9478\textsuperscript{+16.2\%}$^\dagger$ & \cellcolor{gree!30}1.0000\textsuperscript{+87.5\%}$^\dagger$ & \cellcolor{gree!30}0.9057\textsuperscript{+217.5\%}$^\dagger$ & \cellcolor{gree!30}0.7438\textsuperscript{+184.4\%}$^\dagger$ & \cellcolor{gree!30}0.4018\textsuperscript{+57.2\%}$^\dagger$ \\
\textsc{SQLite} & \cellcolor{gree!30}0.9000\textsuperscript{+119.5\%}$^\dagger$ & \cellcolor{gree!30}0.9502\textsuperscript{+107.0\%}$^\dagger$ & \cellcolor{gree!30}0.7295\textsuperscript{+43.5\%}$^\dagger$ & \cellcolor{gree!30}0.9073\textsuperscript{+11.4\%}$^\dagger$ & \cellcolor{gree!30}1.0000\textsuperscript{+150.0\%}$^\dagger$ & \cellcolor{gree!30}0.9446\textsuperscript{+243.7\%}$^\dagger$ & \cellcolor{gree!30}0.5135\textsuperscript{+94.1\%}$^\dagger$ & \cellcolor{gree!30}0.3051\textsuperscript{+18.9\%}$^\dagger$ \\
\textsc{JavaGC} & \cellcolor{gree!30}0.9000\textsuperscript{+86.2\%}$^\dagger$ & \cellcolor{gree!30}0.7052\textsuperscript{+45.7\%}$^\dagger$ & \cellcolor{gree!30}0.6490\textsuperscript{+25.3\%}$^\dagger$ & \cellcolor{gree!30}0.8871\textsuperscript{+8.1\%}$^\dagger$ & \cellcolor{gree!30}1.0000\textsuperscript{+87.5\%}$^\dagger$ & \cellcolor{gree!30}0.6097\textsuperscript{+122.0\%}$^\dagger$ & \cellcolor{gree!30}0.4509\textsuperscript{+76.5\%}$^\dagger$ & \cellcolor{gree!30}0.3038\textsuperscript{+19.4\%}$^\dagger$ \\
\textsc{Polly} & \cellcolor{gree!30}1.0000\textsuperscript{+141.9\%}$^\dagger$ & \cellcolor{gree!30}0.9212\textsuperscript{+93.6\%}$^\dagger$ & \cellcolor{gree!30}0.8433\textsuperscript{+63.2\%}$^\dagger$ & \cellcolor{gree!30}0.9498\textsuperscript{+16.0\%}$^\dagger$ & \cellcolor{gree!30}1.0000\textsuperscript{+150.0\%}$^\dagger$ & \cellcolor{gree!30}0.8921\textsuperscript{+214.9\%}$^\dagger$ & \cellcolor{gree!30}0.7095\textsuperscript{+164.9\%}$^\dagger$ & \cellcolor{gree!30}0.3706\textsuperscript{+43.0\%}$^\dagger$ \\
\hline
\textbf{Average} & \cellcolor{gree!30}\textbf{0.7122}\textsuperscript{+60.3\%}$^\dagger$ & \cellcolor{gree!30}\textbf{0.6720}\textsuperscript{+38.9\%}$^\dagger$ & \cellcolor{gree!30}\textbf{0.6742}\textsuperscript{+30.2\%}$^\dagger$ & \cellcolor{gree!30}\textbf{0.8785}\textsuperscript{+7.2\%}$^\dagger$ & \cellcolor{gree!30}\textbf{0.8389}\textsuperscript{+74.9\%}$^\dagger$ & \cellcolor{gree!30}\textbf{0.5724}\textsuperscript{+85.6\%}$^\dagger$ & \cellcolor{gree!30}\textbf{0.4905}\textsuperscript{+74.0\%}$^\dagger$ & \cellcolor{gree!30}\textbf{0.3108}\textsuperscript{+19.3\%}$^\dagger$ \\
\hline
\end{tabular}
\end{adjustbox}
\end{table*}

%% file: threats.tex
\section{Threats to Validity}
\label{sec:threats}

Here, we elaborate on the potential threats to the validity of this work. 


\subsection{Construct Threats}

   \subsubsection{Landscape Features} 

Indeed, there are many landscape features and metrics \cite{10.1145/3728954, 8832171, DBLP:series/sci/PitzerA12, DBLP:conf/gecco/ThomsonGHB24}. In this work, we use the most representative ones, with respect to both global and local features, and omit those that do not fit our purpose. While those omissions might limit the ability to characterize the configuration landscape, this study remains able to provide reasonable insights to cover the core landscape aspects, especially when fitness landscape analysis for model-based configuration tuning is unexplored. We leave a more comprehensive coverage of other landscape features as future work.


\subsubsection{Accuracy Metrics}

We consider both MAPE and $\mu$RD, which are residual and ranked metrics, respectively. Those are the most common metrics from prior work \cite{flash, DBLP:conf/sigsoft/NairMSA17,DeepPerf, DaL} and come from diverse types.

\subsubsection{Ranking Metrics}

To evaluate \approach, we use NDCG and AP, since they are widely applied to recommendation problems in software engineering. Other metrics, such as Discounted Cumulative Gain \cite{DBLP:journals/tois/JarvelinK02}, Mean Reciprocal Rank \cite{radev2002evaluating}, and Precision@K \cite{DBLP:journals/coling/Vechtomova09}, are neither ill-suited to our domain nor are they rarely used.

\subsubsection{Options Categorization}

To better interpret the landscape information with respect to domain knowledge, we have manually categorized the options. We have done so following a systematic methodology as used in a relevant empirical study in the field~\cite{DBLP:conf/icse/LiangHC25}, including code inspection and document comprehension. However, we agree that unintentional ignorance or cognitive bias is possible.

\subsubsection{Stochastic Bias} 

The surrogate models, tuners, and the proposed \approach~can exhibit stochastic behaviors. To mitigate this, we repeat each experiment $30$ times and use statistical tests, including Wilcoxon Signed Rank/Rank Sum Test and Scott-Knott ESD Test, to validate the significance. While more repetitions can reduce noise further, the 30-run setup balances statistical reliability and computational cost.
    
    
    

\subsection{Internal Threats}

    \subsubsection{Parameters for Landscape Features} 
    
    The calculation of the landscape features requires specific settings. For example, we follow prior work \cite{DBLP:conf/seams/Chen22} to generate sequences via random walks with a length of $50$ for \texttt{CL}; for \texttt{FDC} and others, we use Hamming distance to measure the distance between configurations, consistent with prior work on configurable systems~\cite{8832171, 10.1145/3728954, DBLP:conf/seams/Chen22}. Although these settings might not be optimal, they serve as a readily available rule-of-thumb.

   \subsubsection{Parameters for Models and Tuners}

    For all the settings of surrogate model and tuners, e.g., the $d$ in \texttt{DaL} and the $k$ in \texttt{BOCA}, we follow the default settings that have been used by their authors.

 \subsubsection{Budget, Training, and Testing Sample Size}

 We systematically set the budget for ensuring reasonable convergence of all tuners. For training and testing datasets, we use the common strategy for binary options, while leveraging \texttt{SPLConqueror} \cite{SPL} to generate the size for numeric options---both are standard practice from previous works \cite{DaL,accuracy_can_lie,HINNPerf}.

 Our fixed-budget evaluation compares the final tuning performance after each tuner consumes its allocated resources. Yet, some tuners may be designed to reach an acceptable configuration with fewer measurements, especially when explicit preferences/requirements are available~\cite{DBLP:journals/tosem/ChenL23a,Wang2025LQPR,DBLP:conf/acl/WangC26}. Indeed, it would be interesting to understand the model's usefulness (using dominance landscape) with respect to finding ``good enough'' configurations, e.g., configurations within $1\%$, $5\%$, and $10\%$ of the best available performance, which we leave as a future work.
    

\subsection{External Threats}

    \subsubsection{Configurable Systems} 
    
    We study $18$ configurable systems across diverse domains with configuration spaces ranging from $2.04 \times 10^3$ to $2.67 \times 10^{41}$. Indeed, considering more systems might prove fruitful.
    

 \subsubsection{Surrogate Models} 
 
 Our study considers $10$ surrogate models that are of diverse types and technical foundations. Those are also commonly used ones for model-based configuration tuning. Although such is one of the largest scales in the relevant studies \cite{DBLP:conf/nips/DreczkowskiGB23,DBLP:journals/ase/CaoBZWWSLZ24,DBLP:journals/pvldb/ZhangCLWTLC22,accuracy_can_lie}, we admit that it is still not an exhaustive list.

 \subsubsection{Tuners} 
 
 We examine $16$ most commonly used tuners, including both the sequential and batch model-based ones, spanning across different communities, e.g., database, system, and software engineering. This ensures a good diversity and generalization. Yet, indeed, newer tuners can be used to consolidate our results.

 \subsubsection{Performance Metrics}

\mytag{R2C8}\revision{Our work focuses on tuning a single performance metric. However, we would like to kindly stress that our study and the proposed \approach~do not make assumptions on the target performance metric. For example, over the 18 different systems we study, the performance metrics can differ, e.g., runtime, throughput, and maximum loads. That is, in theory, they are applicable to any performance metric that is quantifiable.}

\mytag{R2C8}\revision{We have not considered the multiple objective case, but our findings and tool can be applicable to every single objective therein, and inform the final conclusion. For example, \approach~can be used to predict the model-tuner pairs for every single objective, and then use them to inform the final choice with respect to the possible trade-offs.}

%% file: related.tex
\section{Related Work}
\label{sec:related}
In this section, we discuss the studies that are relevant to this work.

\subsection{Fitness Landscape Analysis}


\subsubsection{Fitness Landscape Analysis for Software Engineering}

Fitness landscape analysis has been applied to some other optimization problems in Software Engineering. For example, Smith-Miles and Muñoz \cite{DBLP:journals/csur/SmithMilesM23} propose the Instance Space Analysis (ISA), which is part of the fitness landscape analysis, to study average performance bias and test instance imbalance in the landscape of software testing. Similarly, Neelofar et al. \cite{DBLP:journals/tse/NeelofarSMA23} apply ISA to evaluate Search-Based Software Testing (SBST) algorithms, exploring the adequacy of common benchmarks; factors influencing SBST effectiveness; and the strengths and weaknesses of SBST techniques. Albunian et al. \cite{DBLP:conf/gecco/AlbunianFS20} empirically analyze fitness landscapes of Java classes for unit test generation; they find that neutral plateaus caused by private methods, exceptions, and boolean lags harm search while ruggedness helps, and suggest targeted improvements.

While those problems are known to be difficult to optimize as configuration tuning, their characteristics are rather different, e.g., configuration tuning is known as expensive and sparse, but this might not be the case for the others.

\subsubsection{Fitness Landscape Analysis for Configuration}


A limited number of works have considered the concept of landscape for configuration research. Examples are Lustosa et al. propose \texttt{niSNEAK}~\cite{10.1145/3630252}, which uses clustering techniques to explore the configuration space, reducing computational costs while maintaining high predictive accuracy and \texttt{SMOOTHIE} \cite{DBLP:journals/corr/abs-2401-09622}---a tool that emphasizes optimizing hyperparameters by smoothing the loss landscape. Yet, the definition of landscape in those works differs from ours in the sense that they have not exploited any well-established landscape features. The most closely related work is perhaps from Huang et al. \cite{10.1145/3728954}, which models the configuration space as a network as part of a proposed methodology, revealing the ruggedness of the configuration landscape, the distribution of local optima and the influence of feature interactions for a given system. However, they have not studied the role of surrogate models therein.

Overall, the above works differ from ours in the sense that:

\begin{itemize}
    \item they have not distinguished the global and local landscape features;
    \item they mainly focus on studying the landscape of the real system rather than that emulated by the surrogate models, and how it can be considered useful for tuning;
    \item they have not provided a readily available tool, derived from the spatial landscape information, for assisting the practitioners in the field.
\end{itemize}


\subsection{Surrogate Models for Configuration Performance}
Prior works have proposed various surrogate models to guide configuration tuning, ranging from statistical approaches~\cite{SPL, DECART} to more recent deep learning-based methods~\cite{DeepPerf, HINNPerf, DaL, DBLP:journals/pacmse/Gong024}.

For instance, \texttt{SPLConqueror}~\cite{SPL} integrates linear regression with diverse sampling techniques to capture interactions among configuration options. \texttt{DECART}~\cite{DECART} enhances the CART model with hyperparameter tuning and an efficient resampling strategy. However, these statistical models often struggle to accurately capture the sparse configuration data---a common characteristic of configurable software systems~\cite{DBLP:journals/pacmse/Gong024}. To improve accuracy, deep learning-based approaches have gained traction over the past decade for modeling configuration performance. For example, \texttt{HINNPerf} \cite{HINNPerf} utilize the embedding method and hierarchical network blocks to model the complex interactions between configuration options. Similarly, \texttt{DaL}~\cite{DaL} partitions sparse configuration data into distinct, focused subsets, training a dedicated deep neural network for each to achieve superior accuracy.

There are also studies on so-called white-box/interpretable models; however, the explainability therein is mainly about the distribution analysis of the model~\cite{muhlbauer2023analyzing} or correlating the model to the particular codebase of a system~\cite{DBLP:conf/icse/VelezJSAK21,DBLP:journals/ase/VelezJSSAK20}, rather than to the spatial information of the landscape emulated by the model as what we investigated in this work.

The suitability of models for different systems has also been studied. Zhao et al.~\cite{zhao2023automatic} note that Gaussian Processes (\texttt{GP}) effectively model continuous options but struggle with categorical ones, whereas Random Forests (\texttt{RF}) can handle both types but often lack high accuracy in low-dimensional configuration space.

Yet, the above share a common general belief: higher model accuracy can directly enhance configuration tuning performance and vice versa, which has been demonstrated as misleading in previous work~\cite{accuracy_can_lie}.

\subsection{Models-based Configuration Tuning}

Over the past decade, researchers have proposed and used various search heuristics to find better-performing configurations with a limited budget \cite{atconf,bestconfig,BOCA,conex,flash,Ottertune,llamatune,restune}.

\subsubsection{Batch Model-based Tuners}

Batch model-based tuners focus on the tuning/heuristics designs~\cite{conex, gga, bestconfig, Sway}. Among the others, \texttt{BestConfig}~\cite{bestconfig} utilizes the divide-and-diverge sampling method and the recursive bound-and-search algorithm to explore the sample space. The key is to leverage the sparse nature of the configuration space. Some other tuners treat configuration tuning as a black box tuning problem. Examples include \texttt{SWAY}~\cite{Sway}, which is an effort from the software engineering community that seeks to focus on the ``exploration'' of the space first: it selects a large set of configurations and then samples within that set to narrow down to the optimal one. \texttt{MMO}~\cite{DBLP:journals/corr/abs-2112-07303} is a multi-objectivization model that addresses the single performance search by tuning into a multi-objective one through an auxiliary objective, e.g., the other unconcerned performance metric.

The above tuning/heuristics can, but do not have to, seamlessly integrate any pre-trained surrogate model to create batch model-based tuners---a common practice in the field~\cite{DBLP:conf/sigsoft/0001L21,DBLP:journals/corr/abs-2112-07303,DBLP:journals/tosem/ChenLBY18,shi2024efficient}. For instance, Chen et al.~\cite{DBLP:journals/corr/abs-2112-07303} demonstrate that their tuner can be paired with any model to enable cost-effective tuning. Similarly, Shi et al.~\cite{shi2024efficient} employ linear regression as a surrogate to enhance the tuner's exploration capabilities.


\subsubsection{Sequential Model-based Tuners}

Sequential model-based tuners require more sophisticated search strategies, as their models must be updated during the tuning process. Numerous examples exist, primarily differing in the choice of surrogate model and its update mechanism following the Bayesian optimization framework~\cite{Ottertune,restune,flash,BOCA}. For example, \texttt{OtterTune}~\cite{Ottertune} and \texttt{ResTune}~\cite{restune} use Gaussian Processes (\texttt{GP}) as their surrogate models, while \texttt{FLASH}~\cite{flash} leverages Decision Trees (\texttt{DT}) to accelerate the search. Other tuners, such as \texttt{PromiseTune}~\cite{PromiseTune}, use Random Forest to identify promising regions/rules in the configuration landscape, which are then purified by causality inference, and finally incorporate such into a Bayesian optimization process.

Despite the widespread adoption of sequential model-based tuners, they still assume that a more accurate surrogate model is the primary factor in achieving superior tuning performance \cite{DaL,HINNPerf,DeepPerf,SPL,DBLP:conf/asplos/YuBQ18}.

\subsection{Empirical Studies on Configuration}

Various empirical studies exist on configuration from several different perspectives. 

\subsubsection{General Configuration Studies}

In general, most empirical studies on configuration focus on understanding the characteristics of configurable systems. Among others, Xu et al.~\cite{DBLP:conf/sigsoft/XuJFZPT15} demonstrate that the number of configuration options has increased significantly over the years, and as such, most developers have not yet exploited the full benefits offered by those configurations. Zhang et al.~\cite{DBLP:conf/icse/ZhangHLL0X21} investigate how configurations evolve among system versions, from which some patterns were discovered. Other studies focus on understanding the consequences of inappropriate configuration, e.g., those that lead to performance bugs~\cite{DBLP:conf/esem/HanY16}. Huang et al.~\cite{DBLP:conf/icse/LiangHC25} study the identification of performance-sensitive options and the relevant dependencies.

While those works do not target model and tuning, they provide insights about the characteristics of configurable software systems and hence are orthogonal to this study.

\subsubsection{Studies for Configuration Learning}

Since using surrogate models is often considered a promising way to relieve the expansiveness of configuration tuning, many studies have been conducted on different aspects of building such a model~\cite{DBLP:conf/splc/0003APJ21,DBLP:conf/msr/GongC22,DBLP:conf/kbse/JamshidiSVKPA17,chen2019all,DBLP:journals/ase/CaoBZWWSLZ24}. For example, Gong and Chen~\cite{DBLP:conf/msr/GongC22} study the impact of encoding on the accuracy of the model, respectively. Jamshidi et al.~\cite{DBLP:conf/kbse/JamshidiSVKPA17} investigate how a model learned in one environment, e.g. hardware or version, can be transferred to the other while maintaining a good level of accuracy. Chen~\cite{chen2019all} seeks to understand the best way to update a model when a new configuration is measured, considering complete retraining or incremental learning. In fact, those works emphasize the accuracy of the model as a key metric, but are not concerned with how the model can be used for other tasks, e.g., configuration tuning.

\subsubsection{Prior Understandings of Surrogate Models for Tuning}

Some studies have focused explicitly on comparing model-based tuners, covering general configurable systems~\cite{sayyad2013parameter} or specific domains such as database systems~\cite{DBLP:journals/pvldb/ZhangCLWTLC22,van2021inquiry} and defect predictors~\cite{DBLP:conf/icse/LiX0WT20}. For example, Zhang et al.~\cite{DBLP:journals/pvldb/ZhangWLTTLC23} study and train a meta-ranker to recommend suitable tuners for a given workload. However, the proposed method has a completely different goal compared with that of \approach: instead of predicting the best model-tuner pair for an unforeseen system/workload using landscape features, the meta-ranker seeks to infer the best tuner for a workload using its similarity to the other past workloads of the same system, together with standard features such as the number of options etc. Aken et al. \cite{van2021inquiry} replace the model chosen by \texttt{OtterTune} with alternative models and compare their tuning performance, aiming to find the most effective model to be used with \texttt{OtterTune}. Nevertheless, again, their studies neither summarizes how the accuracy of models can impact the tuning nor how/why the model would impact the tuning. Our previous work~\cite{accuracy_can_lie} reveals that the accuracy can lie, but it has not systematically and comprehensively formalized the rationals; more importantly, there still lack guidelines and supports for evaluating model uselessness for tuning beyond accuracy.

Overall, existing research has acknowledged that the differently chosen model can influence the tuning results, but there is still a lack of clear understanding regarding how to assess the usefulness of models; how/why they are important for tuning; and what role do they play therein---all of which are what we seek to disclose in this work through the paradigm of fitness landscape analysis.

%% file: conclusion.tex
\section{Conclusion}
\label{sec:conclusion}

This paper presents the first contribution to assess/explain how and why surrogate models can help (or harm) configuration tuning via fitness landscape analysis---a new perspective beyond the classic accuracy studies. We propose a new theory dubbed, landscape dominance, to better quantify the usefulness of models for tuning, based on which we conduct a large-scale empirical study (up to $27,000$ cases) and reveal that:

\begin{itemize}
    \item The landscape dominance is an effective metric for quantifying model usefulness for tuning;
    \item Landscape features provide largely different aspects of assessment compared with accuracy metrics;
    \item The model-emulated landscape is still far from being close to the real landscape, but this might still be acceptable.
    \item No single model is the most preferred for tuning based on landscape dominance;
    \item Memory and queue-related options are more influential on the landscape emulated by a model. 
\end{itemize}

Drawing on landscape dominance and other findings, we propose \approach, the first tool to predict the usefulness of model-tuner pairs for configuration tuning. Results over $18$ systems demonstrate that \approach~considerably outperforms the random guessing by $79\%-82\%$ of the cases in general with up to $244\%$ improvements.

We envisage that this work not only fills the important gap in understanding the role of surrogate models for tuning, but also provides a readily available tool that can greatly assist practitioners in designing better, more specialized tuners. Most importantly, we hope that our findings will stimulate a batch of new research directions in the field.